\documentclass[11pt,times,letter]{article}
\usepackage[margin=1in]{geometry}

\usepackage{amsmath}
\usepackage{amsfonts}
\usepackage{amssymb}
\usepackage{bbm}
\usepackage{mathtools}

\usepackage{subcaption} 
\usepackage{graphicx}
\usepackage{pgfplots}
\usepackage[all]{nowidow}
\usepackage[utf8]{inputenc}
\usepackage{tikz}
\usepackage{multicol}
\usepackage{algpseudocode,algorithm,algorithmicx}
\usepackage{scalerel}
\usepackage{natbib}
\usepackage[normalem]{ulem}

\usepackage[english]{babel}

\usepackage{mathdots}
\usepackage{yhmath}
\usepackage{cancel}
\usepackage{color}
\usepackage{array}
\usepackage{multirow}
\usepackage{gensymb}
\usepackage{tabularx}
\usepackage{booktabs}
\usetikzlibrary{fadings}
\usetikzlibrary{patterns}
\usetikzlibrary{shadows.blur}

\newtheorem{theorem}{Theorem}
\newtheorem{lemma}{Lemma}
\newtheorem{corollary}{Corollary}

\newtheorem{assumption}{Assumption}
\newtheorem{proposition}{Proposition}
\newtheorem{remark}{Remark}

\newcommand{\onep}{\mathbbm{1}_{t_{k+1}}^+}
\newcommand{\onem}{\mathbbm{1}_{t_{k+1}}^-}
\newcommand{\oned}{\mathbbm{1}_{t_{k+1}}^{\delta}}
\newcommand{\onen}{\mathbbm{1}_{t_{k+1}}}
\newcommand{\cn}{c_{t_{k+1}}}
\newcommand{\cp}{c_{t_{k+1}}^+}
\newcommand{\cm}{c_{t_{k+1}}^-}
\newcommand{\cd}{c_{t_{k+1}}^{\delta}}
\newcommand{\Pn}{p_{t_{k+1}}}
\newcommand{\Pp}{p_{t_{k+1}}^+}
\newcommand{\Pm}{p_{t_{k+1}}^-}
\newcommand{\Pd}{p_{t_{k+1}}^{\delta}}
\newcommand{\lp}{L_{t_{k}}^+}
\newcommand{\lm}{L_{t_{k}}^-}
\newcommand{\ld}{L_{t_{k}}^{\delta}}
\newcommand{\skk}{S_{t_{k+1}}}
\newcommand{\sk}{S_{t_{k}}}
\newcommand{\tkk}{{t_{k+1}}}
\newcommand{\tk}{{t_{k}}}
\newcommand{\F}{\mathcal{F}_{t_{k}}}
\newcommand{\E}{\mathbb{E}}
\newcommand{\muo}{\mu_c}
\newcommand{\muoo}{\mu_{cp}}
\newcommand{\mut}{\mu_{c^2}}
\newcommand{\muto}{\mu_{c^2p}}
\newcommand{\mutt}{\mu_{c^2p^2}}
\newcommand{\Aone}{{}^{\scaleto{(1)}{5pt}}\!A_{\tk}}

\definecolor{blue}{HTML}{1F77B4}
\definecolor{orange}{HTML}{FF7F0E}
\definecolor{green}{HTML}{2CA02C}

\definecolor{Red}{rgb}{1,0,0} 
\definecolor{DRed}{rgb}{0.7,0.3,0} 
\definecolor{Green}{rgb}{0.2,0.5,0.2} \newcommand{\Green}{\color{Green}}
\definecolor{Blue}{rgb}{0,0,1} \newcommand{\Blue}{\color{Blue}}
\definecolor{Orange}{rgb}{1,0.5,0} 
\definecolor{Black}{rgb}{0,0,0} \newcommand{\Black}{\color{Black}}

\definecolor{Grey}{rgb}{0.52, 0.52, 0.51} 
\definecolor{Purple}{rgb}{0.8,0.4, 0.8} 

\usepackage{setspace}
\begin{document}

\title{{Market Making with Stochastic Liquidity Demand: \\ Simultaneous Order Arrival and Price Change Forecasts}}
%
\author{Agostino Capponi\thanks{Department of Industrial Engineering and Operations Research, Columbia University, NY 10027, USA ({\tt ac3827@columbia.edu}).}, Jos\'e E. Figueroa-L\'opez\thanks{Department of Mathematics and Statistics, Washington University in St. Louis, St. Louis, MO 63130, USA ({\tt figueroa-lopez@wustl.edu}). Research supported in part by the NSF Grants: DMS-2015323, DMS-1613016.}, and Chuyi Yu\thanks{Department of Mathematics and Statistics, Washington University in St. Louis, St. Louis, MO 63130, USA ({\tt chuyi@wustl.edu}).}}
%
%
%
\maketitle 
\begin{abstract}
We provide an explicit characterization of the optimal market making strategy in a discrete-time Limit Order Book (LOB). In our model, the number of filled orders during each period depends linearly on the distance between the fundamental price and the market maker's limit order quotes, with random slope and intercept coefficients. The high-frequency market maker (HFM) incurs an end-of-the-day liquidation cost resulting from linear price impact. The optimal placement strategy incorporates in a novel and parsimonious way forecasts about future changes in the asset's fundamental price. We show that the randomness in the demand slope reduces the inventory management motive, and that a positive correlation between demand slope and investors' reservation prices leads to wider spreads. Our analysis reveals that the simultaneous arrival of buy and sell market orders (i) reduces the shadow cost of inventory, (ii) leads the HFM to reduce price pressures to execute larger flows, and (iii)  introduces patterns of nonlinearity in the intraday dynamics of bid and ask spreads. Our empirical study shows that the market making strategy outperforms those which ignores randomness in  demand, simultaneous arrival of buy and sell market orders, and local drift in the fundamental price.
\end{abstract}

\section{Introduction}
Since the last decade, most security trading activities have migrated to electronic markets and, as a result, high frequency trading has become one of the most significant market developments. Estimates of high frequency volumes in the treasury, foreign exchange, equity and index futures markets are typically several deciles of the total traded volume (\cite{staff2015us,markets2011high,securites2010Concept,kirilenko2017flash}). As observed by \cite{MenveldHFT_2013},  market making activities are being predominantly carried out by high-frequency traders. A traditional market maker {provides} liquidity to the exchange by continuously placing bid and ask orders, and {hence, earning profit from the bid-ask spread of her quotes}. {Like} traditional market makers, high frequency market makers (HFMs) make profit from roundtrip transactions  {but, unlike traditional market makers, they also submit} numerous passive orders that are canceled shortly after submission {at} extraordinary high-speed (\cite{securites2010Concept}). More importantly,  compared with traditional market makers, HFMs typically work at privately held firms {and, thus,} inventory control becomes necessary for them to limit the amount of capital tied up in margin accounts (\cite{menkveld2016economics}). {The practice of ending the day close to a flat position is also driven by risk management motives, as it allows the market maker to reduce uncertainty coming from fluctuations in security prices at the beginning of the next trading day.}

Existing literature has analyzed market making control problems with inventory risk. The study of \cite{Ho_Stoll_1981} considers a single period mean-variance utility for a market maker wishing to optimize expected profit from bid-ask spreads, and to find offsetting transactions to minimize inventory risk. \cite{Levi} consider risk-averse market makers with single period mean-variance or exponential utility who use a threshold inventory control policy to reduce the risk from price uncertainty. Due to the static nature of their setup, these studies do not analyze the intraday effects of orders arrival and inventory management on prices and liquidity.

Other studies have considered dynamic models of market making. Some of them aim to penalize intraday inventory holdings (see e.g. \cite{cartea2014buy}, \cite{cartea2015risk} for contributions in this space).\footnote{\cite{Amihud_Mendelson_1980} assume a dynamic model in which dealers are risk neutral and buy and sell orders arrive according to a Poisson process with price-dependent rates. They consider an infinite time horizon and restrict inventory levels to be inside a prespecified interval throughout the entire trading day.} Other studies impose explicit constraints on terminal inventory. Examples of those works include the early contributions of \cite{Bradfield}, who analyze the increasing price variability induced by strategies that target a flat end-of-day inventory level; \cite{OharaOldfield}, who consider a repeated optimal market making problem, in which each day consists of several trading periods, and the market maker maximizes utility over an infinite
number of trading days while facing end-of-day inventory costs; 
\cite{gueant2013dealing}, who sets the penalty for end of day inventory to be proportional to the absolute value of the terminal inventory; and the most recent study of \cite{adrian2016intraday}, who study intraday patterns of prices and volatility induced by inventory management motives with focus on the Treasury market.\footnote{Few empirical studies have analyzed the relationship between trades, prices and bid-ask spreads using transaction data. \cite{Glosten_Harris_1988} and \cite{Hasbrouck_1988} decompose bid-ask spreads into two components, reflecting compensation for inventory costs and adverse selection costs, which arise from the presence of informed traders. They find that, in contrast to the transitory spread component explained by inventory considerations, the permanent component explained by information asymmetries is significant for large trades but not for small ones.}

We solve a {discrete-time} optimal control problem to maximize expected cumulative profits of the market maker, {while incorporating an end-of-the-day} inventory liquidation cost. 
{Such a cost is driven by the assumption of a linear instantaneous price impact: the average price per share in liquidating an inventory of size $I_T$ at time $T$ is $S_T-\lambda I_{T}$, where $S_{T}$ is the fundamental price at time $T$ (typically, the asset's midprice) and $\lambda$ is a constant penalty.} In our framework, the HFM places limit orders (LOs) on the ask and bid sides simultaneously, and cancels the remaining unexecuted quotes shortly {before submitting new quotes in the book}. Her wealth and inventory trajectory {are hence determined by the prices of her quotes and the number of shares that are {filled or lifted} from her orders at these prices}.

{{Modeling} the number of lifted shares between consecutive actions is a key component of our framework. In continuous-time control problems, a common approach is to}  model the probability with which an incoming market order (MO) can lift one share of the HFM's LO in the book (known as `lifting probability'). For instance, \cite{cartea2014buy,cartea2015risk} assume that the {MOs arrive according to a Poisson process and 
model the lifting probability as the exponential of the negative distance of the HFM's quote from the midprice times a constant; 
\cite{cvitanic2010high} instead model the lifting probability with a linear function, assuming the MOs to be uniformly distributed on a preset price interval.  
{An alternative approach, especially predominant in discrete-time control problems, is to directly model the number of lifted} shares via a liquidity demand function.  
For instance, in their work on price pressures, \cite{hendershott2014price} assume that the liquidity demand is normally distributed with a mean that is linear in the bid and ask price. A continuous time stochastic model of a limit order book, which mimics a queuing system where limit orders are executed against market orders, has been proposed by \cite{ContStojov}. Unlike ours, their work does not consider inventory costs.\footnote{A separate stream of literature has analyzed liquidity in a limit order book with endogenous equilibrium dynamics. See, for instance, \cite{Nadotchi}.}

The exponential lifting probabilities used in continuous-time control problems {can be related to the linear demand function used in discrete problems}. Specifically, if $\rho$ is the arrival intensity of MO's and the lifting probability is set to be $\exp(-{\kappa} d)$, where $d$ is the distance between the LO price and the midprice, then, during a time span of $\Delta$, we expect that ${\Delta \rho \exp(-\kappa d)}$ times a MO will lift a LO placed at distance $d$. Since it is typically assumed that only one share of the order is lifted at a time, when $d$ is small (as it is commonly the case), the expected number of shares filled during that time span is approximately equal to $\Delta \rho-\Delta\rho \kappa d$, which is precisely linear in $d$. {Note, however, that the previous argument assumes that a LO with volume $Q\in\mathbb{N}$ is treated as $Q$ independent LOs of volume one with the same lifting probability, which is not the case in practice.}

Our work extends existing models of high-frequency market making in several ways. We assume the demand to be linear when modeling the number of filled shares from the HFM's limit orders. However, unlike \cite{adrian2016intraday} and~\cite{hendershott2014price}, the demand is not {\it deterministic but random}. This means that the actual number of shares bought or sold varies over time, even if the distances of quotes from the fundamental price stay the same. The proposed randomization not only allows for greater flexibility and better fit to empirically observed order flows, but also uncovers novel properties of the resulting optimal placement strategies. For instance, it is known from \cite{adrian2016intraday} that, under a constant demand slope, the inventory adjustment in the optimal placement at any given time decreases with the size of the slope. We show that the variance of the slope further reduces this adjustment. This implies that assets with highly volatile demand profiles require less strict inventory adjustments. We also find that the optimal placement spreads (i.e., the distances between the optimal bid and ask prices and the fundamental price) increase with the correlation between demand slope and investors' reservation price.

{To the best of our knowledge, our framework is the first to incorporate simultaneous arrivals of buy and sell MOs between consecutive market making actions. The time-discretized version of most existing models (e.g. \cite{adrian2016intraday} and \cite{cartea2015risk}), which are obtained from the continuous-time versions via Euler approximations, does not allow for this feature. We first obtain a closed form expression for the optimal strategies that explicitly account for the probability $\pi(1,1)$ of simultaneous arrivals of buy and sell MOs during each small time period. We then perform a comparative statics analysis, and discover novel patterns of optimal placements strategies. 

First, at any given time we find that the optimal bid-ask spread declines if $\pi(1,1)$ increases. The intuition is that a higher likelihood of simultaneous buy and sell orders arrival provides the HFM with more opportunities to manage inventory: the positive net position resulting from the execution of sell MOs offsets the negative net position corresponding to the execution of buy MOs. Second, we show that bid-ask spreads are less sensitive to the passage of time as $\pi(1,1)$ increases. Interestingly, if $\pi(1,1)$ is sufficiently large, the bid-ask spread no longer rises towards the end of the day. This intraday pattern stands in contrast with that identified by \cite{adrian2016intraday}, and can be understood as follows. While the need of reaching a zero
inventory target becomes stronger with passage of time, a larger arrival rate of offsetting buy and sell MOs reduces the shadow cost of {end-of-the-day} inventory and incentivizes the HFM to reduce price pressures for attracting larger flows. Third, we show that the presence of simultaneous arrivals introduces a nonlinearity in the intraday dynamics of bid spread and ask spread. In the absence of simultaneous arrivals, {the} ask spread and bid spread {(i.e., the distances of the HFM's optimal ask and bid quotes from the fundamental price)} are decreasing functions of time. However, if $\pi(1,1) \neq 0$, this monotonicity is broken as we get closer to the end of day. Last, but not least, we observe a novel {\it threshold phenomenon} in the HFM's inventory management: if inventory holdings are below a certain threshold, the HFM  widens her bid and ask spread to dampen trading activity on both sides of the market and preserve the current inventory position, instead of aggressively placing LOs close to the security price to lower her net position.

{Another distinguishing and novel feature of our study, relative to the rest of literature, is that we allow the market maker to incorporate forecasts about the fundamental price of the asset, rather than assuming martingale dynamics. We obtain a parsimonious formula which describes how the investor should adjust her limit order placements based on her asset price forecasts. Intuitively, if the HFM expects future price changes to be negative, she would reduce the bid and ask spread proportionally to the expected price change. The proportionality constant depends on the model parameters in a non-trivial way. This feature also allows the investor to take advantage of sophisticated time series- or machine learning-based forecast procedures of asset prices into the intraday market-making process} (see Section 3 for the technical details).

To the best of our knowledge, our work is unique in that it assesses the performance of the proposed market making strategy against LOB data. Specifically, we first use a rolling window approach to estimate the model parameters. We then test the calibrated model against actual LOB dynamics, allowing for adjustments of LO's placements every second and determining the cash flows and inventory changes generated during the day. At day's close, the HFM submits a MO to liquidate its final inventory, and determine the actual cost taking into account the state of the LOB. We find that the optimal placement yields, on average, larger revenue compared to the situation where $\pi(1,1)=0$ (such as in \cite{adrian2016intraday}), even if $\pi(1,1)$ is estimated to be small and about $0.05$. Our empirical analysis lends strong support to demand stochasticity: the slope coefficient has a standard deviation which is about 200\% larger than the average demand level, and a correlation of about 20\% with the investors' reservation price. Moreover, using real LOB data we estimate the optimal placement strategy based on a simple one-step ahead forecast to outperform the one that presumes a martingale price evolution. 

The solution of the optimal control problem presents nontrivial mathematical challenges. While the first-order optimality conditions involves solving a quadratic equation, establishing the second-order conditions needed for the verification theorem is intricate. It involves establishing several clever estimates, in which we leverage direct inequalities implied by the primitives of our model.

{The rest of the paper is organized as follows. In Section 2, we present the model setup together with our assumptions. Section 3 solves the Bellman equation for the control problem, and proves a verification theorem. In Section 4, we analyze in detail the main economic forces behind the optimal placement strategies. In Section 5, we measure the performance of our market making strategy against real LOB data. We delegate technical proofs to two appendices.} 

\section{{Model Setup}}\label{sec:setup}

In this section we introduce our Limit Order Book (LOB) model and specify the type of considered strategies. We assume the market making strategy runs from time $0$ to a fixed time $T>0$. The HFM places bid and ask limit orders {(LO)} simultaneously on both sides of the LOB for a given asset at preset times $0=t_0<t_1<\dots<t_N<T$. Throughout, we set $t_{N+1}=T$ and $\mathcal{T}=\{t_0,t_{1},\dots, t_{N+1}\}$. All variables introduced below are assumed to be defined on a probability space $(\Omega,\mathbb{P},\mathcal{F})$ equipped with a filtration $\{\mathcal{F}_t\}_{t\in\mathcal{T}}$, which represents the arrival of market makers' available information through time. 

Arrivals of buy and sell market orders (MO) are modeled by two Bernoulli processes.
Specifically, let $\onep$ ($\onem$) be a Bernoulli random variable indicating whether there is at least one buy (sell) market order arriving during {the time period} $[t_k,t_{k+1})$:
\begin{equation}\label{eq:MOind}
    \begin{aligned}
    \onep &= \mathbbm{1}_{\{\text{At least one buy MO arrives during }[t_k,t_{k+1})\}},\\
    \onem &= \mathbbm{1}_{\{\text{At least one sell MO arrives during }[t_k,t_{k+1})\}}.
    \end{aligned}
\end{equation}
We assume that {$\onep,\onem\in$} {$\mathcal{F}_{\tkk}$} and
\begin{equation}
    \mathbbm{P}(\onep=j^+,\onem=j^-|\mathcal{F}_{\tk})={\Black \pi_{\tkk}}(j^+,j^-),
    \label{eq:pi}
\end{equation} 
for $j^\pm\in \{0,1\}$, {where $\pi_{\tkk}:\{0,1\}\times\{0,1\}\to[0,1]$ is a \emph{deterministic} probability distribution}. The marginal conditional probabilities are denoted as
\begin{equation}
    {\Black \pi^\pm_{\tkk}} := \mathbbm{P}(\mathbbm{1}_\tkk^\pm=1|\F),
    \label{eq:pipm}
\end{equation} 
and throughout we assume that $\pi^{+}_{\tkk}>0$ and $\pi^{-}_{\tkk}>0$, for all $k=0,\dots,N$. Concretely, between two consecutive time steps $t_k$ and $t_{k+1}$, the arrival probability of at least one buy (sell) market order is ${\Black \pi^+_{\tkk}}$ ({${\Black \pi^-_{\tkk}}$}).

\begin{remark}
By definition of marginal probabilities, we have that ${\Black \pi^+_{\tkk}} = {\Black \pi_\tkk(1,1)}+{\Black \pi_\tkk(1,0)}$ and ${\Black \pi^-_{\tkk}} = {\Black \pi_\tkk(1,1)}+{\Black \pi_\tkk(0,1)}$. Then the following relation between ${\Black \pi_\tkk(1,1)}$ and ${\Black \pi^\pm_{\tkk}}$ must hold for each $\tkk$:
\begin{equation}
    ({\Black \pi^+_{\tkk}}+{\Black \pi^-_{\tkk}}-1)\vee 0\leq{\Black \pi_\tkk(1,1)}\leq{\Black \pi^+_{\tkk}}\wedge{\Black \pi^-_{\tkk}}.\label{eq:pioo}
\end{equation}
\end{remark}

{The ask (bid) LO is placed at time $t_{k}$, $k=0,\dots,N$, at the price level $a_{t_k}\in\mathcal{F}_{t_{k}}$ ({$b_{t_k}\in\mathcal{F}_{t_{k}}$}).} We parameterize $a_\tk$ and $b_\tk$ as
\begin{equation}
\begin{aligned}
a_{t_k} = S_{t_k}+L_{t_k}^+,\qquad \qquad b_{t_k} = S_{t_k}-L_{t_k}^-,
\end{aligned}\label{eq:abL}
\end{equation}
where $L^{\pm}_{t_{k}}\in \mathcal{F}_{t_{k}}$ are the market maker's spreads and {$S_{t_k}\in \mathcal{F}_{t_{k}}$ is the {fundamental price of the asset at time $t_{k}$. 
{The assumptions on the fundamental prices process $\{S_{t_k}\}_{k=0,\dots,N+1}$ are further specified in Section \ref{sec:BellmanEquation}.} 

The limit orders placed at time $t_k$ may be {fully or partially} executed during the time interval $[t_k,t_{k+1})$, {but} only if there exists at least one arrival of a market order during that period. We assume that the number of filled shares on the bid side during the interval $[t_k,t_{k+1})$ is given by 
\begin{equation}
\begin{aligned}
Q^-_{t_{k+1}}&\triangleq \onem\cm\big[(b_\tk-(\sk-\Pm)\big] =\onem\cm(\Pm-\lm),
\label{eq:Q-}
\end{aligned}
\end{equation}
where {$c_\tkk^-,p_\tkk^-\in\mathcal{F}_{t_{k+1}}$} are positive random variables whose distribution is specified below in Assumption \ref{assump:cp}. When no sell market order arrives during the interval $[t_k,t_{k+1})$, $\onem=0$ and the number of executions on the buy side is $0$. 
{Here,} $\Pm$ is defined such that $\sk-p_{t_{k+1}}^-$ is the lowest price that all sell market orders arriving during $[t_{k},t_{k+1})$ can attain. {In other words, bid limit orders placed by the HFM will not be executed during the interval $[t_k,t_{k+1})$ if the price is smaller than $S_{t_{k}}-p^{-}_{t_{k}}$.} {We refer to $\Pm$ as the reservation price for sellers.}
The demand slope $c_{t_{k+1}}^-$ measures the {rate of increase} in the number of filled shares of the bid order, as the order's bid price $b_{t_{k}}$ gets closer to the fundamental price $\sk$. 
Symmetrically, the number of shares filled by the HFM's ask limit order during $[t_k,t_{k+1})$ is given by 
\begin{equation}
\begin{aligned}
Q^+_{t_{k+1}}&\triangleq \onep\cp\big[(\sk+\Pp)-a_\tk\big] =\onep\cp(\Pp-\lp).
\label{eq:Q+}
\end{aligned}
\end{equation} 

The number of filled shares $Q^\pm_{\tkk}$ is illustrated in Figure \ref{fig:linearQ}. The quantity $Q_\tkk^\pm$ {may be viewed as} {the {``best"} linear fit for the actual demand as shown in Figure \ref{DemandPlotDemo}. More details on the estimation of $Q^\pm_{\tkk}$ are covered in the empirical analysis conducted in Section \ref{RRDATA0}.}

\tikzset{every picture/.style={line width=0.75pt}} 
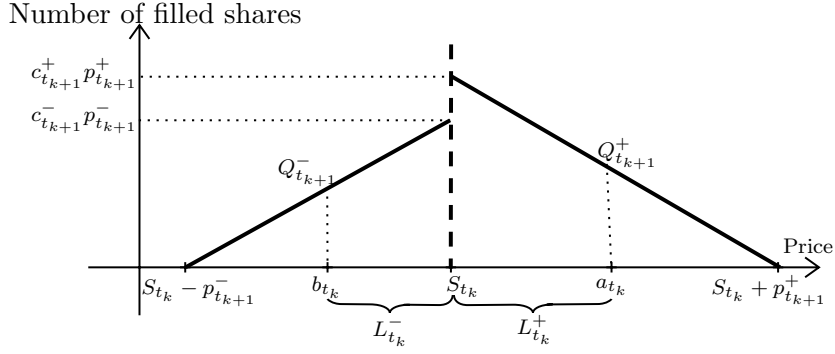
\begin{figure}
\centering

\tikzset{every picture/.style={line width=0.75pt}} 

\begin{tikzpicture}[x=0.75pt,y=0.75pt,yscale=-1,xscale=1]

\draw  (50,142.02) -- (418.54,142.02)(75.83,19.72) -- (75.83,166.22) (411.54,137.02) -- (418.54,142.02) -- (411.54,147.02) (70.83,26.72) -- (75.83,19.72) -- (80.83,26.72)  ;
\draw [line width=1.5]    (232.5,67.8) -- (99.07,142.02) ;
\draw [line width=1.5]    (233.5,45.8) -- (398.92,142.02) ;
\draw   (173.28,142.02) -- (168.05,142.02)(170.67,139.4) -- (170.67,144.63) ;
\draw   (316.52,142.02) -- (311.28,142.02)(313.9,139.4) -- (313.9,144.63) ;
\draw   (170.67,158) .. controls (170.62,158.15) and (172.93,160.51) .. (177.6,160.56) -- (192.15,160.7) .. controls (198.82,160.77) and (202.13,163.14) .. (202.08,165) .. controls (202.13,163.14) and (205.48,160.84) .. (212.15,160.91)(209.15,160.88) -- (226.71,161.06) .. controls (231.38,161.11) and (233.73,158.8) .. (233.78,156) ;
\draw   (234.44,156) .. controls (234.47,158.81) and (236.82,161.12) .. (241.49,161.08) -- (264.23,160.89) .. controls (270.9,160.84) and (274.25,163.14) .. (274.28,165) .. controls (274.25,163.14) and (277.56,160.78) .. (284.22,160.73)(281.22,160.75) -- (306.96,160.54) .. controls (311.63,160.51) and (313.94,158.16) .. (313.9,158) ;
\draw [line width=1.5]  [dash pattern={on 5.63pt off 4.5pt}]  (233,29.5) -- (232.82,142.67) ;
\draw  [dash pattern={on 0.84pt off 2.51pt}]  (311.5,89.8) -- (313.79,141.57) ;
\draw  [dash pattern={on 0.84pt off 2.51pt}]  (170.5,101.8) -- (170.67,141.36) ;
\draw   (235.52,142.02) -- (230.28,142.02)(232.9,139.4) -- (232.9,144.63) ;
\draw   (101.52,142.02) -- (96.28,142.02)(98.9,139.4) -- (98.9,144.63) ;
\draw   (400.52,142.02) -- (395.28,142.02)(397.9,139.4) -- (397.9,144.63) ;
\draw  [dash pattern={on 0.84pt off 2.51pt}]  (75.5,67.8) -- (232.5,67.8) ;
\draw  [dash pattern={on 0.84pt off 2.51pt}]  (76.5,45.8) -- (233.5,45.8) ;

\draw (239,152) node  [inner sep=0.75pt]  [font=\scriptsize,rotate=-359,xslant=0]  {$S_{\tk}$};
\draw (393.63,144.31) node [anchor=north] [inner sep=0.75pt]  [font=\scriptsize]  {$S_{\tk} +p_{\tkk}^{+}$};
\draw (106,144.31) node [anchor=north] [inner sep=0.75pt]  [font=\scriptsize]  {$S_{\tk} -p_{\tkk}^{-}$};
\draw (170.34,152)  node  [font=\scriptsize]  {$b_{\tk}$};
\draw (314.23,152) node  [font=\scriptsize]  {$a_{\tk}$};
\draw (202.39,174.14) node  [font=\scriptsize]  {$L_{\tk}^{-}$};
\draw (275.19,173.28) node  [font=\scriptsize]  {$L_{\tk}^{+}$};
\draw (322.14,84.49) node  [font=\scriptsize]  {$Q_{\tkk}^{+}$};
\draw (160.94,93.09) node  [font=\scriptsize]  {$Q_{\tkk}^{-}$};
\draw (48.14,44.49) node  [font=\scriptsize]  {$c^{+}_{t_{k+1}} p^{+}_{t_{k+1}}$};
\draw (48.14,66.49) node  [font=\scriptsize]  {$c^{-}_{t_{k+1}} p^{-}_{t_{k+1}}$};
\draw (413.07,131.16) node [align=left] {{\scriptsize Price}};
\draw (83.81,13.44) node [align=left] {Number of filled shares};

\end{tikzpicture}
\caption{$S_{\tk}-p_{\tkk}^-$ is the lowest price that a sell market order can attain, and $S_{\tk}+p_{\tkk}^+$ is the highest price that a buy market order can attain during the time interval $[\tk,\tkk)$. The number of filled shares increase as the market maker places limit orders closer to the fundamental price $S_\tk$.}  \label{fig:linearQ}
\vspace{.5 cm}
\end{figure}

\begin{figure}
    \centering
  \includegraphics[width=1\textwidth]{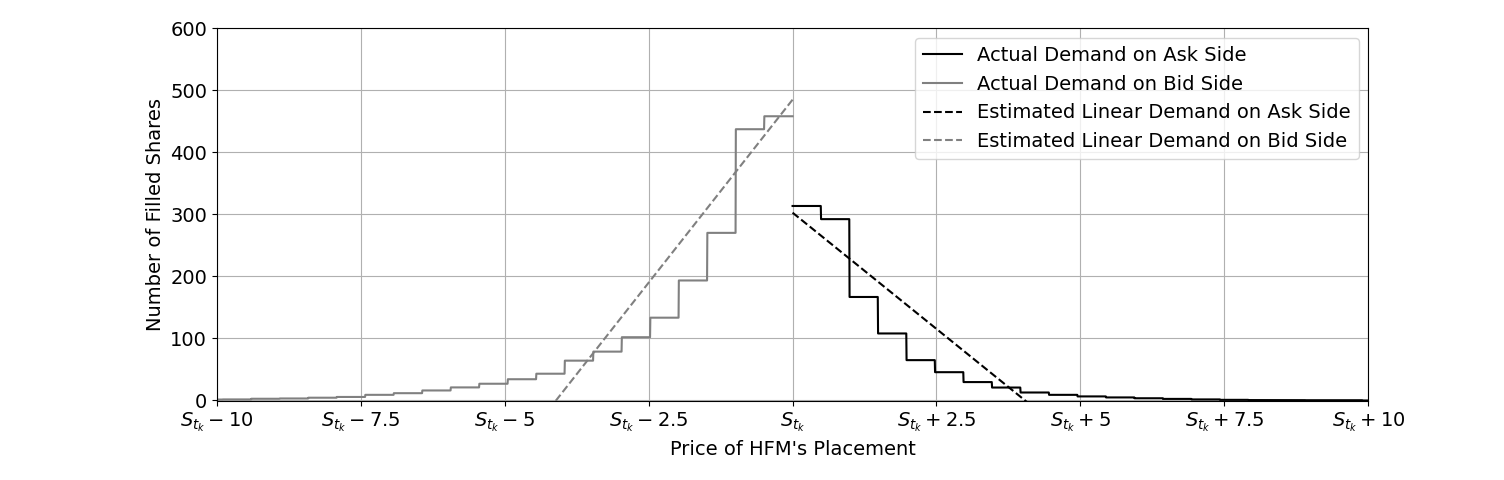}
  \caption{\textbf{{Prototypical Plot of Actual Demand vs. Estimated Linear Demand over Time Interval $[\tk,\tkk)$.}}}
  \label{DemandPlotDemo}
  \vspace{.5 cm}
\end{figure}

Next, we {state} our main assumptions on {$c_.^\pm$ and $p_.^\pm$}.
\begin{assumption}[General Properties of $(c_.^\pm,p_.^\pm)$] {For $k=0,\dots,N$, we have:} 
\begin{enumerate}
    \item $(c_\tkk^\pm,p_\tkk^\pm)$ are $\mathcal{F}_{t_{k+1}}$-measurable,
    \item the {conditional distribution of $(c^+_{t_{k+1}},p^+_{t_{k+1}},c^-_{t_{k+1}},p^-_{t_{k+}})$ given $({\F},\onep,\onem)$} does not depend on $k$ {and is nonrandom},
    \item {$(c^+_{t_{k+1}},p^+_{t_{k+1}})$ and $(c^-_{t_{k+1}},p^-_{t_{k+1}})$} are independent given $({\F},\onep,\onem)$.
\end{enumerate}\label{assump:cp}
\end{assumption}

Next, we introduce {some} further notation related to $(c_.^\pm,p_.^\pm)$:
    \begin{equation}\label{Dfnmucp}
    \begin{split}
     \muo^\pm&:=\E(\cn^\pm|\F,\mathbbm{1}^\pm_\tkk=1),\\
     \mut^\pm&:=\E((\cn^\pm)^2|\F,\mathbbm{1}^\pm_\tkk=1),  \\
     \muoo^\pm&:=\E(\cn^\pm\Pn^\pm|\F,\mathbbm{1}^\pm_\tkk=1),\\
     \muto^\pm&:=\E((\cn^\pm)^2\Pn^\pm|\F,\mathbbm{1}^\pm_\tkk=1),\\
     \mutt^\pm&:=\E((\cn^\pm\Pn^\pm)^2|\F,\mathbbm{1}^\pm_\tkk=1).
    \end{split}
\end{equation}

We consider the following maximization problem for the HFM: 
\begin{equation}\label{ObjProblJ}
\max_{(L_.^+,L_.^-)\in \mathcal{A}}\mathbb{E}[W_T+S_T I_T -\lambda I_T^2],
\end{equation} 
where {$\mathcal{A}$ is the collection of all $\mathcal{F}$-adapted processes, and} $W_T$ and $I_T$ stand for the {market maker's} cash holdings and inventory at the end of period $[0,T]$, respectively. The cash holding and inventory processes, $\{W_\tk\}$ and $\{I_\tk\}$, respectively, satisfy the following equations:
\begin{equation}
\begin{aligned}
    W_{t_{k+1}} &= W_\tk+a_\tk Q_\tkk^+-b_\tk Q_\tkk^-\\
    &=W_\tk+(S_\tk+\lp)\onep\cp(\Pp-\lp) - (S_\tk-\lm)\onem\cm(\Pm-\lm)
\end{aligned}\label{eq:W}
\end{equation}
and
\begin{equation}
\begin{aligned}
    I_{t_{k+1}} &= I_\tk-Q_\tkk^++Q_\tkk^-\\
    &=I_{t_k} -\onep\cp(\Pp-\lp) +\onem\cm(\Pm-\lm),
\end{aligned}\label{eq:I}
\end{equation}
{where $W_{t_{0}}=0$ and $I_{t_{0}}=0$.}

The term $\lambda I^2_T\geq 0$ in the maximization problem {(\ref{ObjProblJ})} is the cost for holding inventory at the terminal time $T$. This formulation captures, in reduced form, the fact that HFMs tend to have de minimis balance sheets, thus making any overnight inventory costly to carry.\footnote{Similar objective criteria have been proposed in earlier studies, such as \cite{cartea2015risk} and \cite{adrian2016intraday}.} 
The penalty term for holding end-of-day inventory can also be interpreted as follows. We can rewrite the last two terms $S_TI_T-\lambda I_T^2$ in the above expectation as $(S_T-\lambda I_T)I_T$.
Then, $S_T-\lambda I_T$ is the average price per share that the {HFM} will get when liquidating her inventory $I_T$ via a MO, under the assumption of a linear instantaneous price impact. For instance, if $I_T>0$ ($I_T<0$), then the {HFM}  will have to submit a sell (buy) MO, which will result in eating into the bid (ask) side of the book. {We will validate the assumption of linear price impact in Section \ref{RRDATA0}.}

\section{The Bellman Equation for the Control Problem}\label{sec:BellmanEquation}\label{sec:Bellman}
At time $t_k$, the value function of the control problem described above {is given by}
\begin{equation}
V_{t_k}=\sup_{(L_.^+,L_.^-)\in \mathcal{A}}\mathbb{E}[W_T+S_T I_T -\lambda I_T^2|\F].
\end{equation} 
By the dynamic programming principle, we obtain it satisfies the following equation
\begin{equation}
V_{t_k}=\sup_{(L_{t_k}^+,L_{t_k}^-)\in \mathcal{A}}\mathbb{E}[V_{t_{k+1}}|\F].\label{eq:DP}
\end{equation}

We now proceed to find the optimal placement strategy for the market maker. {Our objective is to derive it for a general adapted stochastic process of the fundamental price. To illustrate the procedure behind the construction,} we first analyze the setting where the {fundamental price} process is a martingale, and obtain tractable formulas for {the optimal bid and ask prices and the value function} (see Subsection \ref{sec:MGstrategy}). In Subsection \ref{sec:DriftAnaly}, we relax the martingale assumption and provide a general formula for general adapted stochastic price dynamics of the fundamental price.

\subsection{Optimal Strategy {Under a Martingale Fundamental Price Process}}\label{sec:MGstrategy}
{In this subsection,} we assume {that the fundamental price $S_{t_k}\in \mathcal{F}_{t_{k}}$ of the asset is} a martingale:
\begin{equation}\label{MrtCnd0}
	\E(S_\tkk|\F)=S_\tk,\quad {k=0,\dots,N.}
\end{equation}
Furthermore, we assume that {$S_{t_{k+1}}-S_{t_{k}}$}  and $(\onep,\onem,c^+_{t_{k+1}},p^+_{t_{k+1}},c^-_{t_{k+1}},p^-_{t_{k+}})$ are conditionally independent given ${\F}$.
We {start by making} the following ansatz for the value function $V_\tk$: \begin{equation}
V_\tk=v(t_k,\sk,W_\tk,I_\tk) := W_\tk+\alpha_{\tk}I_\tk^2+\sk I_\tk+h_{\tk}I_\tk+g_{\tk},\label{eq:V1}
\end{equation} where 
{$\alpha:\mathcal{T}\to\mathbb{R}$, $h:\mathcal{T}\to\mathbb{R}$, and $g:\mathcal{T}\to\mathbb{R}$ are deterministic functions  defined on $\mathcal{T}=\{t_0,t_{1},\dots,t_{N+1}\}$ (recall that we set $t_{N+1}=T$)}. Since $V_T=W_T+S_TI_T-\lambda I_T^2$, we obtain the terminal conditions $\alpha_T=-\lambda$, $g_T=0$, and $h_T=0$.

We can determine the functions $\alpha_.$, $h_.$, and $g_.$ by plugging the { ansatz~(\ref{eq:V1})} back into Eq.~(\ref{eq:DP}), and then using Eqs.~(\ref{eq:W})-(\ref{eq:I}). This yields the following iterative representation for the value function
\begin{align}
    \begin{split}
       & v(t_k,\sk,W_\tk,I_\tk)\\
        &\quad =\sup_{(L_{t_k}^+,L_{t_k}^-)\in \mathcal{A}}\mathbb{E}\big[v(t_{k+1},\skk,W_{\tkk},I_{\tkk})\big|\F\big]\\
        & \quad = \sup_{(L_{t_k}^+,L_{t_k}^-)\in \mathcal{A}}\mathbb{E}\big[v\big(t_{k+1},\skk,W_\tk+a_\tk Q_\tkk^+-b_\tk Q_\tkk^-, I_\tk-Q_\tkk^++Q_\tkk^-\big)\big|\F\big]{\Green.}
    \end{split}
    \label{eq:dprogram}
\end{align}
\normalsize
From the construction of $a_{\tk}$, $b_{\tk}$, and $Q_{\tkk}^\pm$ (see Eqs.~(\ref{eq:abL})-(\ref{eq:Q+})), we know that $a_{\tk}$ and $Q_{\tkk}^+$ are linear in $L_{\tk}^+$, while $b_{\tk}$ and $Q_{\tkk}^-$ are linear in $L_{\tk}^-$.
{Also, by our ansatz,} $v(t_k,\sk,W_\tk,I_\tk)$ is linear in $W_\tk$ and quadratic in $I_\tk$. {Denoting} the expectation on the right-hand side of Eq.~(\ref{eq:dprogram}) as $f(L_\tk^+,L_\tk^-)$, {we} can then conclude that $f(L_\tk^+,L_\tk^-)$ is quadratic in $L_{\tk}^+$ and $L_{\tk}^-$. Therefore, we can use the first-order conditions to find the candidates $L_{\tk}^{\pm,*}$.  
 We can then evaluate the second partial derivative, and establish that the critical point $(L_{\tk}^{+,*},L_{\tk}^{-,*})$ is indeed a maximum point. We state this fact in the following {proposition}, whose proof is given in {Appendix \ref{sec:OptimalL}.
 \begin{proposition}[Optimal Controls]\label{prop:optimalcontrol}
The optimal controls {that solve the optimization problem (\ref{eq:dprogram}) using the ansatz {(\ref{eq:V1})} and state dynamics (\ref{eq:W})-(\ref{eq:I})} are given, {for $k=0,\dots,N$}, by
\begin{align}
\begin{split}
     L_{\tk}^{+,*}&={}^{\scaleto{(1)}{5pt}}\!A^+_{\tk}I_\tk+{}^{\scaleto{(2)}{5pt}}\!A^+_{\tk}+{}^{\scaleto{(3)}{5pt}}\!A^+_{\tk},\\
     L_{\tk}^{-,*}&=-{}^{\scaleto{(1)}{5pt}}\!A^-_{\tk}I_\tk-{}^{\scaleto{(2)}{5pt}}\!A^-_{\tk}+{}^{\scaleto{(3)}{5pt}}\!A^-_{\tk}.\label{eq:Ltilde}
\end{split}
\end{align}
\normalsize
where the coefficients above are specified as 
\begin{align}
\label{eq:A1}
    &{}^{\scaleto{(1)}{5pt}}\!A^\pm_{\tk}=\frac{\beta^{\pm}_{t_{k}}\alpha_{t_{k+1}}}{\gamma_{t_{k}}},\quad 
    {}^{\scaleto{(2)}{5pt}}\!A^\pm_{\tk}=\frac{\beta^{\pm}_{t_{k}}h_{t_{k+1}}}{2\gamma_{t_{k}}},\\
    \nonumber
    &{}^{\scaleto{(3)}{5pt}}\!A^\pm_{\tk}=\frac{\pi^\mp_\tkk}{2\gamma_{t_{k}}}(\alpha_{\tkk}\mut^\mp-\muo^\mp)\big[\pi^\pm_{\tkk}(\muoo^\pm-2\alpha_{\tkk}\muto^\pm)
        +2\alpha_{\tkk}\pi_\tkk(1,1)\muo^\pm\muoo^\mp\big]\\
            \nonumber
        &\qquad +\pi_\tkk(1,1)\frac{\alpha_{\tkk}}{2\gamma_{t_{k}}}\muo^+\muo^-\big[\pi^\mp_\tkk(\muoo^\mp-2\alpha_{\tkk}\muto^\mp)+2\alpha_{\tkk}\pi_\tkk(1,1)\muo^\mp\muoo^\pm\big]{\Green,}
\end{align}
and 
\begin{align*}
\gamma_{t_{k}}&:=\big[{\pi_\tkk(1,1)}\alpha_{\tkk}\muo^+\muo^-\big]^2-{\pi^+_{\tkk}}{ \pi^-_{\tkk}}(\alpha_{\tkk}\mut^+-\muo^+)(\alpha_{\tkk}\mut^--\muo^-),\\
\beta_{t_{k}}^{\pm}&:={\pi^+_{\tkk}}{\pi^-_{\tkk}}\muo^\pm(\alpha_{\tkk}\mut^\mp-\muo^\mp)-{\pi^\mp_\tkk}{\Black \pi_\tkk(1,1)}\alpha_\tkk\muo^\pm(\muo^\mp)^2{.}
\end{align*}
In the expressions above, $\alpha:\mathcal{T}\to\mathbb{R}$ and {$h:\mathcal{T}\to\mathbb{R}$}
are specified using the following backward equations: $\alpha_T=-\lambda$,
$h_T=0$ at $T=t_{N+1}$ and, for $k=0,\dots, N$:
\begin{align}\label{eq:alpha}
   \begin{split}
    \alpha_{\tk}&=\alpha_{\tkk}+\sum_{\delta=\pm}{\Black \pi^\delta_\tkk}\big[(\alpha_{\tkk}\mut^\delta-\muo^\delta)({}^{\scaleto{(1)}{5pt}}\!A^\delta_{\tk})^2+2\alpha_{\tkk}\muo^\delta ({}^{\scaleto{(1)}{5pt}}\!A^\delta_{\tk})\big]\\
   &\qquad  \qquad+2\alpha_{\tkk}{\Black \pi_\tkk(1,1)}\muo^+\muo^-({}^{\scaleto{(1)}{5pt}}\!A^+_{\tk}{}^{\scaleto{(1)}{5pt}}\!A^-_{\tk}),
          \end{split}
\end{align}   
and
\begin{align}\nonumber
        h_{\tk}&= h_{\tkk}+\sum_{\delta=\pm}{\Black \pi^\delta_\tkk}\Big\{2(\alpha_{\tkk}\mut^\delta-\muo^\delta) \big[{}^{\scaleto{(1)}{5pt}}\!A^\delta_{\tk}({\delta\;{\Black{}^{\scaleto{(3)}{5pt}}\!A^\delta_{\tk}+ {}^{\scaleto{(2)}{5pt}}\!A^\delta_{\tk}}})\big]\\
        \label{eq:MG_h}   
        &\quad +2\alpha_{\tkk}\muo^\delta({\delta\,{\Black{}^{\scaleto{(3)}{5pt}}\!A^\delta_{\tk}+ {}^{\scaleto{(2)}{5pt}}\!A^\delta_{\tk}}})-2\alpha_{\tkk}({\delta\muoo^\delta})
        +({\delta} {}^{\scaleto{(1)}{5pt}}\!A^\delta_{\tk})(\muoo^\delta+\delta h_{\tkk}\muo^\delta-2\alpha_{\tkk}\muto^\delta)\Big\}\\
        \nonumber
        &\quad -2\alpha_{\tkk}{\Black \pi_\tkk(1,1)}\muo^+\muo^-\Big[{}^{\scaleto{(1)}{5pt}}\!A^+_{\tk}({\Black {}^{\scaleto{(3)}{5pt}}\!A^-_{\tk}-{}^{\scaleto{(2)}{5pt}}\!A^-_{\tk}})-{}^{\scaleto{(1)}{5pt}}\!A^-_{\tk}({}^{\scaleto{(2)}{5pt}}\!A^+_{\tk}+{}^{\scaleto{(3)}{5pt}}\!A^+_{\tk})+\frac{\muoo^+}{\muo^+}({}^{\scaleto{(1)}{5pt}}\!A^-_{\tk})-\frac{\muoo^-}{\muo^-}({}^{\scaleto{(1)}{5pt}}\!A^+_{\tk})\Big].
\end{align}
\end{proposition}
The} following {key} lemma will be needed {to show that the critical point $(L_{\tk}^{+,*},L_{\tk}^{-,*})$ of Proposition \ref{prop:optimalcontrol} is indeed a maximum point and also} in the analysis of the optimal placement's properties in Section \ref{PrpOptPlc2b}. Its proof is {intricate and} deferred to Appendix \ref{sec:proof_PostLmm}.
\begin{lemma}\label{lemma:alpha}
The quantity $\alpha_{\tk}$ defined in Eq.~(\ref{eq:alpha}) is strictly decreasing with $t_k$ and negative for every $t_k$.
 \end{lemma}

%
Recall $(L_{\tk}^{+,*},L_{\tk}^{-,*})$ are admissible if $(L_{\tk}^{+,*},L_{\tk}^{-,*})\in\F$. It is easy to check that $(L_{\tk}^{+,*},L_{\tk}^{-,*})\in\F$ since $I_\tk\in\F$ and ${}^{\scaleto{(1)}{5pt}}\!A^\pm_{\tk},{}^{\scaleto{(2)}{5pt}}\!A^\pm_{\tk},{{}^{\scaleto{(3)}{5pt}}\!A^\pm_{\tk}}$ are deterministic functions.}
{From the previous result and the expression for bid and ask prices in Eq.~\eqref{eq:abL}, we deduce that} the optimal placements at time $\tk$ are
\begin{align}
   a_\tk^* &= \sk+{}^{\scaleto{(1)}{5pt}}\!A^+_{\tk}I_\tk+{}^{\scaleto{(2)}{5pt}}\!A^+_{\tk}+{}^{\scaleto{(3)}{5pt}}\!A^+_{\tk},\label{eq:tilde_a}\\ 
   b_\tk^* &= \sk+{}^{\scaleto{(1)}{5pt}}\!A^-_{\tk}I_\tk{\Black +}{}^{\scaleto{(2)}{5pt}}\!A^-_{\tk}{\Black -}{}^{\scaleto{(3)}{5pt}}\!A^-_{\tk},\label{eq:tilde_b}
\end{align}
where $a_\tk^*$ is the price for the ask limit order and $b_\tk^*$ is the price for the bid limit order.

Recall that 
\begin{equation}
	V_{t_k}=\sup_{(L_.^+,L_.^-)\in \mathcal{A}}\mathbb{E}[W_T+S_T I_T -\lambda I_T^2|\F].
	\label{eq:DPbb}
\end{equation} 
We next prove a verification theorem for the optimal placements given in Eq.~(\ref{eq:Ltilde}). {Its proof is given in Appendix \ref{ProofVerifyH}.}
\begin{theorem}[Verification Theorem]\label{VeriThrm1}
{The optimal value function $V_\tk$ {of} the control problem (\ref{eq:DPbb}) is given by 
\begin{equation*}
	V_\tk=v(t_k,\sk,W_\tk,I_\tk),
\end{equation*} 
where, for $\tk\in \mathcal{T}$,  
\[
	v(\tk,s,{\mathsf{w}},i)= {\mathsf{w}}+\alpha_{\tk}i^{2}+s i+h_{\tk}i+g_{\tk},
\]
{with $\alpha_\tk$ and $h_\tk$ given in {Proposition} \ref{prop:optimalcontrol}, and $g_\tk$ defined as $g_{T}=0$ and, for $k=0,\dots,N$,
\begin{align*}       
    \begin{split}
        g_{\tk}&=g_{\tkk}+\sum_{\delta=\pm}{\pi^\delta_\tkk}\Big[(\alpha_{\tkk}\mut^\delta-\muo^\delta)({ {}^{\scaleto{(3)}{5pt}}\!A^\delta_{\tk}+(\delta\, {}^{\scaleto{(2)}{5pt}}\!A^\delta_{\tk}}))^2+\alpha_{\tkk}\mutt^\delta-{(\delta h_{\tkk})}\muoo^\delta\\
        &\quad\qquad\qquad\qquad\qquad\quad+(\muoo^\delta+{(\delta h_{\tkk})}\muo^\delta-2\alpha_{\tkk}\muto^\delta)({ {}^{\scaleto{(3)}{5pt}}\!A^\delta_{\tk}+{(\delta\,} {}^{\scaleto{(2)}{5pt}}\!A^\delta_{\tk}}))\Big]\\
        &\qquad\qquad-2\alpha_{\tkk}{\pi_\tkk(1,1)}\muo^+\muo^-\Big[({}^{\scaleto{(2)}{5pt}}\!A^+_{\tk}+{}^{\scaleto{(3)}{5pt}}\!A^+_{\tk})({ {}^{\scaleto{(3)}{5pt}}\!A^-_{\tk}-{}^{\scaleto{(2)}{5pt}}\!A^-_{\tk}})\\
        &\quad\qquad\qquad\qquad\qquad\qquad\qquad\qquad-\frac{\muoo^+}{\muo^+}({{}^{\scaleto{(3)}{5pt}}\!A^-_{\tk}-{}^{\scaleto{(2)}{5pt}}\!A^-_{\tk}})-\frac{\muoo^-}{\muo^-}({}^{\scaleto{(2)}{5pt}}\!A^+_{\tk}+{}^{\scaleto{(3)}{5pt}}\!A^+_{\tk})+\frac{\muoo^+\muoo^-}{\muo^-\muo^+}\Big]{.}
    \end{split}\label{eq:g}
\end{align*}

Furthermore,} the optimal controls are given by $L_.^{\pm,*}$ as defined in (\ref{eq:Ltilde}).}
\end{theorem}

\subsection{Optimal Strategy with {a} {General Adapted {Fundamental Price} Process}}\label{sec:DriftAnaly}
In this subsection, we relax the martingale assumption on the fundamental price process $\{S_{t_k}\}_{t_{k}\in\mathcal{T}}$ made in the previous subsection, {and consider a general adapted process. {Furthermore, we assume that, conditionally on ${\F}$,  $\{{S_{t_{j+1}}-S_{t_j}}\}_{{j\geq{}k}}$ and $(\onep,\onem,c^+_{t_{k+1}},p^+_{t_{k+1}},c^-_{t_{k+1}},p^-_{t_{k+}})$ are independent.
Let us} introduce the notation:
\[
	\Delta_\tk:=\E(S_\tkk-S_\tk|\F).
\] 
{The variable} {$\Delta_\tk$ reflects the {HFM's forecast about the asset price's change} in the interval $[\tk,\tkk)$ {based on her information available at $t_{k}$}. Including this term makes our model more flexible and, {as we shall see in Section \ref{RRDATA0}, the resulting optimal placement strategies achieve better empirical performance.}
We leave the rest of the model setup as in Section \ref{sec:setup}}. 

{We define the price change forecasts
\begin{align}\label{Forecastsbb}
	\Delta_{t_j}^{t_k}:=\E(\Delta_{t_j}|\F)=\E(S_{t_{j+1}}-S_{t_{j}}|\F),\quad j\geq{}k,
\end{align}
and recall the standard convention $\prod_{\ell=k}^{k-1}=1$. The following result gives the optimal placement spreads for an arbitrary adapted price process $\{S_{t_{k}}\}_{t_k\in\mathcal{T}}$ in terms of the forecasts (\ref{Forecastsbb}) and the optimal placement strategy ${L}_{\tk}^{\pm,*}$ of {Proposition} \ref{prop:optimalcontrol}. The proof is provided {in Appendix \ref{sec:OptimalL}}.
\begin{theorem}[Optimal Controls with a General Adapted Fundamental Price Process]\label{prop:optimalcontrolNMG}
The optimal controls which solve the dynamic optimization problem (\ref{eq:dprogram}) are given, for $k=0,\dots,N$, by
\begin{align}
\begin{split}
     \widetilde{L}_{\tk}^{+,*}&={L}_{\tk}^{+,*}+\frac{\beta_{t_k}^{+}}{2\gamma_{t_{k}}}\Delta_{t_{k}}+\Big(\frac{\beta^{+}_{t_{k}}}{2\gamma_{t_{k}}}\Big)
     \sum_{j=k+1}^{N}{\prod_{\ell=k+1}^{j}\xi_{\ell}}\Delta_{t_j}^{t_k},\\
    \widetilde{L}_{\tk}^{-,*}&={L}_{\tk}^{-,*}-\frac{\beta_{t_k}^{-}}{2\gamma_{t_{k}}}\Delta_{t_{k}} -\Big(\frac{\beta^{-}_{t_{k}}}{2\gamma_{t_{k}}}
  \Big)  \sum_{j=k+1}^{N}{\prod_{\ell=k+1}^{j}\xi_{\ell}}\Delta_{t_j}^{t_k},\label{eq:LtildeNMGbb}
\end{split}
\end{align}
{where $\beta^{\pm}_{t_k}$ and $\gamma_{t_k}$ are the deterministic sequences introduced in {Proposition} \ref{prop:optimalcontrol}, and ${L}_{\tk}^{\pm,*}$ are the optimal spreads defined therein. The quantity $\xi_{k}$ is defined as:
\begin{align}
	\xi_{k}&=  1+\frac{\alpha_{t_{k+1}}}{\gamma_{t_k}}\sum_{\delta=\pm}
	{\pi^\delta_\tkk}
	\beta^{\delta}_{t_k}
	\Big\{\frac{\beta^{\delta}_{t_{k}}}{\gamma_{t_{k}}}(\alpha_{\tkk}\mut^\delta-\muo^\delta)+2\muo^\delta\Big\}+2\frac{\alpha_{\tkk}^2}{\gamma_{t_k}^2}{\pi_\tkk(1,1)}\muo^+\muo^-\beta^{+}_{t_{k}}\beta^{-}_{t_{k}}.
	\label{eq:xi}
\end{align}}
\end{theorem}
The optimal {placement strategies} at time $\tk$ with a non-martingale dynamics for the fundamental price process can then be written as:
\begin{align}
    \widetilde{a}_\tk^* = \sk+\widetilde{L}_{\tk}^{+,*}, \qquad \qquad 
    \widetilde{b}_\tk^* &= \sk-\widetilde{L}_{\tk}^{-,*},\label{eq:tilde_bNMG}
\end{align}
where $\widetilde{a}_\tk^*$ is the price for the ask limit order and $\widetilde{b}_\tk^*$ is the price for the bid limit order. {Eq.~(\ref{eq:LtildeNMGbb}) highlights that we can split the problem of finding the optimal trading strategy into two subproblems. First, we compute the {recursive expressions} {(\ref{eq:A1})}-(\ref{eq:MG_h}). This is done ``offline'' at the beginning of each trading day. {That is to say, all parameters needed to {compute}  ${L}_{\tk}^{\pm,*}$ (i.e., the optimal controls with a martingale price process) are predetermined at the beginning of the day.} Second, we solve the forecasting problem of determining $\{\Delta_{t_j}^{t_k}\}_{j=k,\dots,N}$, {and compute $\widetilde{L}_{\tk}^{\pm,*}$ using the expression of ${L}_{\tk}^{\pm,*}$ as in Eq.~(\ref{eq:LtildeNMGbb})}. This task is done ``online'' at each time $t_{k}$.} Thus, under a general adapted fundamental price process, the optimal strategy $\widetilde{L}_{\tk}^{\pm,*}$ {incorporates the views of the HFM about {changes in the fundamental based on her information available} at time $\tk$.}

\begin{remark}
	As shown in 
Appendix \ref{sec:OptimalL}, for the case of a general adapted price process $\{S_{t_{k}}\}_{t_k\in\mathcal{T}}$, the ansatz for the value function $V_\tk$ takes the form: 
\begin{equation}
V_\tk=v(t_k,\sk,W_\tk,I_\tk) := W_\tk+\alpha_{\tk}I_\tk^2+\sk I_\tk+\widetilde{h}_{\tk}I_\tk+\widetilde{g}_{\tk}\label{eq:V}
\end{equation} 
where, as in Subsection \ref{sec:MGstrategy}, $\alpha:\mathcal{T}\to\mathbb{R}$ is a deterministic function, but $\{\widetilde{h}_{t_{k}}\}_{t_{k}\in\mathcal{T}}$ and $\{\widetilde{g}_{t_{k}}\}_{t_{k}\in\mathcal{T}}$ are now processes adapted to the filtration $\{\mathcal{F}_t\}_{t\in\mathcal{T}}$. The precise expressions for $\widetilde{h}$ and $\widetilde{g}$ are given in {Eqs.~(\ref{eq:hNMGc})-(\ref{eq:gNMGb}) using notation (\ref{DfnFrcgh})}. The proof of the corresponding verification theorem proceeds along similar lines as the proof of Theorem \ref{VeriThrm1}.
\end{remark}

In} the next proposition, we provide conditions under which the bid-ask spread is guaranteed to be positive (i.e., $a_{t_{k}}>b_{t_{k}}$). We defer the proof to Appendix \ref{sec:proof_adm}.
\begin{proposition}[Conditions for a Optimal Positive Spread]  \label{remark:admb}
Under both martingale and non-martingale price processes, the optimal placement strategy yields positive spreads {at all times} (i.e., $a_{t_{k}}>b_{t_{k}}$, for all $k=0,\dots,N$), provided that the following three conditions hold:
\begin{enumerate}
\item[(1)] The first and second conditional moments of $c^\pm$ defined in Eq.(\ref{Dfnmucp}) satisfy
\begin{equation}\label{Cnd1PosSpr}
\muo := \muo^+=\muo^-,\quad \mut := \mut^+=\mut^-.
\end{equation}
\item[(2)] Buy and sell market orders arrive with the same probability:
\begin{equation}\label{Cnd2PosSpr}
	{\Black \pi^+_{\tkk}}={\Black \pi^-_{\tkk}}.
\end{equation}
\item[(3)] The conditional expectations of $(cp)^{\pm}$ and $(c^2p)^{\pm}$ defined in Eq.(\ref{Dfnmucp}) satisfy
\begin{equation}\label{Cnd3PosSpr}
	\mu_{cp}^\pm=\mu_{c}^\pm\mu_{p}^\pm,\quad \mu_{c^2 p}^\pm=\mu_{c^2}^\pm\mu_p^\pm,
\end{equation}
where  $\mu_p^{\pm}:=\E(p_{\tkk}^\pm|\F,\mathbbm{1}^\pm_\tkk=1)$.
\end{enumerate}
\end{proposition}

Conditions (\ref{Cnd1PosSpr}) and (\ref{Cnd2PosSpr}) imply a symmetric market. Under Condition (\ref{Cnd1PosSpr}), both mean and variance of the bid demand slope $c_{t_{k+1}}^-$ are the same as those on the ask side. Condition (\ref{Cnd2PosSpr}) postulates that buy and sell MOs arrive with the same probability within each time interval $[\tk,\tkk)$. Condition (\ref{Cnd3PosSpr}) postulates that the demand slope $c_{t_{k+1}}^\pm$ and the reservation price $p_{t_{k+1}}^\pm$ are {uncorrelated}. {These assumptions are empirically supported by the analysis of Section \ref{RRDATA0}.}

\section{Properties of the Optimal Placement Strategies}\label{PrpOptPlc2b}
In this section, we will discuss {the behavior of the optimal placement strategies and their sensitivities to model parameters, such as the arrival rate $\pi_\tk(1,1)$, the inventory level $I$,  and the penalty $\lambda$ on the terminal inventory.

\subsection{Sensitivity of Optimal Strategies to $\pi_\tk(1,1)$}\label{SubSec41a}
\subsubsection{Case $\pi_\tk(1,1)\equiv 0$.}\label{sec:pioo0} We first consider the situation where only one type of MOs (buy or sell) can arrive between {two times.} Recall $\mathbbm{P}(\onep=j^+,\onem=j^-|\mathcal{F}_{\tk})=\pi_\tkk(j^+,j^-)$, {$j^\pm\in\{0,1\}$, where {\Black $\onen^+(\onen^-)$} indicates} whether there are arrivals of buy {\Black (sell)} MOs during {$[\tk,\tkk)$}. 
If ${\Black \pi_\tkk(1,1)}=0$,
it follows from Eq.~\eqref{eq:tilde_bNMG} that the best placement strategies take the following form:
\begin{align}\label{EqaSE1}
   \widetilde{a}_\tk^{*,0} & = \sk+\overbrace{\dfrac{\alpha^0_{\tkk}\muo^+}{\muo^+-\alpha^0_{\tkk}\mut^+}I_\tk+\dfrac{\muoo^+-2\alpha^0_{\tkk}\muto^+}{2[\muo^+-\alpha^0_{\tkk}\mut^+]} +\dfrac{ (\Delta_\tk+{ \widetilde{h}^0_{\tk}})\muo^+}{2[\muo^+-\alpha^0_{\tkk}\mut^+]}}^{\widetilde{L}_{\tk}^{+,*,0}}\\
    \label{EqaSE2}
       \widetilde{b}_\tk^{*,0} & = \sk+\overbrace{\dfrac{\alpha^0_{\tkk}\muo^-}{\muo^--\alpha^0_{\tkk}\mut^-}I_\tk-\dfrac{\muoo^--2\alpha^0_{\tkk}\muto^-}{2[\muo^--\alpha^0_{\tkk}\mut^-]}+\dfrac{ (\Delta_\tk+{ \widetilde{h}^0_{\tk}})\muo^-}{2[\muo^--\alpha^0_{\tkk}\mut^-]}}^{-\widetilde{L}_{\tk}^{-,*,0}}{\Green,}
\end{align}
where 
\begin{align}
\begin{split}
    \alpha^0_{\tk}&=\alpha^0_{\tkk}+\sum_{\delta=\pm}{\Black \pi^\delta_\tkk}\dfrac{(\alpha^0_{\tkk}\muo^\delta)^2}{\muo^\delta-\alpha^0_{\tkk}\mut^\delta}\\
    { \widetilde{h}^0_{\tk}}&= { \sum_{j=k+1}^{N}\prod_{\ell=k+1}^{j}{\xi^0_{\ell}}\Delta_{t_j}^{t_k}},
\end{split}\label{eq:specialcasealpha}
\end{align}
{and $\xi^0_{k}$ is given by Eq.~(\ref{eq:xi}), setting $\pi_\tkk(1,1)=0$ therein.}

\begin{remark}[Weaker Conditions For {a} Positive Spread ($\widetilde{a}_\tk^{*,0}>\widetilde{b}_\tk^{*,0}$)]
{Under the symmetry condition (\ref{Cnd1PosSpr}), it follows from (\ref{EqaSE1})-(\ref{EqaSE2}) that} 
\begin{equation}
{ \widetilde{a}_\tk^{*,0}-\widetilde{b}_\tk^{*,0}}= \widetilde{L}_{\tk}^{+,*,0}+\widetilde{L}_{\tk}^{-,*,0}
        =\dfrac{\muoo^+-2\alpha^0_{\tkk}\muto^+}{2(\muo^+-\alpha^0_{\tkk}\mut^+)}+\dfrac{\muoo^--2\alpha^0_{\tkk}\muto^-}{2(\muo^--\alpha^0_{\tkk}\mut^-)},\label{eq:Lspread}
\end{equation}
which is positive, because $\alpha^0_\tk\leq{}0$ as shown in Lemma \ref{lemma:alpha}. 
\end{remark}

{The second term in (\ref{EqaSE1})-(\ref{EqaSE2}) is the adjustment for inventory holdings, whose coefficient is negative because {$\alpha_{\tkk}^{0}<0$} per Lemma \ref{lemma:alpha}. This is intuitive, because when the inventory is positive (negative), the bid and ask placements decrease (increase) to {attract more sales (purchases) of the stock.} 
{The shadow cost of inventory is low for most of the day and only becomes significant} {near the end}. This is because, as shown in Fig.~\ref{fig:AlphaHProto}, {$\alpha^0_k$} is close to $0$ for most of the day {and {decreases rapidly} to $-\lambda$ near the end}. It is also interesting to note that the variance of $c$, {which is the random slope in the linear demand function,} reduces the shadow cost of inventory because the coefficient of $I_{t_k}$ can be written {as}:
\[
	\dfrac{\alpha^0_{\tkk}\muo^{\pm}}{\muo^{\pm}-\alpha^0_{\tkk}\mut^{\pm}}=
	\dfrac{\alpha^0_{\tkk}}{1-\alpha^0_{\tkk}\muo^{\pm}- \alpha^0_{\tkk}{\rm Var}(c^{\pm}_{\tkk}{|\F})/\muo^{\pm}}.
\]
{As ${\rm Var}(c^{\pm}_{\tkk}{|\F})$ becomes larger, the HFM tends to act less `aggressively' in order to zero out her inventory; i.e., when the inventory is positive (negative), {the second terms of $\widetilde{a}_\tk^{*,0}$ and $\widetilde{b}_\tk^{*,0}$ in Eqs.~(\ref{EqaSE1})-(\ref{EqaSE2}) become larger (smaller), thus} the ask (bid) placement is not that close to $S_\tk$ and the bid (ask) placement is less deep into the book.} {We can explain this phenomenon as follows. Consider two LOB dynamics with the same value of $\mu_{c}^\pm$, but one of those having larger ${\rm Var}(c^{\pm}_{\tkk}{|\F})$. Because $c^{\pm}\geq{}0$, the book with larger variance will generally have larger demand slope $c^{\pm}_{t_{k}}$. 
As a result, more shares of the HFM's LOs will be executed (see Fig.~\ref{fig:linearQ}) and, hence, the HFM can {act less `aggressively'} when attempting to zero out her inventory.}
 The feature just described cannot be captured by linear demand functions with deterministic slope as in  \cite{hendershott2014price} and \cite{adrian2016intraday}. We provide further analysis on the sensitivity of the optimal strategy to the inventory cost  in Subsection \ref{sec:InvAnalysis}.

\begin{figure}
\centering
\begin{subfigure}{.5\textwidth}
  \centering
  \includegraphics[width=.8\linewidth]{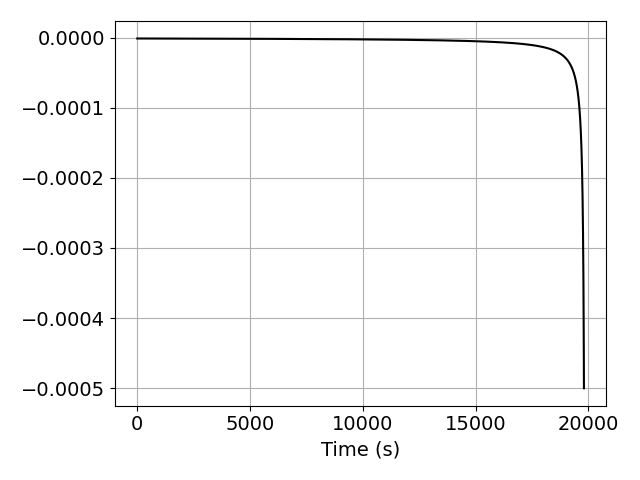}
  \caption{$\alpha^0_{\tk}$}
\end{subfigure}%
\begin{subfigure}{.5\textwidth}
  \centering
  \includegraphics[width=.8\linewidth]{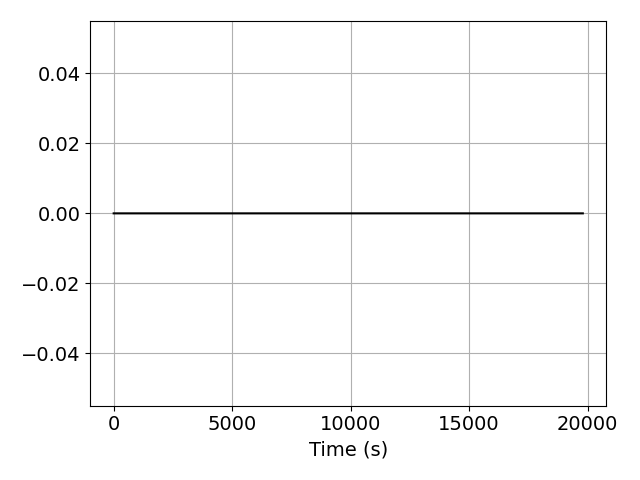}
  \caption{$\widetilde{h}^{0}_{\tk}$}
\end{subfigure}
\caption{{\textbf{{Paths of $\alpha^0_{\tk}$ and $\widetilde{h}^{0}_{\tk}$ in a Symmetric Market.}} The action times are {chosen} to be {each second from time $0$ to $19800$ seconds (5.5 hours). $\lambda = 0.0005$. We choose the parameters so that Conditions~(\ref{Cnd1PosSpr})-(\ref{Cnd2PosSpr}) and $\mu_{cp}^+=\mu_{cp}^-$ hold. Specifically,} we set $\muo^\pm=100$ , $\mut^\pm = 1\times 10^4$, $\mu_{cp}^\pm = 500$, $\mu_{c^2p}^\pm = 5\times 10^4$, $\mu_{c^2p^2}^\pm = 1\times 10^5$, $\pi_\tk^+=\pi_\tk^-\equiv 0.2$ and $\pi_\tk(1,1)\equiv 0$.}}
\label{fig:AlphaHProto}
\vspace{.5 cm}
\end{figure}

If {$\tk$} is far from the terminal time $T$ and {the market is reasonably ``symmetric"}\footnote{{That is, the { fundamental price} process is a martingale and {the} Conditions~(\ref{Cnd1PosSpr})-(\ref{Cnd2PosSpr}) of Proposition~\ref{remark:admb} are satisfied as well as $\mu_{cp}^+=\mu_{cp}^-$. Under these conditions, $\widetilde{h}^{0}_{\tk}\equiv{}0$.}},  
{the sequences} {$\alpha^0_{\tk}$ and $\widetilde{h}^{0}_{\tk}$} defined in Eq.~(\ref{eq:specialcasealpha}) are close to zero most of the time {(see Fig.~\ref{fig:AlphaHProto})}. The optimal strategy is then mainly dependent on the second term of $\widetilde{L}^{\pm,*,0}_\tk$ and the drift $\Delta_{t_k}$ in the {price dynamics of $S_\tk$}. It follows from  (\ref{EqaSE1})-(\ref{EqaSE2}) that 
\begin{align}
	\widetilde{a}_\tk^{*,0}&\approx \sk+\frac{\mu_{cp}^{+}}{2\mu_{c}^{+}}+\frac{1}{2}\Delta_{t_k}=\sk+\dfrac{\mu_{p}^{+}}{2}+\dfrac{{\rm Cov}(c^{+}_\tkk,p^{+}_{\tkk}|\F)}{2\mu_{c}^{+}} +\frac{1}{2}\Delta_{t_k},\label{eq:askpioo0}\\
	\widetilde{b}_\tk^{*,0}&\approx \sk-\frac{\mu_{cp}^{-}}{2\mu_{c}^{-}}+\frac{1}{2}\Delta_{t_k}=\sk-\dfrac{\mu_{p}^{-}}{2}-\dfrac{{\rm Cov}(c^{-}_\tkk,p^{-}_{\tkk}|\F)}{2\mu_{c}^{-}} +\frac{1}{2}\Delta_{t_k},\label{eq:bidpioo0}
\end{align}
where recall that $\mu_p^{\pm}=\E(p_{\tkk}^\pm|\F,\mathbbm{1}^\pm_\tkk=1)$. The correlation between $c$ and $p$ now plays a key role in the optimal placements.
Under the martingale {Condition (\ref{MrtCnd0}) and under Condition (\ref{Cnd3PosSpr})}, the last two terms in the optimal placements become zero. 
The optimal placements {are then} near the midpoint between $S_\tk$ and $\sk\pm{\mu_{p}^\pm}$ for most of the time. However, when the correlation between $c$ and $p$ is positive, 
{instead of placing LOs around $\sk\pm \mu_{p}^\pm/2$, the HFM will tend to go deeper into the book. Roughly, a larger realization of $c$ also implies a large value of $p$, resulting in a larger demand function and, hence, greater opportunity for the HFM to obtain better prices for her filled LOs. Another way to understand (\ref{eq:askpioo0})-(\ref{eq:bidpioo0}) is to recall that $c^{\pm}p^{\pm}$ is the y-intercept of the demand functions (see Fig.~\ref{fig:linearQ}) and, thus, the larger $\mu_{cp}^{\pm}$, the larger the demand function and the deeper the HFM could place her LOs.}
The discussion above holds for most of the day. 
However, when $t_k$ gets closer to $T$, the second term of (\ref{EqaSE1})-(\ref{EqaSE2}) will 
play }a more important role in the best strategy because $\alpha^0_{\tkk}$ is no longer close to zero by end of the day. Hence, the optimal strategy is mostly influenced by the inventory level towards the day's end.

\subsubsection{Case $\pi_\tk(1,1)\not\equiv 0$.} 
{The probability $\pi_\tk(1,1)$ of simultaneous arrivals of buy and sell MOs during a time step is typically small at high-frequency trading (say, at {1 seconds} or less). For the empirical analysis conducted in Section \ref{RRDATA0}, we find that $\pi_\tk(1,1)\approx 0.05$ for a trading period of 1 second. However, this is no longer the case if the trading frequency is smaller (say, at 5 seconds or more). In that case, it is important to account for the event of joint arrivals. 
The following corollary sheds light on the optimal placement's behavior under conditions (\ref{Cnd1PosSpr})-(\ref{Cnd3PosSpr}) plus additional conditions (which are {reasonably} met by our data} in Section \ref{RRDATA0}).

\begin{corollary}\label{cor:spread_t} 
{Under Assumptions (\ref{Cnd1PosSpr})-(\ref{Cnd2PosSpr}), the optimal spreads are
\begin{itemize}
	\item invariant to the local drifts $\{\Delta_{t_{k}}\}_{k=0,\dots,N}$;
	\item independent on the inventory level.
\end{itemize}
Suppose that, in addition to (\ref{Cnd1PosSpr})-(\ref{Cnd2PosSpr}), the condition (\ref{Cnd3PosSpr}) as well as the following conditions hold:
\begin{align}\label{Cnd1Pioo}
	&\mu^2_c = \mu_{c^2},\qquad
\pi^\pm_{\tk}\equiv \pi^\pm,\quad  \pi_{\tk}(1,1)\equiv\pi(1,1),
\end{align}
for some constants $\pi^{\pm}\in (0,1)$ and $\pi(1,1)\in{[(\pi^++\pi^--1)\vee 0,\pi^+\wedge\pi^-]}$.
Then,  the optimal spreads are}
\begin{itemize}
\item non-decreasing with time and, if $\pi(1,0)=\pi(0,1)=0$, they are flat throughout the trading horizon;
\item decreasing with $\pi(1,1)$ at {any given} time point.
\end{itemize}
\end{corollary}
We prove Corollary~\ref{cor:spread_t} in Appendix~\ref{pf:spread_t}. We know that while the drift in the mid-price process and the HFM's net inventory position can affect the {optimal} bid and ask prices {at any given time}, the optimal spread is invariant to the specific value of the drift and inventory position. It can be seen from Fig.~\ref{fig:OptSprdPioo} that, as $t_k$ approaches the terminal time $T$, the optimal spread widens due to the penalty placed on the terminal inventory. By widening the optimal spread, the HFM attempts to trade predominantly on one side of the book (say, sell side if inventory is positive), so to control the inventory level. As $\pi(1,1)$ increases, the probability {of simultaneous arrivals of buy and sell MOs increases, hence, providing} more opportunities for the HFM to manage {her} inventory. This is because the positive net position resulting from the execution of sell MOs and the negative net position corresponding to the execution of buy MOs are more likely to be canceled out with each other {when $\pi(1,1)$ is positive}. 
{Thus, the HFM tends} to narrow the spread to get more LOs filled on both sides of the book and gain larger profit.

\begin{figure}[h]
    \centering
    \includegraphics[width=.4\textwidth]{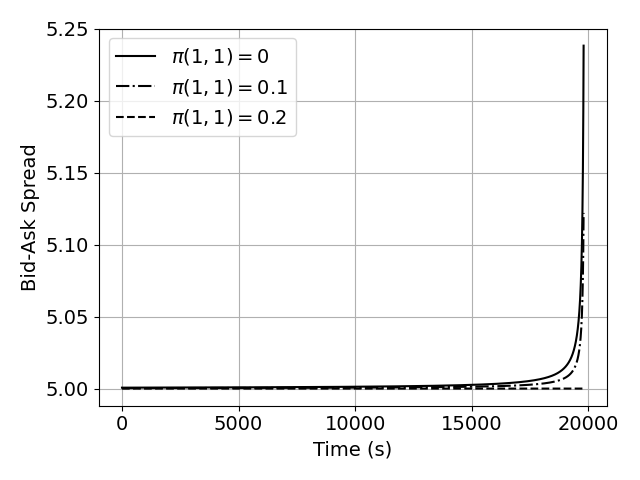}
    \caption{\textbf{Optimal Bid-Ask Spread within a Trading Horizon.} The action times are {chosen} to be {each second from time $0$ to $19800$ seconds (5.5 hours). $\lambda = 0.0005$. We choose the parameters so that Conditions~(\ref{Cnd1PosSpr})-(\ref{Cnd3PosSpr}) and {(\ref{Cnd1Pioo})} hold. Specifically,} we set $\muo^\pm=100$ , $\mu_p^\pm = 5$, $\mu_{c^2p^2}^\pm = 1\times 10^5$, $\pi^+=\pi^-= 0.2$. $\pi(1,1)$ ranges from $0$ to $\pi^\pm$. {These parameter values are consistent with the empirical estimates obtained from our data and given} in Section \ref{RRDATA0}.}
    \label{fig:OptSprdPioo}
\end{figure}

\subsection{Sensitivity of the Optimal Strategies to Inventory Holdings}\label{sec:InvAnalysis}
{We {now} generalize the sensitivity analysis of optimal placements on inventory levels developed earlier for the case  $\pi(1,1)=0$. We account for the nonzero probability of joint arrivals, i.e. for $\pi(1,1) > 0$.}
\begin{corollary}\label{cor:ABI}
The optimal ask price $\widetilde{a}_\tk^*$ and bid price $\widetilde{b}_\tk^*${,} as defined in Eq.~\eqref{eq:tilde_bNMG}, are strictly decreasing with inventory $I_\tk$.
\end{corollary}
The proof is given in Appendix~\ref{pf:ABI}. Corollary~\ref{cor:ABI} reflects the HFM's ability to control inventory through the optimal  {placement} strategies under a general adapted fundamental price process. 
When the HFM has a large net long inventory position, she puts ask and bid quotes at lower price to accelerate selling and dampen buying activities. If instead the inventory position becomes large but net short, she will raise the bid and ask prices to accelerate buying and dampen selling.

Fig.~\ref{fig:OptAskBidInv} plots the distance of the HFM's {optimal} ask and bid {quotes} from the {fundamental price $S_\tk$} 
within the last 500-seconds before the end of trading, {for different inventory levels}. As we mentioned in Subsection~\ref{sec:pioo0}, if the market is reasonably ``symmetric" and {under the assumption that $c$ and $p$ are independent,} the {agent's optimal} ask spread, {$\widetilde{a}_\tk^{*}-\sk$}, and bid spread, {$\sk-\widetilde{b}_\tk^{*}$}, are close to $\mu_p^+/2$ and $\mu_p^-/2$, respectively, for most of the time, {regardless the inventory level.} {We remark that $\mu_p^{\pm}/2$ equals $2.5$ ticks under the parameter specification used to produce Fig.~\ref{fig:OptAskBidInv}. By the end of trading, the optimal {ask and bid} prices become sensitive to {the} inventory level.} The HFM {then} chooses {from two} different strategies {depending} on {her} inventory level:
{ 
\begin{itemize}
\item If {her} inventory level {is low} ({e.g.,} between $-250$ to $250$ shares), the HFM will widen both bid and ask { spreads} to decrease both buying and selling, as {$t_k$} approaches $T$ and, hence, keep the inventory low till the end;
\item When {her} inventory level is high ({e.g.,} outside $-250$ to $250$ shares), the HFM will narrow {her} ask (bid) spread to facilitate selling (buying) {of} shares, while widening {the} bid (ask) spread to dampen buying (selling) with a large positive (negative) net position.
\end{itemize}}
{Thus, under the parameter specification used to produce Fig.~\ref{fig:OptAskBidInv}, the inventory levels $\pm250$ shares are boundaries for the different end-of-horizon behaviors as described above.} 

\begin{figure}[h]
    \centering
    \includegraphics[width=.7\textwidth]{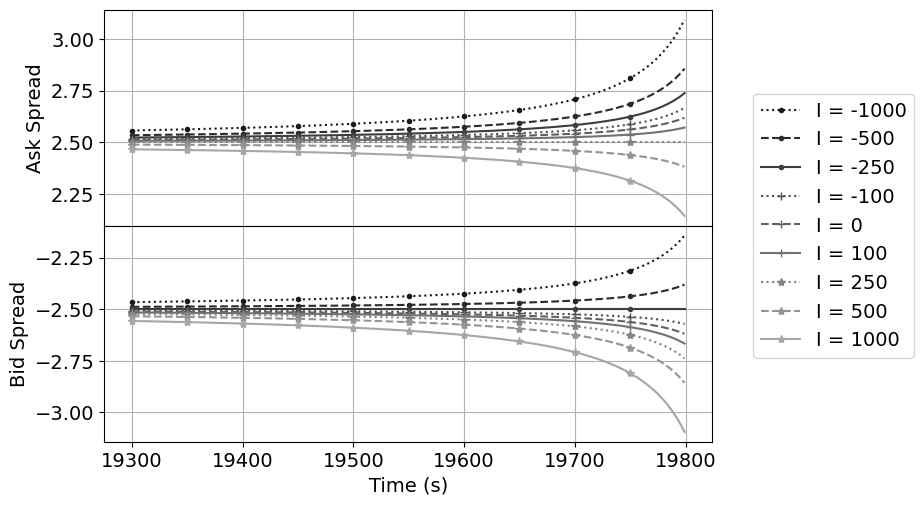}
    \caption{\textbf{{Optimal Bid and Ask Spreads in the Last 500-Seconds for Various Inventory Levels.}} The action times are {chosen} to be {each second.} {We choose the parameters so to satisfy  Conditions~(\ref{Cnd1PosSpr})-(\ref{Cnd3PosSpr}) and {(\ref{Cnd1Pioo})}. Specifically, we set $\muo^\pm=100$ , $\mu_p^\pm = 5$, $\mu_{c^2p^2}^\pm = 1\times 10^5$, $\pi^+=\pi^-= 0.2$, $\pi(1,1)=0$, and $\lambda = 0.0005$. These parameter values are consistent with the estimates obtained from real data in Section} \ref{RRDATA0}.}
    \vspace{.5 cm}
    \label{fig:OptAskBidInv}
\end{figure}

\subsection{Sensitivity of the Optimal Strategies to Inventory Penalty}
We first consider the case of $\pi_\tkk(1,1)=0$ under a symmetric market and martingale dynamics for the {fundamental price} process. The following result then characterizes the optimal placements relative to the baseline price level 
{$S_{t_{k}}\pm\mu_{p}/2$.} The proof is given in Appendix~\ref{pf:InvThreshold}. 
\begin{corollary}\label{cor:InvThreshold} Assume the market is symmetric (i.e., Conditions~(\ref{Cnd1PosSpr})-{(\ref{Cnd3PosSpr})} of Proposition~\ref{remark:admb} hold and that 
{$\mu_p:=\mu_{p}^+=\mu_{p}^-$}), and $\pi_\tkk(1,1)=0$ (i.e., only one type of MOs can arrive during each subinterval). Then, under a martingale {fundamental price} process, there exists a threshold for the inventory level,
{
$$
\overline{I}^\pm=\pm\dfrac{\mut\mu_p}{2\muo},
$$ }
such that the following statements hold for every penalty term {$\lambda>0$}:
\begin{itemize}
\item When the inventory level $I_\tk\in(\overline{I}^-,\overline{I}^+)$, the optimal strategy is to place the ask and bid quotes deeper in the LOB {relative} to the levels 
{${S_{t_{k}}+\mu_{p}/2}$} and {${S_{t_{k}}-\mu_{p}/2}$}, respectively; 
\item When the inventory level $I_\tk>\overline{I}^+$ ($I_\tk<\overline{I}^-$), the optimal strategy is to place the ask (bid) quote closer to {$S_{t_{k}}$} than {to} { ${S_{t_{k}}+\mu_{p}/2}$} ( {${S_{t_{k}}-\mu_{p}/2}$}), and the bid (ask) quote {farther} from {$S_{t_{k}}$} than {from}  {${S_{t_{k}}-\mu_{p}/2}$} ({${S_{t_{k}}+\mu_{p}/2}$}) into the LOB.
\end{itemize}

\end{corollary}

It can be seen from Fig.~\ref{fig:OptABLmbdaPioo0} that if there is no inventory holding cost, the optimal strategy is to keep the ask and bid prices constant throughout the day, no matter how {much inventory} the HFM holds. With {a} larger penalty, {a} HFM with {a} positive net position will {place} {her} bid LOs deeper into the book near the end of the trading horizon to avoid more  purchases {of stock}. For the ask side, she {{will} pick one of three different strategies{\Blue :} (a) place the ask LO further {from} {$S_\tk$}, (b) place the ask LO closer to {$S_\tk$}, or (c) keep the ask LO at the same price level as earlier in the trading day. {According to Corollary~\ref{cor:InvThreshold}, the selection between (a), (b), or (c) depends on whether $I_{t_k}>\overline{I}^{+}$, $I_{t_k}=\overline{I}^{+}$, or $I_{t_k}<\overline{I}^{+}$, respectively}. Under the parameter specification used to produce Fig.~\ref{fig:OptABLmbdaPioo0}, we find that the threshold inventory is $\overline{I}^\pm=\pm250$.}

The left panel of Fig.~\ref{fig:OptABLmbdaPioo0} {shows} the optimal placements when the position is net long. As the inventory approaches $100$ holdings, the HFM places both bid and ask orders deeper into the book near the end of trading horizon in order to maintain the present inventory level. {The larger the penalty for terminal inventory, the deeper she will go into the LOB on both sides.}
If the inventory consists of $250$ shares, the left panel of the second row in Fig.~\ref{fig:OptABLmbdaPioo0} indicates that the optimal bid placement goes even deeper into the book{\Green,} while the price trajectory of the optimal ask remains flat till the end of the trading horizon. {If the inventory reaches the threshold {$\bar{I}^{\pm}$}, the strategy of the HFM is insensitive to the inventory penalty, no matter how large it is.} However, when the inventory level reaches $500$, the HFM needs to lower the optimal ask price in order to {get} more ask LOs executed {and, hence, lower the} inventory level. {The higher the inventory, the closer she puts her ask quote to the mid-price.} {A similar discussion applies to the case of negative inventory positions, as} shown in the right panel of Fig.~\ref{fig:OptABLmbdaPioo0}.
\begin{figure}[h]
    \centering
    \includegraphics[width=0.8\textwidth]{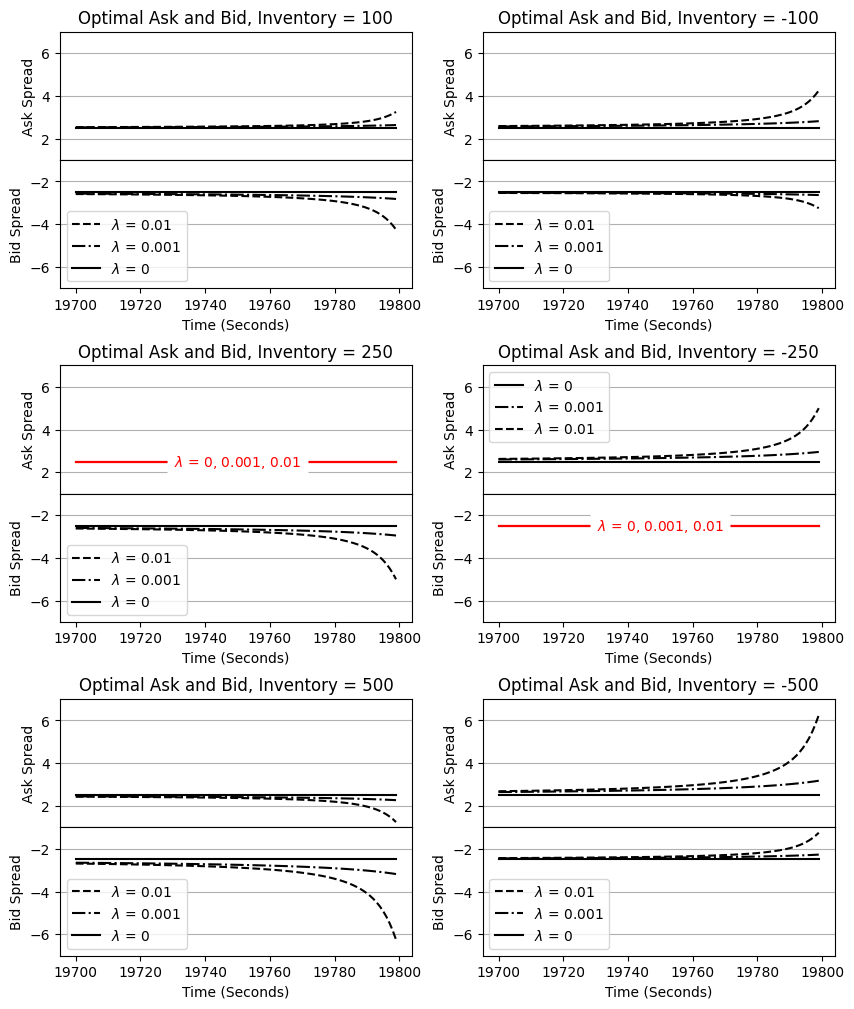}
    \caption{\textbf{Optimal strategies in the last 100 seconds, for various inventory and penalty levels.} The action times are {chosen} to be each second. We let $\lambda$ range from $0$ to $0.01$. We set $\muo^\pm=100$ , $\mu_p^\pm = 5$, $\mu_{c^2p^2}^\pm = 1\times 10^5$, $\pi^+=\pi^-= 0.2$, $\pi(1,1)=0$. These parameters satisfy Conditions~(\ref{Cnd1PosSpr})-(\ref{Cnd3PosSpr}) and {(\ref{Cnd1Pioo})}. The parameter values are consistent with the estimates obtained from real data in Section \ref{RRDATA0}.}
    \label{fig:OptABLmbdaPioo0}
    \vspace{.5 cm}
\end{figure}

{However, if we allow for joint arrivals, i.e. {$\pi_\tk(1,1)> 0$}, we can observe significant differences in the} optimal strategies. Fig.~\ref{fig:OptABLmbdaPiooNot0} {illustrates that, if $\pi_\tk(1,1)>0$, there does not exist an inventory level such that the optimal prices are flat throughout the trading horizon}. Furthermore, for some inventory levels, the optimal strategies {are no longer monotonic in time.} We observe a valley (peak) pattern in the optimal ask (bid) placement for some positive (negative) intermediate inventory level. {Our intuition for this trading pattern is as follows. Consider, for example,} the left {panel} in Fig.~\ref{fig:OptABLmbdaPiooNot0}, where $I_\tk = 250$. {If there is still enough trading time left, it is of higher priority for the HFM to sell more and lower the inventory level because there will be opportunities to buy later and profit from the roundtrip transaction.} {However, as} the time gets closer to the terminal time $T$, it {becomes} more important to profit directly from less, but wider, roundtrip transactions because there is not enough time left for the market maker to conclude the roundtrip transaction. {If buy and sell MOs can arrive simultaneously, roundtrip transactions {are more likely} to happen within two consecutive actions. Notice that as the penalty $\lambda$ gets larger, the optimal strategy becomes more `aggressive' because of the stronger incentives to make higher profits and compensate for the cost of holding terminal inventory.}

\begin{figure}[h]
    \centering
    \includegraphics[width=1\textwidth]{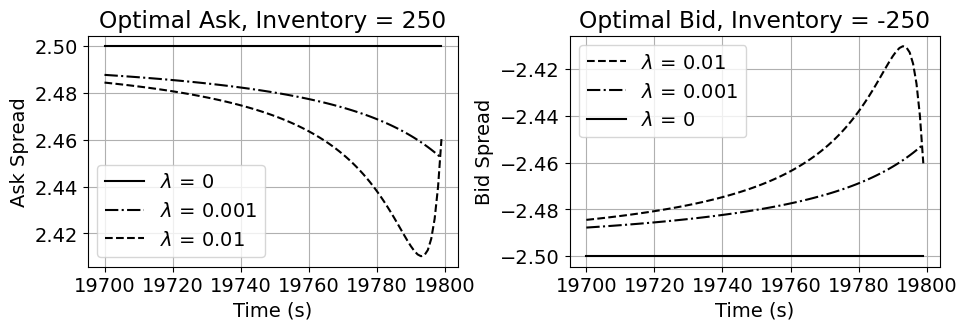}
    \caption{\textbf{The last 100 seconds optimal strategies with various inventory and penalty levels.} The action times are {chosen} to be every 1 second.  $\lambda$ ranges from $0$ to $0.01$. $\muo^\pm=100$ , $\mu_p^\pm = 5$, $\mu_{c^2p^2}^\pm = 1\times 10^5$, $\pi^+=\pi^-= 0.2$, $\pi(1,1)=0.05$ and the parameters satisfy Conditions~(\ref{Cnd1PosSpr})-(\ref{Cnd3PosSpr}) and {(\ref{Cnd1Pioo})}. The values of the parameters are consistent with the estimates from real data given in Section \ref{RRDATA0}.}
    \label{fig:OptABLmbdaPiooNot0}
    \vspace{.5 cm}
\end{figure}


\section{{Data Calibration and Performance Analysis}}\label{RRDATA0}
This section studies the performance of the optimal strategies derived in Section \ref{sec:Bellman} using real LOB data. We first describe the data set and the parameter estimation procedure. We then present the performance analysis, and additionally compare the optimal strategy  against ``benchmark'' strategies that place limit orders on fixed levels in the limit order book.

\medskip
\noindent
\textbf{Data.} We use of LOB data of the MSFT stock during the year of 2019 (252 days in total). Our data set is obtained from Nasdaq TotalView-ITCH 5.0, which is a direct data feed product offered by The Nasdaq Stock Market, LLC\footnote{\tt{http://www.nasdaqtrader.com/Trader.aspx?id=Totalview2}}. TotalView-ITCH uses a series of event messages to track any change to the state of the LOB. For each message, we observe the timestamp, type, direction, volume, and price. We reconstruct the dynamics of the top 20 levels of the LOB directly from the event message data. We treat each day as an independent sample.

\medskip
\noindent
\textbf{Actions.} We assume no latency in the HFM's actions and the HFM's order is always ahead of the queue of the LOs with same price in the LOB. We fix the action times for the HFM to be every second of a trading period running from 10:00 a.m. to 15:30 p.m. Thus, the HFM acts $19800$ times in a regular trading day. At the beginning of each 1-second subinterval, the HFM places an ask and a bid LO, each of a fixed volume. The volume is set to be 500 shares, roughly matching the average volume of MOs arriving within 1-second intervals. The tick size of MSFT stock is one cent. The calculated optimal ask (bid) price is round-up (down) to the nearest tick such that, in reality, the order can be executed at a better price while the filled size remains unchanged.

\medskip
\noindent
\textbf{{Historical Window Size for Parameter Estimation.}} {The parameters plugged into the optimal strategy for the current day are estimated via historical averages including the prior 20 trading days. Recall that those parameters are the arrival probabilities $\pi_\tkk^\pm, \pi_\tkk(1,1)$ defined in Eqs.~(\ref{eq:pi})-(\ref{eq:pipm}), and the conditional expectations related to $(c_.^\pm,p_.^\pm)$, defined in Eq.~(\ref{Dfnmucp}). Because there is {a total of} 252 trading days in year 2019, we {compute terminal} {revenues} for 232 days, i.e., starting from the 21st trading day.}
\subsection{Parameter Estimation}\label{sec:paraest}

\medskip
\noindent
\textbf{Frequencies of MOs.} During a typical trading day, sell and buy MOs usually arrive more frequently near the opening or closing of the stock {market}. To capture this ‘U’ shape intraday pattern, we model {the parameters {$\pi^\pm_\tkk$ and $\pi_\tkk(1,1)$} defined} in Eqs.~(\ref{eq:pi})-(\ref{eq:pipm}) as quadratic deterministic functions of {the} time {$\tkk$}. {More specifically, for} the i-th trading day, we {first compute}
 \begin{align}
\bar{\pi}_\tkk^{\pm,i}&= \dfrac{1}{20}\sum_{{j} = 1}^{20}\mathbbm{1}_{t_{k+1}}^{\pm,i-{j}},\\
\bar{\pi}_\tkk^{i}(1,1)&= \dfrac{1}{20}\sum_{{j} = 1}^{20}(\mathbbm{1}_{t_{k+1}}^{+,i-{j}}\cdot\mathbbm{1}_{t_{k+1}}^{-,i-{ j}}),\label{EstimatePiH}
\end{align}
where $\mathbbm{1}_{t_{k+1}}^{\pm,i}$ are the MO indicators defined in Eq.~(\ref{eq:MOind}) for trading day $i$. By conducting {a least-squares quadratic fit to the} time series of arrival probabilities $\bar{\pi}_\tkk^{\pm,i}$ and $\bar{\pi}_\tkk^{i}(1,1)$, we obtain the estimates of $\pi^\pm_\tkk$ and $\pi_\tkk(1,1)$ for the i-th trading day. We denote these estimates by  $\hat{\pi}^{\pm,i}_\tkk$ and $\hat{\pi}^{i}_\tkk(1,1)$, respectively. Fig.~\ref{fig:pipath} shows the prototypical intraday patterns of { $\hat{\pi}^\pm_\tkk$ and $\hat{\pi}_\tkk(1,1)$}.

\begin{figure}[h]
    \centering
    \includegraphics[width=.5\textwidth]{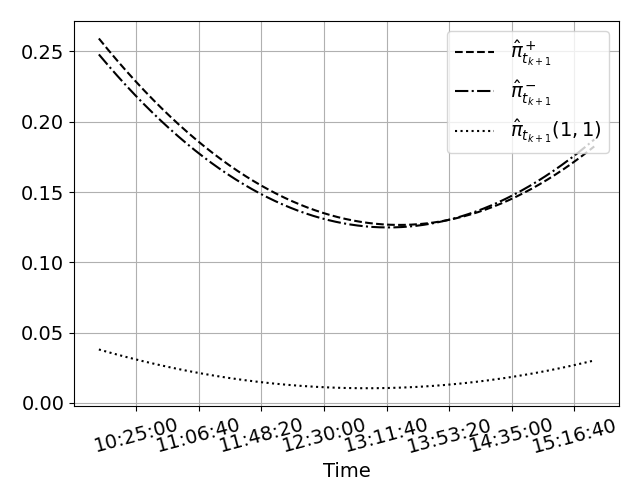}
    \caption{\textbf{Prototypical Trajectories of MOs Frequencies {within} a Trading Day.} During a typical trading day, sell and buy MOs arrive more frequently when the time is closer to the opening or closing of the {market}. We model {$\pi^\pm_\tkk$ and $\pi_\tkk(1,1)$} as quadratic deterministic functions of time.}\label{fig:pipath}
    \vspace{.5 cm}
\end{figure}

\medskip
\noindent
\textbf{Demands Function.} For each {1-second} subinterval within {a day {used to estimate the parameters of the model}}, {we {first} compute the actual demand at each price level. 
{Suppose the HFM places ask LOs at price level $P_l$ at time $\tk$. At time $t_i\in[\tk,\tkk)$, we observe a buy MO$_i$ with volume $V_{MO_{i}}$ submitted to the market, and the volume of existing ask LOs in the book with prices lower than $P_l$ at this moment is $V_{LO_{i}}$. Then the number of shares to be filled with this buy MO in the HFM's placement equals to $(V_{MO_{i}}-V_{LO_{i}})\vee0$. We compute this quantity for all buy MOs arriving during the interval $[\tk,\tkk)$, and use $\sum_{i}\big((V_{MO_{i}}-V_{LO_{i}})\vee0\big)$ to quantify the actual demand at price level $P_l$ during $[\tk,\tkk)$. The computation on the bid side is symmetric (see the piecewise constant {graph} in Fig.~\ref{DemandPlot0} for an example of the actual demand during a {1-second} subinterval). Then we conduct a weighted linear regression on each side of the book, with the actual demand being the response variable and the price level (specifically, its distance to $S_\tk$) being the predictor, to estimate $Q_\tkk^\pm$.}}
We place higher weight on price levels closer to $S_\tk$ and smaller weights on the price levels which are deep in the book.
{Fig.~\ref{DemandPlot0} shows the prototypical linear fit to the {actual} demand function in one subinterval.} Fig.~\ref{fig:CPstationary} {plots} {the} estimated {time series} $(c^\pm_\tk,p^\pm_\tk)$ throughout a trading day. By {virtue of the} augmented Dickey-Fuller test (ADF), all $(c_.^\pm,p_.^\pm)$-related time series {defined} in Eq.~(\ref{Dfnmucp}) are reasonably stationary.

\begin{figure}
    \centering
  \includegraphics[width=.5\textwidth]{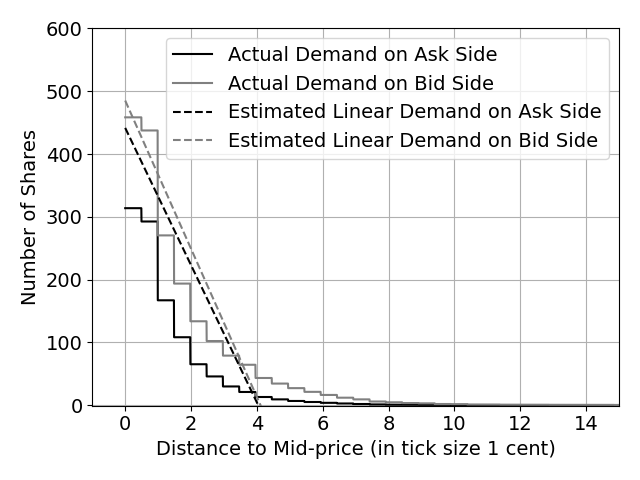}
  \caption{\textbf{{Prototypical Plot of Actual Demand} vs. Estimated Linear Demand over a 1-Second Trading Interval}}%
  \label{DemandPlot0}
  \vspace{.5 cm}
\end{figure}

\begin{figure}
    \centering
    \includegraphics[width=1\textwidth]{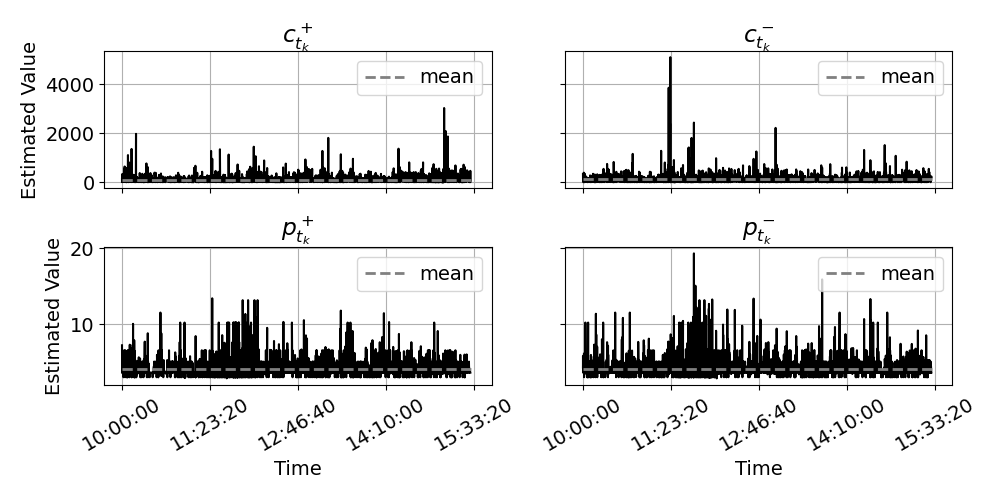}
     \caption{\textbf{Estimated Values of $(c^\pm_\tk,p^\pm_\tk)$ throughout a Prototypical Trading Day.} For each coefficient, the average value is shown in gray dashed line.}
    \label{fig:CPstationary}
    \vspace{.5 cm}
\end{figure}

We then proceed to {estimate the $i$-day conditional expectations $\mu^{\pm,i}_{\{c,p\}}$ defined in Eq.~(\ref{Dfnmucp})} by averaging {the} corresponding regression parameters over {all} subintervals within that day. {We denote these estimates by $\hat{\mu}^{\pm,i}_{\{c,p\}}$}. Table \ref{tab:ParameterEst} shows the average of {$\hat{\mu}^{\pm,i}_{\{c,p\}}$} over all $252$ trading days in 2019. {These results suggest that the {symmetry} assumption imposed on the demand of buy and sell orders (Eq.~(\ref{Cnd1PosSpr})), and the assumption of independence between $c^\pm$ and $p^\pm$ (Eq.~(\ref{Cnd1PosSpr})) are largely satisfied. In our implementation, the estimates of $\mu^{\pm}_{\{c,p\}}$ used {to compute the optimal strategies in} the i-th trading day are obtained by averaging {$\hat{\mu}^{\pm,i-j}_{\{c,p\}}\ (j = 1,\dots,20)$} {over the previous} 20 training days.}
\
\
\begin{table}[ht]
\centering
\setlength{\tabcolsep}{12pt}
\begin{tabular}{ll}
\hline
$\bar{\mu}_c^+ = 94.86$ & $\bar{\mu}_c^- = 98.801$ \\ \hline
{$\bar{\mu}_p^+ = 3.977$} & {$\bar{\mu}_p^- = 3.981$} \\ \hline
$\bar{\mu}_{cp}^+ = 432.06$ & $\bar{\mu}_{cp}^- = 451.53$ \\ \hline
$\bar{\mu}_{c^2}^+ = 6.74\times 10^4$ & $\bar{\mu}_{c^2}^- = 3.94\times 10^4$ \\ \hline
${\bar{\mu}_{p^2}^+ = 17.25}$ & ${\bar{\mu}_{p^2}^- = 17.30}$ \\ \hline
$\bar{\mu}_{c^2p}^+ = 3.22\times 10^5$ & $\bar{\mu}_{c^2p}^- = 1.90\times 10^5$ \\ \hline
$\bar{\mu}_{c^2p^2}^+ = 1.73\times 10^6$ & $\bar{\mu}_{c^2p^2}^- = 1.08\times 10^5$ \\ \hline
\end{tabular}
\vspace{.2cm}
\caption{{$\bar{\mu}^{\pm}_{\{c,p\}}$: Average Values of $\hat{\mu}^{\pm,i}_{\{c,p\}}$ over 252 Trading Days in 2019. }}
\label{tab:ParameterEst}
\end{table}

\medskip
\noindent
\textbf{Drift of the Midprice Process.} {Following standard conventions in the literature, we set the fundamental price $S_{t_{k}}$ to be the midprice, i.e., the average of the best bid and best ask prices (see also \cite{hendershott2014price}).} Recall, from Section \ref{sec:DriftAnaly}, the definition $\Delta_\tk = \E(S_\tkk-S_\tk|\F)$, where $S_\tk$ is the midprice at time {$\tk$}. Since the optimal strategies are  computed using {a} backward induction algorithm, we need to estimate $\Delta_\tk$, and additionally make predictions on future price changes conditioned on the present information (see Eq.~\eqref{eq:LtildeNMGbb})}. For computational efficiency {(see also Remark \ref{CommentsADrift0} below for further discussion)}, we {hereafter} assume that
\begin{equation}\label{AsmpDriftF}
{{\Delta_{t_{j}}^{t_k}}=\E(S_{t_{j+1}}-S_{t_{j}}|\mathcal{F}_{t_{k}})=0, \quad j\geq{}k+1}.
\end{equation}
Under this assumption, {Eq.~(\ref{eq:LtildeNMGbb}) simplifies as \small
\begin{align*}
\     \widetilde{L}_{\tk}^{\pm,*}&={L}_{\tk}^{\pm,*}\pm\Big({
     {\Black \frac{\pi^{+}_\tkk \pi^{-}_{\tkk}}{2\gamma_{t_{k}}}}(\alpha_{\tkk}\mut^\mp-\muo^\mp)\muo^\pm
	    \mp{\Black \pi_\tkk(1,1)}\frac{\alpha_{\tkk}}{2\gamma_{t_{k}}}\muo^+\muo^-\pi^\mp_{\tkk}\muo^\mp}\Big)\Delta_{t_{k}},
\end{align*}
\normalsize
where ${L}_{\tk}^{\pm,*}$ are the optimal spreads defined in {Proposition} \ref{prop:optimalcontrol}. The above expression indicates that we only need to predict the immediate midprice change to compute the optimal strategy. In our implementation with real data, we estimate $\Delta_{t_k}$ by taking {the} average over the last 5 increments in the midprice:
\begin{equation}
    \hat{\Delta}_{t_k}= \frac{1}{5}\sum_{i = 1}^5 (S_{t_{k-i+1}}-S_{t_{k-i}})=\dfrac{S_{t_{k}}-S_{t_{k-5}}}{5}.
\end{equation}
In this way{\Green,} the optimal strategy with $\Delta_\tk$ is able to respond {quicker} to {local midprice trends}.

\begin{remark}\label{CommentsADrift0}
	{In practice, one could expect ${\Delta_{t_{j}}^{t_k}}=\E(S_{t_{j+1}}-S_{t_{j}}|\mathcal{F}_{t_{k}})$ to quickly decrease to $0$ as $j$ is farther away from $k$, otherwise, statistical arbitrage opportunities would appear. Furthermore, the estimation error of the forecasts $\Delta_{t_{j}}^{t_k}$ increases quickly as $t_{j}$ is farther away from $t_k$. Hence, the reduction in the misspecification error (the error in assuming that $\Delta_{t_{j}}^{t_k}=0$ when they are not) will be offset by the estimation error of the forecasts $\Delta_{t_{j}}^{t_k}$. Therefore, in practice, it is better to consider very few steps ahead forecasts in formula \eqref{eq:LtildeNMGbb}. The assumption (\ref{AsmpDriftF}) appears to be a good compromise between accuracy and computational efficiency.}
\end{remark}

\subsection{Results}\label{sec:232result}
This section shows the performance of optimal strategies on the MSFT stock during the year 2019. {We compute the terminal cash flow $W_T$ and inventory $I_T$ for each trading day by executing the optimal strategy over a time period against the observed market data. Within each subinterval $[\tk,\tkk)$, the change in inventory is given by $I_{t_{k+1}}-I_\tk = -\widetilde{Q}_\tkk^++\widetilde{Q}_\tkk^-$, where $\widetilde{Q}_\tkk^+$ and  $\widetilde{Q}_\tkk^-$, are the actual numbers of filled shares in the HFM's placement on ask and bid side, respectively, and computed from transaction data (in the same way as we compute the actual demand described in Section \ref{sec:paraest}). The change in cash flows are given by $W_{t_{k+1}} - W_\tk = a_\tk \widetilde{Q}_\tkk^+-b_\tk \widetilde{Q}_\tkk^-$, where $a_\tk, b_\tk$ are, respectively, the ask and bid prices implied by the strategy.} As a comparison benchmark, we {also} consider fixed-level strategies, which always quote at some fixed level in the LOB (e.g., always quote at level I, level II, etc...){.}

\medskip
\noindent
\textbf{Control on Terminal Inventory.}
Fig.~\ref{fig:OptimalPaths}  shows the intraday price and inventory paths of the optimal {strategy} compared with the `Level 1'- `Level 6' strategies for a prototypical trading day. As we can see from Fig. \ref{fig:PricePath}, the optimal prices {typically} swing between {the levels 2 and 3} in the LOB at the beginning of the trading {period}. {During the last portion of} the trading horizon, the optimal ask prices go down from level 3 to level 2, and the optimal bid prices go down from level 3 to level 6. {This is the case because}, as the HFM gains {a} positive net position during the trading process ({see} Fig. \ref{fig:InvPath}), {she} gradually lowers both {her} ask and bid prices to buy less and sell more {and, hence, to revert} the net position towards zero. 
 {As shown in Fig. \ref{fig:InvPath},} from 10:00 am-12:30 pm, the level of the net position goes positive under each strategy, {likely because of the decreasing midprice trend early in the day}. However, if the HFM executes {according} to the optimal {strategy}, the penalty on the terminal inventory {prevents} the inventory from {exploding} and pulls it back close to zero by the end. This shows that the effectiveness of the liquidation penalty $-\lambda I_T^2$ in controlling inventory and avoiding large end of the day costs.

\begin{figure}
\centering
\begin{subfigure}[b]{1\textwidth}
\centering
   \includegraphics[width=.9\textwidth]{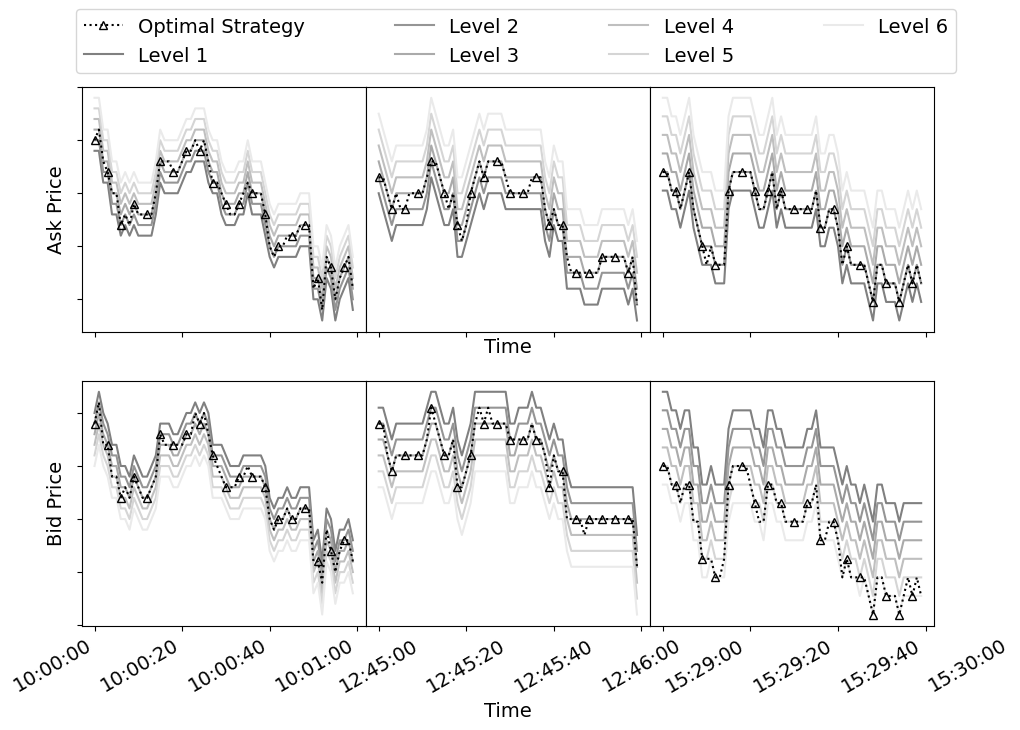}
   \caption{The Intraday Prices Paths.}
   \label{fig:PricePath} 
\end{subfigure}

\begin{subfigure}[b]{1\textwidth}
\centering
   \includegraphics[width=.5\textwidth]{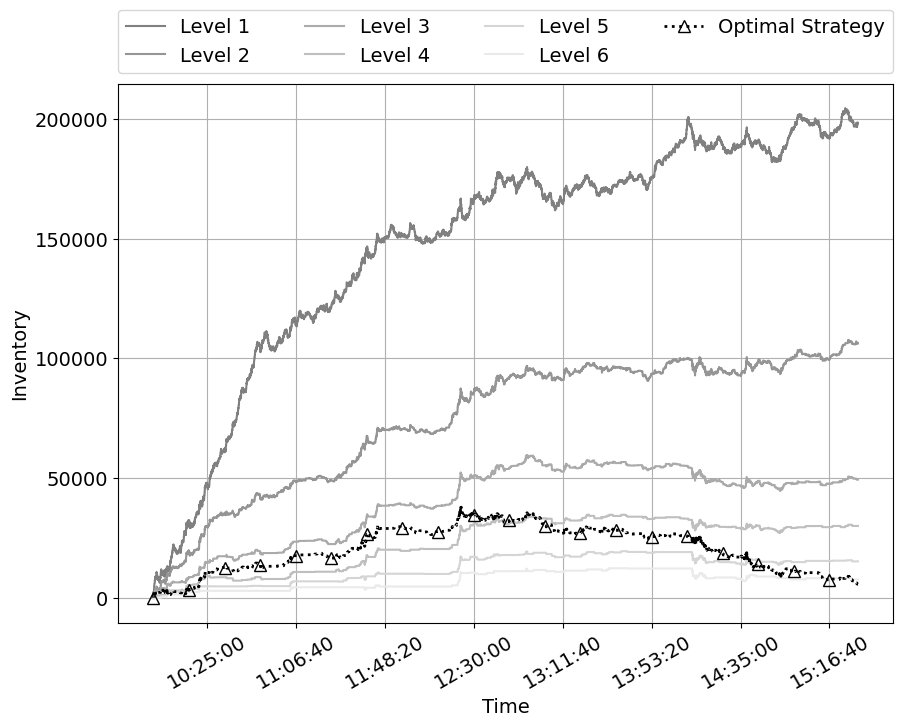}
   \caption{The Intraday Inventory Paths.}
    \label{fig:InvPath}
\end{subfigure}

\caption[Optimal Paths]{\textbf{The Intraday Price and Inventory Paths of the Optimal {Strategy} Compared with the Benchmark Strategies for a Prototypical Trading Day.} `Optimal Strategy' corresponds to the optimal strategy under the non-martingale {price} assumption. `Level 1'- `Level 6' represent the benchmark strategies that place LOs at a fixed level (i.e. level 1- level 6, respectively) in the LOB. (a) Upper row shows prices on the ask side and lower row shows prices on the bid side. Three columns from left to right represent three 1-minute time windows, which are at the beginning of the trading horizon 10:00$-$10:01, in the middle of the trading horizon 12:45$-$ 12:46, and at the end of the trading horizon 15:29$-$15:30, respectively.
}
\label{fig:OptimalPaths}
\end{figure}

\medskip
\noindent
\textbf{Probability Distribution of {Terminal Value.}} 
Table~\ref{tab:TerminalReward} reports the {means and standard deviations} of the terminal {objective values $W_T+S_TI_T-\lambda I_T^2$} under different strategies. `Level 1'- `Level 6' represent the benchmark strategies that place LOs at a fixed level (i.e., level 1- level 6, respectively) in the LOB. {For comparison,} Table~\ref{tab:TerminalRewardLiquid} presents the means and standard deviations of the terminal {values} $W_{T}+\bar{S}_{T}I_{T}$, computed using the actual average price $\bar{S}_{T}$ per share that the HFM will get when liquidating her inventory $I_T$ {with MOs} based on the state of the book at time $T$. {We refer to $\bar{S}_{T}I_{T}$ as the liquidation proceeds.}  We do not observe significant differences with the results presented in Table~\ref{tab:TerminalReward}. This suggests that {the} penalty parameter $\lambda$, fixed to be $0.0005$, guarantees that the term $(S_T-\lambda I_T)I_T$ in the objective function matches well the realized average {proceeds} of liquidating all net positions using MOs at end of the trading horizon.

\begin{table}[t]
\centering
\setlength{\tabcolsep}{8pt}
\resizebox{\textwidth}{!}{%
\begin{tabular}{@{}ccccccc@{}}
\toprule\midrule
 & \multicolumn{2}{c}{\begin{tabular}{c}Optimal Strategy\\ with Non-Martingale {Fundamental Price}\\ {and $\pi_\tk(1,1)\geq 0$}\end{tabular}} & \multicolumn{2}{c}{\begin{tabular}{c}Optimal Strategy\\ with Martingale {Fundamental Price}\\ {and $\pi_\tk(1,1)\geq 0$}\end{tabular}}&
 \multicolumn{2}{c}{\begin{tabular}{c}{Optimal Strategy}\\ {with Non-Martingale Fundamental Price} \\{and $\pi_\tk(1,1)\equiv 0$}\end{tabular}} \\ \midrule
Mean & \multicolumn{2}{c}{$6.13\times10^4$} & \multicolumn{2}{c}{$5.80\times10^4$}& \multicolumn{2}{c}{{ $6.11\times10^4$}}                                               \\ \midrule
Std. & \multicolumn{2}{c}{$1.22\times10^6$} & \multicolumn{2}{c}{$1.30\times10^6$} & \multicolumn{2}{c}{{ $1.22\times10^6$}}                                              \\ \toprule\midrule
     & Level 1             & Level 2             & Level 3            & Level 4              & Level 5              & Level 6              \\ \midrule
Mean & $-7.78\times10^{6}$        &  $-9.99\times10^{5}$       &    $	-1.14\times10^5$     &   $-3.64\times10^4$      &     $-5.16\times10^4 $   &$-3.69\times10^4  $                       \\ \midrule
Std. & $1.52\times10^{7}$        &   $4.49\times10^{6}$      &   $2.01\times10^{6}$      &   $1.07\times10^{6}$      &    $7.16\times10^{5}$     &$4.81\times10^{5}$                       \\ \bottomrule
\end{tabular}%
}
\vspace{.1mm}
\caption{Mean and Std. of the Terminal Objective {Values} $W_T+S_TI_T-\lambda I_T^2$ over 232 Days. We fix $\lambda = 0.0005$. {We control cash holdings and inventory processes using both Optimal and Benchmark Strategies.}}\label{tab:TerminalReward}
\vspace{.5 cm}
\end{table}

\
\
\begin{table}[]
\centering
\setlength{\tabcolsep}{8pt}
\resizebox{\textwidth}{!}{%
\begin{tabular}{@{}ccccccc@{}}
\toprule\midrule
 & \multicolumn{2}{c}{\begin{tabular}{c}Optimal Strategy\\ with Non-Martingale {Price}\\ {and $\pi_\tk(1,1)\geq 0$}\end{tabular}} & \multicolumn{2}{c}{\begin{tabular}{c}Optimal Strategy\\ with Martingale {Price}\\ {and $\pi_\tk(1,1)\geq 0$}\end{tabular}}&
 \multicolumn{2}{c}{\begin{tabular}{c}{Optimal Strategy}\\ {with Non-Martingale Price} \\{and $\pi_\tk(1,1)\equiv 0$}\end{tabular}} \\ \midrule
Mean & \multicolumn{2}{c}{$6.00\times10^4$}& \multicolumn{2}{c}{$5.56\times10^4$}& \multicolumn{2}{c}{{$5.97\times10^4$}}                                               \\ \midrule
Std. & \multicolumn{2}{c}{$1.22\times10^6$} & \multicolumn{2}{c}{$1.30\times10^6$}& \multicolumn{2}{c}{{$1.22\times10^6$}}                                               \\ \bottomrule
\end{tabular}%
}
\vspace{.1mm}
\caption{Mean and Std. of the Terminal {Values $W_{T}+\bar{S}_{T}I_{T}$ (Terminal Cash Holdings plus Liquidation Proceeds) over 232 Days} {using} {Different} Strategies.}
\label{tab:TerminalRewardLiquid}
\vspace{.5 cm}
\end{table}

{Based on the results of Table~\ref{tab:TerminalReward} and Table~\ref{tab:TerminalRewardLiquid}, we can conclude that the optimal strategies outperform the fixed level 1-level 6 strategies.} With the incorporation of the drift term $\Delta_{t_k}$ in the midprice process, we achieve a higher average and a lower standard deviation of the terminal {values}. {We observe that allowing for simultaneous arrivals of buy and sell MOs also leads to a higher average of the terminal values.} {Hereafter we focus on the optimal strategy computed using non-martingale fundamental price dynamics and assuming $\pi_\tk(1,1)\geq0$.} The optimal {strategy yields a} positive average terminal {value}. However, as shown in Fig.~\ref{fig:232RevenueHist}, the distributions of terminal {values} appear to exhibit heavy tails on both sides with {kurtosis larger than $14$.} Such a large kurtosis results in high standard deviation estimates. We use the subsample bootstrap method proposed by \cite{hall1996bootstrap} to construct a confidence interval for the mean. 
From Fig.~\ref{fig:232RevenueHistObjFunc}, we can see that the {95\%-}confidence interval for the mean of the terminal objective $W_T+S_TI_T-\lambda I_T^2$ is $[-1.09\times 10^5,2.32\times 10^5]$. {Fig.~\ref{fig:232RevenueHistLiquid} shows that the 95\%-confidence interval for the mean of {the terminal values $W_T+\bar{S}_TI_T$ is $[-1.11\times 10^5,2.31\times 10^5]$.} We remark, however, that bootstrap based CIs tend to be highly conservative for heavy tailed distributions as shown in the simulations of \cite{peng2004empirical}.
 In {Section}~\ref{sec:ExtremeNeg}, we will show that these extreme negative revenues are due to {large} structural breaks over time, and discuss how to identify and potentially exclude atypical days from the analysis.

\begin{figure}
\centering
\begin{subfigure}{1\textwidth}
\centering
   \includegraphics[width=.6\textwidth]{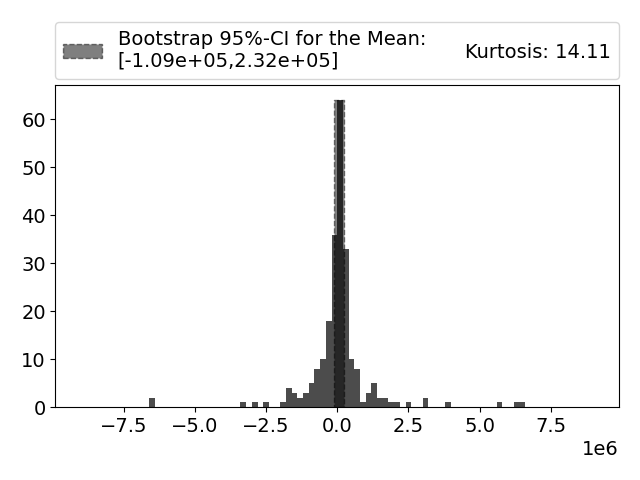}
   \caption{{Terminal Objective Values: $W_T+S_TI_T-\lambda I_T^2$ ($\lambda = 0.0005$).}}
   \label{fig:232RevenueHistObjFunc} 
\end{subfigure}
\begin{subfigure}[b]{1\textwidth}
\centering
   \includegraphics[width=.6\textwidth]{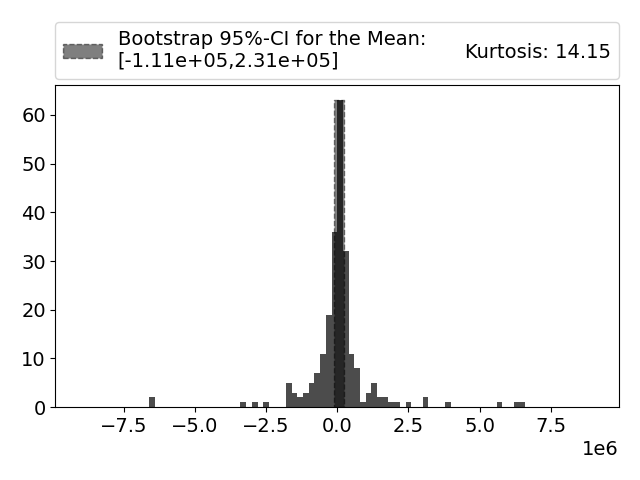}
   \caption{{Terminal Values: $W_{T}+\bar{S}_{T}I_{T}$ (Cash Holdings + Liquidation Proceeds).}}
    \label{fig:232RevenueHistLiquid}
\end{subfigure}
    \caption{{\textbf{Histogram of the Terminal Values Obtained From the Optimal {Strategy} in Year 2019 (232 Trading Days Included).} We compute the terminal values achieved by the optimal strategy for each trading day of the year, starting from the 21st trading day. In each day, we use the prior 20 days to estimate parameters. This results in a total of 232 trading days used to estimate the probability distribution.}}
\label{fig:232RevenueHist}\vspace{.5 cm}
\end{figure}

\subsection{{Days with Extreme Negative Revenues}}\label{sec:ExtremeNeg}
On some days, the market experiences `atypical' demand and supply due to various factors (e.g. non-scheduled news arrival, entry of new market participants, etc.), which are not predictable from recent market data. These `atypical' patterns can result in structural parameter breaks, and constitute the main reason for the observance of extreme negative revenues in Section \ref{sec:232result}. In our case, this means that the parameter values estimated based on the last 20 days can differ by a large extent from the {actual parameter values of the current trading day} when the strategy is implemented.

One key parameter that significantly affects the performance of the optimal strategies is $\muoo^\pm$, defined in Eq.~(\ref{Dfnmucp}). {Recall the value of $c^{\pm}p^{\pm}$ is the $y$-intercept of the demand functions (see Fig.~\ref{fig:linearQ}) and {a biased estimate} of $\muoo^\pm$ can lead to misleading {predictions} of filled shares near the midprice, which are the most critical ticks.}
{For each trading day $i$, we therefore compute the difference between the average of historical estimated values of $\muoo^\pm$ based on 20 past days and the estimated values from the current trading day $i$:}
{
\[err_{cp}^{\pm,i} := \frac{1}{20}\sum_{j = 1}^{20}\hat{\mu}^{\pm,i-j}_{cp}-\hat{\mu}^{\pm,i}_{cp},\]
}{where $\hat{\mu}^{\pm,i}_{cp}$ is defined in Section \ref{sec:paraest}. The empirical distributions of $err_{cp}^{\pm,i}$ are heavy right-tailed which means that $\muoo^\pm$ are much overestimated for some trading days.} We {therefore} identify days when {either $\muoo^+$ or $\muoo^-$} are overestimated, and mark {days with error larger than} the 0.95 quantile {of the empirical distributions of $err_{cp}^{\pm,i}$} as days with large structural parameter break.

Another critical parameter in our analysis is the value of $\pi_\tk(1,1)$, which represents the probability of simultaneous arrival of buy and sell MOs within a 1-second subinterval. To determine whether a structural break has occurred in the estimate of such a parameter on a given trading day $i$, we first compute the historical estimate of $\pi_\tk(1,1)$ for the day $i$ as:
\[
	\bar{\hat{\pi}}^{i}(1,1)= \frac{1}{N}\sum_{k=1}^{N}\hat{\pi}_\tkk^{i}(1,1),
\]
where $\{\hat{\pi}_\tkk^{i}(1,1)\}_{k=0,\dots,N}$ are the least-squares estimates of $\bar{\pi}_\tkk^{i}(1,1)$ defined in (\ref{EstimatePiH}). We then compute the difference between the historical estimates $\bar{\hat{\pi}}^{i}(1,1)$ and the estimated probability $\tilde{\pi}^{i}(1,1):=\sum_{k = 0}^{N-1}(\mathbbm{1}_{t_{k+1}}^{+,i}\cdot\mathbbm{1}_{t_{k+1}}^{-,i})/N$ for day $i$, {and set $err_{\pi(1,1)}^{i}:=\bar{\hat{\pi}}^{i}(1,1) - \tilde{\pi}^{i}(1,1)$.}
We mark days for which the absolute value of {$err_{\pi(1,1)}^{i}$} is greater than the 0.95 quantile {of its empirical distribution} as days with large structural parameter breaks.

\medskip
\noindent
\textbf{Results after Excluding Days with Large Structural Parameter Break.} 
Using the criteria {described above}, we identify days with a large structural break in the estimate of either $\muoo^\pm$ or $\pi_\tk(1,1)$. Our analysis indicates that there are 22 out of 232 trading days for which this occurs. Table~\ref{tab:DiscardingDaysFormula} and Table~\ref{tab:DiscardingDaysActualLiquidating} compare the terminal objective values before and after excluding those 22 `atypical' days. The { 95\%} confidence intervals\footnote{The confidence intervals {are constructed using the} standard normal approximation method and the subsample bootstrap method proposed by \cite{hall1996bootstrap}{\Green,} as mentioned in Section \ref{sec:232result}. \cite{peng2004empirical} show that the subsample bootstrap method provides a more conservative estimate for the confidence interval of the mean if the distribution is heavy-tailed.} for the {mean of these terminal values} only {consist of} positive values once we exclude `atypical' trading days. The average and standard deviation of {terminal values} are also significantly increased and reduced, respectively. As shown in the histograms of Fig.~\ref{fig:RevenueHistAtypical}, the selection criterion discussed above {effectively} excludes days where revenues are extreme and negative.

\begin{table}[]
\centering
\setlength{\tabcolsep}{8pt}
\resizebox{.8\textwidth}{!}{%
\begin{tabular}{@{}ccccccc@{}}
\toprule\midrule
 & \multicolumn{3}{c}{{With all Days}} & \multicolumn{3}{c}{{Excluding `Atypical' Days}} \\ \midrule
Mean (Std.) & \multicolumn{3}{c}{$6.13\times10^4\ (1.22\times10^6)$}                                       & \multicolumn{3}{c}{$1.56\times10^5\ (1.04\times10^6)$}                                               \\ \midrule
\begin{tabular}{c}{95\%} Confidence Interval of Mean\\Normal Approximation\end{tabular} & \multicolumn{3}{c}{$[-9.72\times10^4,2.20\times10^5]$}                                           & \multicolumn{3}{c}{$[1.41\times10^4,2.98\times10^5]$}                                               \\ \midrule
\begin{tabular}{c}{95\%} Confidence Interval of Mean\\Subsample Bootstrap\end{tabular}& \multicolumn{3}{c}{$[-1.09\times 10^5,2.32\times 10^5]$}                                           & \multicolumn{3}{c}{$[2.56\times 10^3,3.10\times 10^5]$}                                               \\ \bottomrule
\end{tabular}%
}
\vspace{.1mm}
\caption{{Terminal objective value $W_T+S_TI_T-\lambda I_T^2$. We consider both the inclusion and exclusion of the 22 `Atypical' Days.} {We set $\lambda = 0.0005$.}}
\label{tab:DiscardingDaysFormula}
\vspace{.5 cm}
\end{table}

\
\
\begin{table}[]
\centering
\setlength{\tabcolsep}{8pt}
\resizebox{.8\textwidth}{!}{%
\begin{tabular}{@{}ccccccc@{}}
\toprule\midrule
 & \multicolumn{3}{c}{{With all Days}} & \multicolumn{3}{c}{{Excluding `Atypical' Days}} \\ \midrule
Mean (Std.) & \multicolumn{3}{c}{$6.00\times10^4\ (1.22\times10^6)$}                                       & \multicolumn{3}{c}{$1.55\times10^5\ (1.04\times10^6)$}                                               \\ \midrule
\begin{tabular}{c}{95\%} Confidence Interval of Mean\\Normal Approximation\end{tabular} & \multicolumn{3}{c}{$[-9.84\times10^4,2.18\times10^5]$}                                           & \multicolumn{3}{c}{$[1.29\times10^4,2.97\times10^5]$}                                               \\ \midrule
\begin{tabular}{c}{95\%} Confidence Interval of Mean\\Subsample Bootstrap\end{tabular}& \multicolumn{3}{c}{$[-1.11\times 10^5,2.31\times 10^5]$}                                           & \multicolumn{3}{c}{$[1.35\times 10^3,3.08\times 10^5]$}                                               \\ \bottomrule
\end{tabular}%
}
\vspace{.1mm}
\caption{{Terminal value $W_{T}+\bar{S}_{T}I_{T}$ (terminal cash holdings plus liquidation proceeds).} We consider both the inclusion and exclusion of the 22 `Atypical' Days.}
\label{tab:DiscardingDaysActualLiquidating}
\vspace{.5 cm}
\end{table}

\begin{figure*}[t!]
\centering
\begin{subfigure}[b]{1\textwidth}
\centering
   \includegraphics[width=.8\textwidth]{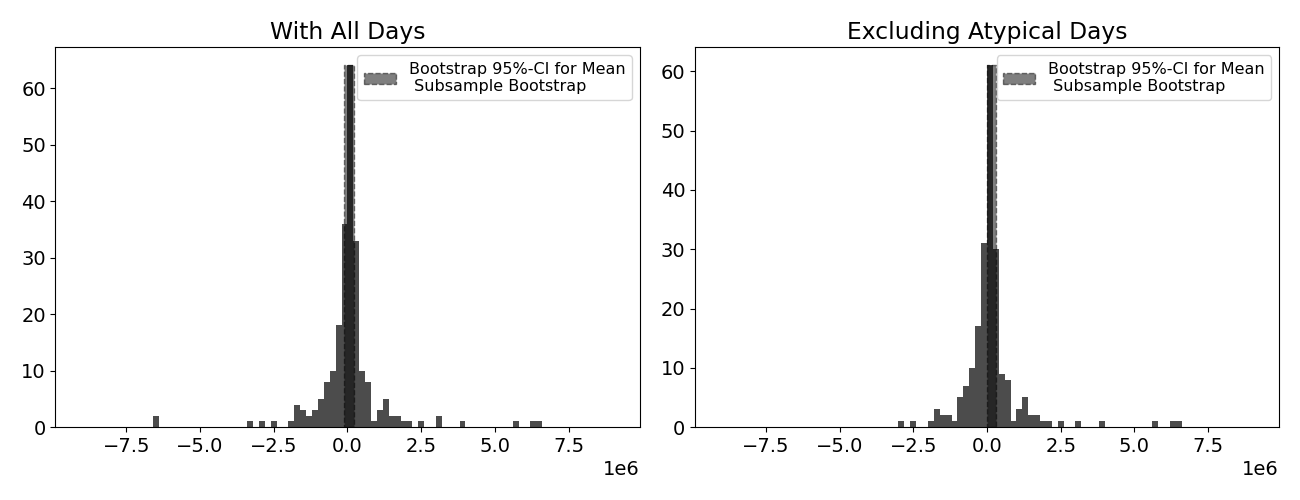}
   \caption{Terminal {Objective Values}: $W_T+S_TI_T-\lambda I_T^2$ ($\lambda = 0.0005$)}
   \label{fig:HistAtypicalFormula} 
\end{subfigure}

\begin{subfigure}[b]{1\textwidth}
\centering
   \includegraphics[width=.8\textwidth]{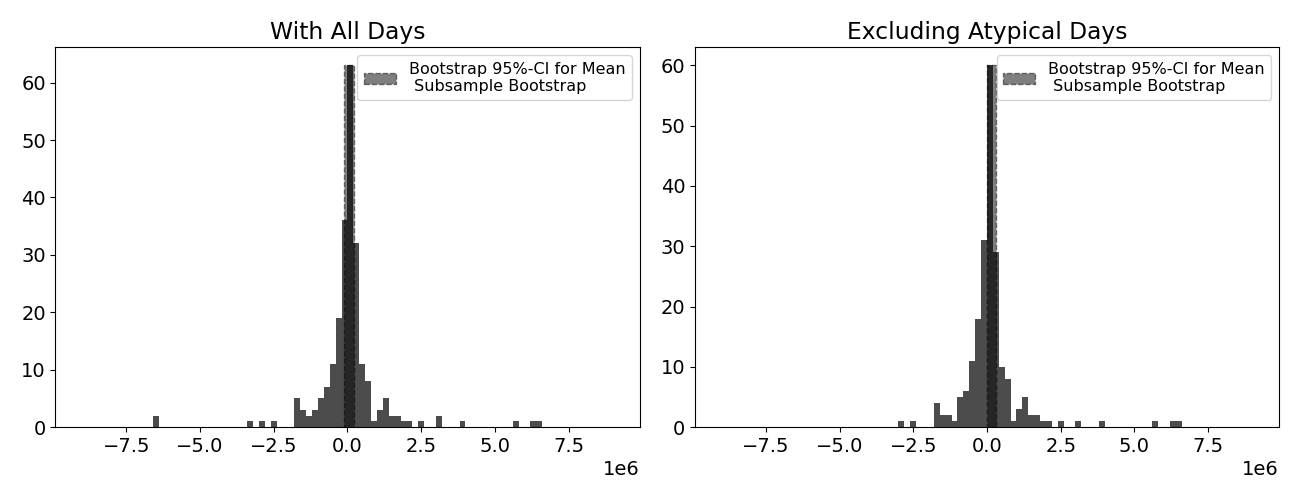}
   \caption{{Terminal Values: $W_{T}+\bar{S}_{T}I_{T}$ (Cash Holdings + Liquidation Proceeds)}}
    \label{fig:HistAtypicalActualLiquidating}
\end{subfigure}
    \caption{{\textbf{Histogram of Terminal {Values} in Year 2019 Before (Left Panels) and After Excluding `Atypical' Days (Right Panels).} We compute the terminal values achieved by the optimal strategy for each trading day of the year, starting from the 21st trading day. For each day, we use the prior 20 days to estimate parameters. This results in a total of 232 trading days, which are used to estimate the probability distribution  shown in the left panel. There are 22 out of 232 trading days identified as `atypical' days. The right panel shows the distribution of terminal values after excluding those 22 `atypical' days.}}\label{fig:RevenueHistAtypical}
    \vspace{.5 cm}
\end{figure*}

\appendix

 \renewcommand{\theequation}{A-\arabic{equation}}
\setcounter{equation}{0}  

\section{Proofs of Section \ref{sec:Bellman}}

\subsection{Proofs of Proposition \ref{prop:optimalcontrol} and Theorem  \ref{prop:optimalcontrolNMG}.}\label{sec:OptimalL}

We prove {Proposition} \ref{prop:optimalcontrol} and {Theorem} \ref{prop:optimalcontrolNMG} through four steps:

\smallskip
\noindent
\textbf{Step 1.} 
We start by proposing the following ansatz for the value function $V_\tk$:
\begin{equation}
V_\tk=\widetilde{v}(t_k,\sk,W_\tk,I_\tk) := W_\tk+\alpha_{\tk}I_\tk^2+\sk I_\tk+\widetilde{h}_{\tk}I_\tk+\widetilde{g}_{\tk},\label{eq:V_NMGb}
\end{equation} 
where $\alpha:\mathcal{T}\to\mathbb{R}$ is a deterministic function defined on $\mathcal{T}=\{t_0,t_{1},\dots,t_{N+1}\}$ (recall that we set $t_{N+1}=T$) and $\{\widetilde{h}_{t}\}_{t\in\mathcal{T}}$, $\{\widetilde{g}_{t}\}_{t\in\mathcal{T}}$ are some processes adapted to the filtration $\{\mathcal{F}_t\}_{t\in\mathcal{T}}$. Since $V_T=W_T+S_TI_T-\lambda I_T^2$, we have the terminal conditions $\alpha_T=-\lambda$, $\widetilde{g}_T=0$, and $\widetilde{h}_T=0$. In what follows, we
will use the following notation:
\begin{align}\label{DfnFrcgh}
{\widetilde{h}_\tkk^\tk:=\mathbb{E}[\widetilde{h}_\tkk|\F],\quad
\widetilde{g}_\tkk^\tk:=\mathbb{E}[\widetilde{g}_\tkk|\F]}.
\end{align}

By plugging Eq.~(\ref{eq:W}), (\ref{eq:I}), and (\ref{eq:V_NMGb}) into the right-hand side of the Bellman equation (\ref{eq:DP}), we get 
\begin{align}\label{BgnBlEq}
    \begin{split}
    V_{t_{k}}&=\sup_{L_{t_k}^\pm}\mathbb{E}\bigg\{\sum_{\delta=\pm}(\sk+\delta\ld)\delta\oned\cd(\Pd-\ld)\\
    &\qquad\qquad+\alpha_{\tkk}\big[I_\tk-\sum_{\delta=\pm}\delta\oned\cd(\Pd-\ld)\big]^2\\
    &\qquad\qquad+\skk\big[I_\tk-\sum_{\delta=\pm}\delta\oned\cd(\Pd-\ld)\big]\\
    &\qquad\qquad+ \widetilde{h}_{\tkk}\big[I_\tk-\sum_{\delta=\pm}\delta\oned\cd(\Pd-\ld)\big]\\
    &\qquad\qquad+\widetilde{g}_{\tkk}\bigg|\F\bigg\}
    \end{split}
\end{align}
We expand the squares inside the expectation above and arrange the terms as follows:
\begin{align}
    \label{eq:expNMG-0}
   &\sum_{\delta=\pm}\oned\big[-\cd(\ld)^2+(\cd\Pd-\delta\cd\sk)\ld+\delta\cd\Pd\sk\big]\\
   \nonumber
    &+\alpha_{\tkk}\bigg\{I_\tk^2+\sum_{\delta=\pm}\oned\Big\{(\cd)^2(\ld)^2+\big[2\delta I_\tk\cd-2(\cd)^2\Pd\big]\ld\\
        \label{eq:expNMG-2}
    &\qquad\qquad\quad\qquad\qquad\qquad+(\cd\Pd)^2-2\delta I_\tk\cd\Pd\Big\}\\
    \nonumber
    &\qquad\qquad+2\onep\onem\cp\cm(-\lp\lm+\Pp\lm+\Pm\lp-\Pp\Pm)\bigg\}\\
        \label{eq:expNMG-1}
    &+\skk\bigg[I_\tk+\sum_{\delta=\pm}\oned(-\delta\cd\Pd+\delta\cd\ld)\bigg]
    \\
    &+ \widetilde{h}_{\tkk}I_\tk+\sum_{\delta=\pm}\oned(-\delta \widetilde{h}_{\tkk}\cd\Pd+\delta \widetilde{h}_{\tkk}\cd\ld)+\widetilde{g}_{\tkk}
\label{eq:expNMGb}
\end{align}
The conditional expectations of most terms above are easy to compute from the conditions in Assumption \ref{assump:cp} and the adaptability of the controls $\{L_{t_k}^\pm\}$, $\{S_{t_k}\}$, and $\{I_{t_k}\}$. For instance, we can easily see that
\begin{align*}
	&\mathbb{E}\big[\onep\onem\cp\cm\Pp{\Pm}\big|\F\big]\\
	&\quad =\mathbb{E}\big[\onep\onem\mathbb{E}\big[\cp\Pp\big|\F,\onep\onem\big]\mathbb{E}\big[\cm{\Pm}\big|\F,\onep\onem\big]\big|\F\big]\\
	&\qquad =\mu_{cp}^{+}\mu_{cp}^{-}\mathbb{E}\big[\onep\onem\big|\F\big]=\mu_{cp}^{+}\mu_{cp}^{-}\pi_{t_{k+1}}(1,1).
\end{align*}
For the terms in (\ref{eq:expNMG-1}), using the conditional independence of $(\mathbbm{1}_{t_{k+1}}^{\pm},c^{\pm}_{t_{k+1}},p^{\pm}_{t_{k+1}})$  and $S_{t_{k+1}}-S_{t_{k}}$ given ${\F}$ stated after (\ref{MrtCnd0}), we have:
\begin{align*}
	&\mathbb{E}\big[(\skk-S_{t_k})\oned\cd\Pd\big|\F\big]\\
	&\quad =
	\mathbb{E}\big[\skk-S_{t_k}\big|\F\big]\mathbb{E}\big[\oned\cd\Pd\big|\F\big]=\Delta_{t_{k}}\pi_{t_{k+1}}^{\delta}\mu_{cp}^{\delta},
\end{align*}
and, thus, $\mathbb{E}\big[\skk\oned\cd\Pd \big|\F\big]=(S_{t_{k}}+\Delta_{t_{k}})\pi_{t_{k+1}}^{\delta}{\mu_{cp}^\delta}$.  Similarly, we can show that \\$\mathbb{E}\big[\skk\oned\cd \big|\F\big]=(S_{t_{k}}+\Delta_{t_{k}})\pi_{t_{k+1}}^{\delta}\mu_{c}^{\delta}$. 
For the terms in (\ref{eq:expNMGb}), let us assume for now that:
\begin{align}\label{TrickyNdInd1}
	 \mathbb{E}\big[\widetilde{h}_{\tkk}\oned\cd\big|\F\big]&=
	  \widetilde{h}_{\tkk}^{t_{k}}\mathbb{E}\big[\oned\cd\big|\F\big],\\
	  	  \label{TrickyNdInd2}
	 \mathbb{E}\big[\widetilde{h}_{\tkk}\oned\cd\Pd\big|\F\big]&=
	  \widetilde{h}_{\tkk}^{t_{k}}\mathbb{E}\big[\oned\cd\Pd\big|\F\big].
\end{align}
The above identities will be proved below in Step 4. Using the previous arguments, we can compute the conditional expectation $\mathbb{E}\big[\cdot\big|\F\big]$ of the terms in Eqs.~(\ref{eq:expNMG-0})-(\ref{eq:expNMGb}), and plug them in the right-hand side of Eq.~(\ref{BgnBlEq}) to get:
\begin{align}
\nonumber
    &\alpha_{\tk}I_{t_k}^2+S_{t_k} I_{t_k}+\widetilde{h}_{\tk}I_{t_k}+\widetilde{g}_{\tk}\\ \nonumber
    &=\sup_{L_{t_k}^\pm}\sum_{\delta=\pm}{\pi^\delta_\tkk}\Big\{(\alpha_{\tkk}\mut^\delta-\muo^\delta)(\ld)^2+\big[\muoo^\delta+\delta \widetilde{h}_{\tkk}^\tk\muo^\delta+\alpha_{\tkk}(2\delta\muo^\delta I_\tk-2\muto^\delta)+\delta\muo^\delta\Delta_\tk\big]\ld\\ \nonumber
    &\qquad\qquad\qquad\qquad\qquad+\alpha_{\tkk}(\mutt^\delta-2\delta\muoo^\delta I_\tk)-\delta \widetilde{h}_{\tkk}^\tk\muoo^\delta-\delta\muoo^\delta\Delta_\tk\Big\}\\ \nonumber
    &\qquad\quad+\alpha_{\tkk}I_\tk^2+2\alpha_{\tkk}{\pi_\tkk(1,1)}(-\muo^+\muo^-\lp\lm+\muoo^+\muo^-\lm+\muo^+\muoo^-\lp-\muoo^+\muoo^-)\\
    &\qquad\quad+I_\tk(\sk+\Delta_\tk)+ \widetilde{h}_{\tkk}^\tk I_\tk+\widetilde{g}_{\tkk}^\tk
\label{eq:L_NMG}
\end{align}
Denote the right hand side of above equation as $\sup_{L_{t_k}^\pm}\widetilde{f}(\lp,\lm)$. As we can see $\widetilde{f}(\lp,\lm)$ is a quadratic function of $\lp $ and $\lm$. Setting the partial derivatives with respect to $\lp$ and $\lm$, respectively, equal to $0$, we have
\begin{align*}
    \partial_{\lp} \widetilde{f}&=2{\pi^+_{\tkk}}(\alpha_{\tkk}\mut^+-\muo^+)\lp
    +{\pi^+_{\tkk}}\big[\muoo^++ \widetilde{h}_{\tkk}^\tk\muo^++\alpha_{\tkk}(2\muo^+I_\tk-2\muto^+)+\muo^+\Delta_\tk\big]\\
    &-2\alpha_{\tkk}{\pi_\tkk(1,1)}\muo^+\muo^-\lm+2\alpha_{\tkk}{\pi_\tkk(1,1)}\muo^+\muoo^-=0,\\
    \partial_{\lm} \widetilde{f}&=2{\pi^-_{\tkk}}(\alpha_{\tkk}\mut^--\muo^-)\lm
    +{\pi^-_{\tkk}}\big[\muoo^-- \widetilde{h}_{\tkk}^\tk\muo^-+\alpha_{\tkk}(-2\muo^-I_\tk-2\muto^-)-\muo^-\Delta_\tk\big]\\
    &-2\alpha_{\tkk}{\pi_\tkk(1,1)}\muo^+\muo^-\lp+2\alpha_{\tkk}{\pi_\tkk(1,1)}\muo^-\muoo^+=0.
\end{align*}
Solving for $\lp$ and $\lm$, we get the expressions 
\begin{align}\label{DfnOptLbb}
    &\widetilde{L}_{\tk}^{+,*}={}^{\scaleto{(1)}{5pt}}\!A^+_{\tk}I_\tk+{}^{\scaleto{(2)}{5pt}}\!\widetilde{A}^+_{\tk}+{}^{\scaleto{(3)}{5pt}}\!\widetilde{A}^+_{\tk},&\widetilde{L}_{\tk}^{-,*}=-{}^{\scaleto{(1)}{5pt}}\!A^-_{\tk}I_\tk{\Black-}{}^{\scaleto{(2)}{5pt}}\!\widetilde{A}^-_{\tk}{\Black+}{}^{\scaleto{(3)}{5pt}}\!\widetilde{A}^-_{\tk}
\end{align}
where $\Aone^\pm,  {}^{\scaleto{(2)}{5pt}}\!\widetilde{A}^\pm_{\tk}, {}^{\scaleto{(3)}{5pt}}\!\widetilde{A}^\pm_{\tk}$ are given as
\begin{align}
\label{eq:A1NMGb}
    &{}^{\scaleto{(1)}{5pt}}\!A^\pm_{\tk}=\frac{\beta^{\pm}_{t_{k}}\alpha_{t_{k+1}}}{\gamma_{t_{k}}},\quad 
    {}^{\scaleto{(2)}{5pt}}\!\widetilde{A}^\pm_{\tk}=\frac{\beta^{\pm}_{t_{k}}\widetilde{h}_{t_{k+1}}^{\tk}}{2\gamma_{t_{k}}},\\
\label{eq:A1NMGbc}
&{}^{\scaleto{(3)}{5pt}}\!\widetilde{A}^\pm_{\tk}={\frac{\pi^\mp_\tkk}{2\gamma_{t_{k}}}}(\alpha_{\tkk}\mut^\mp-\muo^\mp)\big[{\pi^\pm_{\tkk}}(\muoo^\pm-2\alpha_{\tkk}\muto^\pm)+2{\pi_\tkk(1,1)}\alpha_{\tkk}\muo^\pm\muoo^\mp\pm\pi^\pm_{\tkk}\Delta_\tk\muo^\pm\big]\nonumber \\
        &\quad \qquad+{\pi_\tkk(1,1)}\frac{\alpha_{\tkk}}{2\gamma_{t_{k}}}\muo^+\muo^-\big[{\pi^\mp_\tkk}(\muoo^\mp-2\alpha_{\tkk}\muto^\mp)+2\alpha_{\tkk}{\pi_\tkk(1,1)}\muoo^\pm\muo^\mp\mp\pi^\mp_{\tkk}\Delta_\tk\muo^\mp\big].
\end{align}
By plugging $\widetilde{L}_{\tk}^{\pm,*}$ back into Eq.~(\ref{eq:L_NMG}) and matching terms with respect to $I_\tk$, we obtain the following recursive expressions for $\alpha_{\tk},\tilde{h}_{\tk}$, and $\tilde{g}_{\tk}$:
\begin{align}\label{eq:alphaNMGb}
   \begin{split}
    \alpha_{\tk}&=\alpha_{\tkk}+\sum_{\delta=\pm}{\pi^\delta_\tkk}\big[(\alpha_{\tkk}\mut^\delta-\muo^\delta)({}^{\scaleto{(1)}{5pt}}\!A^\delta_{\tk})^2+2\alpha_{\tkk}\muo^\delta ({}^{\scaleto{(1)}{5pt}}\!A^\delta_{\tk})\big]\\
   &\qquad  \qquad+2\alpha_{\tkk}{\pi_\tkk(1,1)}\muo^+\muo^-({}^{\scaleto{(1)}{5pt}}\!A^+_{\tk}{}^{\scaleto{(1)}{5pt}}\!A^-_{\tk}),
          \end{split}
\end{align}   
\begin{align}   
        \widetilde{h}_{\tk}&= \widetilde{h}_{\tkk}^\tk+\sum_{\delta=\pm}{\pi^\delta_\tkk}\Big\{2(\alpha_{\tkk}\mut^\delta-\muo^\delta)\big[{}^{\scaleto{(1)}{5pt}}\!A^\delta_{\tk}({(\delta\, {\Black{}^{\scaleto{(3)}{5pt}}\!\widetilde{A}^\delta_{\tk})}+ {}^{\scaleto{(2)}{5pt}}\!\widetilde{A}^\delta_{\tk}})\big] +2\alpha_{\tkk}\muo^\delta({(\delta\, {\Black{}^{\scaleto{(3)}{5pt}}\!\widetilde{A}^\delta_{\tk})}+{}^{\scaleto{(2)}{5pt}}\!\widetilde{A}^\delta_{\tk}})\nonumber \\
        &\qquad\qquad\qquad\qquad\qquad\qquad-2\alpha_{\tkk}{(\delta\muoo^\delta)}+({\delta} \,{}^{\scaleto{(1)}{5pt}}\!A^\delta_{\tk})(\muoo^\delta+{(\delta \widetilde{h}_{\tkk}^\tk)}\muo^\delta-2\alpha_{\tkk}\muto^\delta)\Big\}\nonumber \\
        &\quad-2\alpha_{\tkk}{\pi_\tkk(1,1)}\muo^+\muo^-\Big[{}^{\scaleto{(1)}{5pt}}\!A^+_{\tk}({{}^{\scaleto{(3)}{5pt}}\!\widetilde{A}^-_{\tk}-{}^{\scaleto{(2)}{5pt}}\!\widetilde{A}^-_{\tk}})-{}^{\scaleto{(1)}{5pt}}\!A^-_{\tk}({}^{\scaleto{(2)}{5pt}}\!\widetilde{A}^+_{\tk}+{}^{\scaleto{(3)}{5pt}}\!\widetilde{A}^+_{\tk})+\frac{\muoo^+}{\muo^+}({}^{\scaleto{(1)}{5pt}}\!A^-_{\tk})-\frac{\muoo^-}{\muo^-}({}^{\scaleto{(1)}{5pt}}\!A^+_{\tk})\Big]\nonumber \\
        &\quad+{\Delta_\tk\big[{}^{\scaleto{(1)}{5pt}}\!A^+_{\tk}\pi_\tkk^+\muo^+ +{{}^{\scaleto{(1)}{5pt}}\!A^-_{\tk}}\pi_\tkk^-\muo^-+1\big]}
           \label{eq:hNMGc}
\end{align}
\begin{align} \nonumber
        \widetilde{g}_{\tk}&=\widetilde{g}_{\tkk}^\tk+\sum_{\delta=\pm}{\pi^\delta_\tkk}\big[(\alpha_{\tkk}\mut^\delta-\muo^\delta)({{}^{\scaleto{(3)}{5pt}}\!\widetilde{A}^\delta_{\tk}+{(\delta\, {}^{\scaleto{(2)}{5pt}}\!\widetilde{A}^\delta_{\tk}})})^2+\alpha_{\tkk}\mutt^\delta-{(\delta\, \widetilde{h}_{\tkk}^\tk)}\muoo^\delta\\
         \nonumber
        &\quad\qquad\qquad\qquad\qquad\quad+(\muoo^\delta+{(\delta \widetilde{h}_{\tkk}^\tk)}\muo^\delta-2\alpha_{\tkk}\muto^\delta)({{}^{\scaleto{(3)}{5pt}}\!\widetilde{A}^\delta_{\tk}+{(\delta\, {}^{\scaleto{(2)}{5pt}}\!\widetilde{A}^\delta_{\tk}})})\big]\\
         \nonumber
        &\qquad\qquad-2\alpha_{\tkk}{\pi_\tkk(1,1)}\muo^+\muo^-\Big[({}^{\scaleto{(2)}{5pt}}\!\widetilde{A}^+_{\tk}+{}^{\scaleto{(3)}{5pt}}\!\widetilde{A}^+_{\tk})({{}^{\scaleto{(3)}{5pt}}\!\widetilde{A}^-_{\tk}-{}^{\scaleto{(2)}{5pt}}\!\widetilde{A}^-_{\tk}})\\
        \nonumber
        &\quad\qquad\qquad\qquad\qquad\qquad\qquad\qquad-\frac{\muoo^+}{\muo^+}({{}^{\scaleto{(3)}{5pt}}\!\widetilde{A}^-_{\tk}-{}^{\scaleto{(2)}{5pt}}\!\widetilde{A}^-_{\tk}})-\frac{\muoo^-}{\muo^-}({}^{\scaleto{(2)}{5pt}}\!\widetilde{A}^+_{\tk}+{}^{\scaleto{(3)}{5pt}}\!\widetilde{A}^+_{\tk})+\frac{\muoo^+\muoo^-}{\muo^-\muo^+}\Big]\\
        &\qquad\qquad+{\Delta_{t_{k}}}\Big[({}^{\scaleto{(3)}{5pt}}\!\widetilde{A}^\delta_{\tk}+ {}^{\scaleto{(2)}{5pt}}\!\widetilde{A}^\delta_{\tk})\pi_\tkk^+\muo^+-({}^{\scaleto{(3)}{5pt}}\!\widetilde{A}^\delta_{\tk}- {}^{\scaleto{(2)}{5pt}}\!\widetilde{A}^\delta_{\tk})\pi_\tkk^-\muo^--\pi_\tkk^+\muoo^++\pi_\tkk^-\muoo^-\Big]\label{eq:gNMGb}
\end{align}

\smallskip
\noindent
\textbf{Step 2.}  We next prove that $\widetilde{L}_{\tk}^{\pm,*}$ are indeed the maximum point of the function $\tilde{f}(L^+_\tk,L^-_\tk)$. To this end, we will use Lemma \ref{lemma:alpha}, which states that  $\alpha_\tk<0$. Indeed, for every $\tk$, we have:
\begin{align*}
    D& = (\partial_{\lp}^2 \tilde{f})(\partial_{\lm}^2\tilde{f})-(\partial_{\lp\lm}\tilde{f})^2 \\
    &= 4{\pi^+_{\tkk}}{\pi^-_{\tkk}}(\alpha_{\tkk}\mut^+-\muo^+)(\alpha_{\tkk}\mut^--\muo^-)-4\big[{\pi_\tkk(1,1)}\alpha_{\tkk}\muo^+\muo^-\big]^2>0,\\
	\partial_{\lp}^2 \tilde{f}&=\alpha_{\tkk}\mut^+-\muo^+<0.
\end{align*}
By the second derivative test, $\tilde{f}(\lp,\lm)$ takes its maximum value at $\widetilde{L}_{\tk}^{\pm,*}$.

\medskip
\noindent
\textbf{Step 3.} We now show  that (\ref{eq:LtildeNMGbb}) holds.
Note that, by plugging the expressions of ${}^{\scaleto{(2)}{5pt}}\!\widetilde{A}^\pm_{\tk}$ and ${}^{\scaleto{(3)}{5pt}}\!\widetilde{A}^\pm_{\tk}$ given in (\ref{eq:A1NMGb})-(\ref{eq:A1NMGbc}) into (\ref{eq:hNMGc}), $\widetilde{h}_{\tk}$ can be written as 
\begin{align}\label{Repr2Us}
	\widetilde{h}_{\tk}=d_{k}+\xi_{k} (\widetilde{h}_{\tkk}^\tk +\Delta_\tk),
\end{align}
for some deterministic constant $d_{k}$ and
\begin{align*}
	\xi_{k}&=  1+\frac{\alpha_{t_{k+1}}}{\gamma_{t_k}}\sum_{\delta=\pm}
	{\pi^\delta_\tkk}
	\beta^{\delta}_{t_k}
	\Big\{\frac{\beta^{\delta}_{t_{k}}}{\gamma_{t_{k}}}(\alpha_{\tkk}\mut^\delta-\muo^\delta)+2\muo^\delta\Big\}+2\frac{\alpha_{\tkk}^2}{\gamma_{t_k}^2}{\pi_\tkk(1,1)}\muo^+\muo^-\beta^{+}_{t_{k}}\beta^{-}_{t_{k}}.
\end{align*}
Note also that $h_\tk$ defined in Eq.~(\ref{eq:MG_h}) can also be written as
\begin{align}
	h_{\tk}=d_{k}+\xi_{k} h_{\tkk},
\end{align}
where $d_{k},\xi_k$ are the same as those in (\ref{Repr2Us}).
Since $h_{t_{N+1}}=0$ and $\widetilde{h}_{t_{N+1}}=0$, for the time point $t_N$, we have that $\widetilde{h}_{t_N}=d_{N}+ \xi_{N}\Delta_{t_N}$ and $h_{t_N}=d_{N}$. By induction, we get
$$
	{h}_{\tk}=\sum_{j=k}^{N}\prod_{\ell=k}^{j-1}\xi_{\ell}d_j,
$$
where $\prod_{\ell=k}^{k-1}\xi_{\ell}:=1$, and
\begin{equation}\label{Dertildeh2}
\widetilde{h}_{\tk}=\sum_{j=k}^{N}\prod_{\ell=k}^{j-1}\xi_{\ell}(d_j+\xi_{j}\Delta_{t_j}^{t_k})=h_{\tk}+\sum_{j=k}^{N}\prod_{\ell=k}^{j}\xi_{\ell}\Delta_{t_j}^{t_k}.
\end{equation}
In particular, we have 
$$
\widetilde{h}_{\tkk}^{\tk}=\mathbb{E}\Big[h_{t_{k+1}}+\sum_{j=k+1}^{N}\prod_{\ell=k+1}^{j}\xi_{\ell}\Delta_{t_j}^{t_{k+1}}\Big|\mathcal{F}_{t_{k}}\Big]
=h_{t_{k+1}}+\sum_{j=k+1}^{N}\prod_{\ell=k+1}^{j}\xi_{\ell}\Delta_{t_j}^{t_{k}}.
$$
Plugging the above expression into ${}^{\scaleto{(2)}{5pt}}\!\widetilde{A}^\pm_{\tk}$ defined in (\ref{eq:A1NMGb}) and, then, plugging ${}^{\scaleto{(1)}{5pt}}\!{A}^\pm_{\tk}$, ${}^{\scaleto{(2)}{5pt}}\!\widetilde{A}^\pm_{\tk}$, and ${}^{\scaleto{(3)}{5pt}}\!\widetilde{A}^\pm_{\tk}$ into (\ref{DfnOptLbb}), we deduce that
\begin{align}
\begin{split}
     \widetilde{L}_{\tk}^{+,*}&={L}_{\tk}^{+,*}+\frac{\beta^{+}_{t_{k}}}{2\gamma_{t_{k}}}\Delta_{t_{k}}+\Big(\frac{\beta^{+}_{t_{k}}}{2\gamma_{t_{k}}}\Big)
     \sum_{j=k+1}^{N}\prod_{\ell=k+1}^{j}\xi_{\ell}\Delta_{t_j}^{t_k},\\
    \widetilde{L}_{\tk}^{-,*}&={L}_{\tk}^{-,*}-\frac{\beta^{+}_{t_{k}}}{2\gamma_{t_{k}}}\Delta_{t_{k}} -\Big(\frac{\beta^{-}_{t_{k}}}{2\gamma_{t_{k}}}
  \Big)  \sum_{j=k+1}^{N}\prod_{\ell=k+1}^{j}\xi_{\ell}\Delta_{t_j}^{t_k}.\label{eq:LtildeNMGcc}
\end{split}
\end{align}
This proves {Proposition} \ref{prop:optimalcontrol} and {Theorem} \ref{prop:optimalcontrolNMG} at once. 

\medskip
\noindent
\textbf{Step 4.} It remains to show the validity of the identities (\ref{TrickyNdInd1})-(\ref{TrickyNdInd2}). First note that the formula (\ref{Dertildeh2}) can be derived directly from the equations (\ref{eq:A1NMGb})-(\ref{eq:hNMGc}) regardless (\ref{DfnOptLbb}) holds true or not. Using (\ref{Dertildeh2}), we then have
\begin{align*}
	 \mathbb{E}\big[\widetilde{h}_{\tkk}\oned\Pd\big|\F\big]
	 &=
	 \mathbb{E}\Big[\Big(h_{\tkk}+\sum_{j=k+1}^{N}\prod_{\ell=k+1}^{j}\xi_{\ell}\Delta_{t_j}^{t_{k+1}}\Big)\oned\cd\Big|\F\Big]\\
	 &=h_{\tkk}\pi_{t_{k+1}}^{\delta}\mu^{\delta}_{c}+\sum_{j=k+1}^{N}\prod_{\ell=k+1}^{j}\xi_{\ell} \mathbb{E}\Big[\Delta_{t_j}^{t_{k+1}}\oned\cd\Big|\F\Big].
\end{align*}
Next, using the conditional independence of  $(\mathbbm{1}_{t_{k+1}}^{\pm},c^{\pm}_{t_{k+1}},p^{\pm}_{t_{k+1}})$ and $\{S_{t_{j+1}}-S_{t_{j}}\}_{j\geq{}k}$ given ${\F}$, for $j\geq{}k$, 
\begin{align*}
	\mathbb{E}\Big[\Delta_{t_j}^{t_{k+1}}\oned\cd\Big|\F\Big]&=	
	\mathbb{E}\Big[(S_{t_{j+1}}-S_{t_{j}})\oned\cd\Big|\F\Big]\\
	&=	
	\mathbb{E}\Big[S_{t_{j+1}}-S_{t_{j}}\Big|\F\Big]
	\mathbb{E}\Big[\oned\cd\Big|\F\Big]\\
	&=\Delta_{t_j}^{t_{k}}\pi_{t_{k+1}}^{\delta}\mu^{\delta}_{c}.
\end{align*}
We then deduce that $\mathbb{E}\big[\widetilde{h}_{\tkk}\oned\cd\big|\F\big]=\widetilde{h}_{\tkk}^{t_{k}}\mathbb{E}\big[\oned\cd\big|\F\big]$. The proof of (\ref{TrickyNdInd2}) is the same.

\subsection{Proof of Lemma \ref{lemma:alpha}}\label{sec:proof_PostLmm}
From the terminal condition we have $\alpha_T=-\lambda<0$. So, we only need to prove that $0<\alpha_{\tk}/\alpha_{\tkk}<1$ whenever $\alpha_{\tkk}<0$. By plugging ${}^{\scaleto{(1)}{5pt}}\!{A}^\pm_{\tk}$ defined in Eq.~(\ref{eq:A1}) into Eq.~(\ref{eq:alpha}), we can write $\alpha_{\tk}/\alpha_{\tkk}=1+N_k/D_k$.
where 
\begin{align*}
	N_k&= \pi^+_{\tkk}\pi^-_{\tkk}\alpha_{\tkk}[(\muo^+)^2\pi^+_{\tkk}(\alpha_{\tkk}\mut^--\muo^-)+(\muo^-)^2\pi^-_{\tkk}(\alpha_{\tkk}\mut^+-\muo^+)]\\
	&\quad -2{\pi^+_{\tkk}}{\pi^-_{\tkk}}\pi_\tkk(1,1)\alpha^2_{\tkk}(\muo^+\muo^-)^2,\\
	D_k&={\big[{\pi_\tkk(1,1)}\alpha_{\tkk}\muo^+\muo^-\big]^2-{\pi^+_{\tkk}}{\pi^-_{\tkk}}(\alpha_{\tkk}\mut^+-\muo^+)(\alpha_{\tkk}\mut^--\muo^-)}.
\end{align*}
Therefore, it suffices to show that $N_{k}/D_k\in{(-1,0)}$ whenever $\alpha_{\tkk}<0$. 
First, we prove that $D_k<0$ and $N_k>0$. Indeed, the first term in $D_k$ satistifies:
\[
\big[{\pi_\tkk(1,1)}\alpha_{\tkk}\muo^+\muo^-\big]^2\leq{\alpha_{\tkk}^{2}}{\pi^+_{\tkk}}{\pi^-_{\tkk}}(\muo^+\muo^-)^2
\leq {\alpha_{\tkk}^{2}}{\pi^+_{\tkk}}{\pi^-_{\tkk}}\mut^+\mut^-,
\] 
by using the facts that $\pi_\tkk(1,1)\leq\pi^+_{\tkk}\wedge\pi^-_{\tkk}$ and $\mut^\pm\geq(\muo^\pm)^2$. Combined with the second term in $D_k$, we have
\begin{equation}
D_k\leq {\pi^+_{\tkk}}{\pi^-_{\tkk}}\big[\alpha_{\tkk}(\mut^+\muo^-+\muo^+\mut^-)-\muo^+\muo^-\big]<0, \label{eq:Dneg}
\end{equation}
since $\alpha_{\tkk}<0$ by assumption and $\mu_.^\pm\geq 0$. To prove that $N_k>0$, note that, since $\alpha_{\tkk}<0$ and $\pi_\tkk(1,1)\leq\pi^+_{\tkk}\wedge\pi^-_{\tkk}$,  the first term in $N_k$ satisfies
\begin{align*}
&\pi^+_{\tkk}\pi^-_{\tkk}\alpha_{\tkk}[(\muo^+)^2\pi^+_{\tkk}(\alpha_{\tkk}\mut^--\muo^-)+(\muo^-)^2\pi^-_{\tkk}(\alpha_{\tkk}\mut^+-\muo^+)]\\
&\quad\geq \pi^+_{\tkk}\pi^-_{\tkk}\pi_\tkk(1,1)\alpha_{\tkk}[(\muo^+)^2(\alpha_{\tkk}\mut^--\muo^-)+(\muo^-)^2(\alpha_{\tkk}\mut^+-\muo^+)].
\end{align*}
Combining the formula above with the second term in $N_k$, we have
\begin{align*}
N_k&\geq \alpha_{\tkk}\pi^+_{\tkk}\pi^-_{\tkk}\pi_\tkk(1,1)\big\{\alpha_{\tkk}(\muo^+)^2[\mut^--(\muo^-)^2]\\
&\qquad\qquad\qquad\qquad\qquad\qquad\qquad+\alpha_{\tkk}(\muo^-)^2[\mut^+-(\muo^+)^2]-\muo^+\muo^-({\muo^++\muo^-})\big\}\\
&\geq -\alpha_{\tkk}\pi^+_{\tkk}\pi^-_{\tkk}\pi_\tkk(1,1)\muo^+\muo^-{(\muo^++\muo^-)}>0.
\end{align*}
The second inequality holds since $\mut^\pm\geq(\muo^\pm)^2$ and $\alpha_\tkk<0$. Thus $N_k/D_k<0$, which implies that $\alpha_{\tk}$ is always larger than $\alpha_{\tkk}$ whenever  $\alpha_{\tkk}<0$.
 Next we prove that, whenever $\alpha_{\tkk}<0$, $N_k/D_k>-1$ or, equivalently, $D_k+N_k<0$. Note {that
\begin{align}
    D_k+N_k&={\Black \pi_\tkk(1,1)}(\alpha_{\tkk}\muo^+\muo^-)^{2}(\pi(1,1)-2\pi^+_{\tkk}\pi^-_{\tkk})+\alpha_{\tkk}(\pi^+_{\tkk})^{2}\pi^-_{\tkk}(\muo^+)^2(\alpha_{\tkk}\mut^--\muo^-)\nonumber \\
    &\quad+\alpha_{\tkk}\pi^+_{\tkk}(\pi^-_{\tkk})^{2}(\muo^-)^2(\alpha_{\tkk}\mut^+-\muo^+)-{\Black \pi^+_{\tkk}}{\Black \pi^-_{\tkk}}(\alpha_{\tkk}\mut^+-\muo^+)(\alpha_{\tkk}\mut^--\muo^-). \label{TermNTTBN}
\end{align}
Let} us first see $N_k+D_k$ as a linear function of $\mut^+$ and note that
\begin{align}\nonumber
\partial_{\mut^+} (N_k+D_k)&= \pi^+_{\tkk}(\pi^-_{\tkk})^2\alpha^2_{\tkk}(\muo^-)^2-\pi^+_{\tkk}\pi^-_{\tkk}\alpha^2_{\tkk}\mut^-+\pi^+_{\tkk}\pi^-_{\tkk}\alpha_{\tkk}\muo^-\nonumber\\
&\leq \pi^+_{\tkk}(\pi^-_{\tkk})^2\alpha^2_{\tkk}\mut^--\pi^+_{\tkk}\pi^-_{\tkk}\alpha^2_{\tkk}\mut^-+\pi^+_{\tkk}\pi^-_{\tkk}\alpha_{\tkk}\muo^-\label{eq:ND1}\\
& = \pi^+_{\tkk}\pi^-_{\tkk}\alpha^2_{\tkk}\mut^-(\pi^-_{\tkk}-1)+\pi^+_{\tkk}\pi^-_{\tkk}\alpha_{\tkk}\muo^-<0\label{eq:ND2},
\end{align}
where (\ref{eq:ND1}) holds from $\mut^-\geq(\muo^-)^2$ while (\ref{eq:ND2}) holds since $\pi^-_{\tkk}<1$ and $\alpha_\tkk<0$. Thus $N_k+D_k$ decrease with $\mut^+$. Since $\mut^+\geq(\muo^+)^2$, substituting $\mut^+$ with $(\muo^+)^2$, we have
\begin{align}
\begin{split}
    D_k+N_k&\leq{\pi^+_{\tkk}}{\pi^-_{\tkk}}\alpha_{\tkk}(\muo^+)^2\big[{\pi^+_{\tkk}}(\alpha_{\tkk}\mut^--\muo^-)-{\pi_\tkk(1,1)}\alpha_{\tkk}(\muo^-)^2\big]\\
    &\quad+{\pi^+_{\tkk}}{\pi^-_{\tkk}}\alpha_{\tkk}(\muo^-)^2\big\{{\pi^-_{\tkk}}[\alpha_{\tkk}(\muo^+)^2-\muo^+]-{\pi_\tkk(1,1)}\alpha_{\tkk}(\muo^+)^2\big\}\\
    &\quad +\big[{\pi_\tkk(1,1)}\alpha_{\tkk}\muo^+\muo^-\big]^2-{\pi^+_{\tkk}}{\pi^-_{\tkk}}[\alpha_{\tkk}(\muo^+)^2-\muo^+](\alpha_{\tkk}\mut^--\muo^-).
\end{split}\label{eq:NDtemp}
\end{align}
Similarly, the RHS of (\ref{eq:NDtemp}) can be seen as a linear decreasing function of $\mut^-$ since the coefficient of $\mut^-$ is $\pi^+_{\tkk}\pi^-_{\tkk}\alpha^2_{\tkk}(\muo^+)^2(\pi^+_{\tkk}-1)+\pi^+_{\tkk}\pi^-_{\tkk}\alpha_{\tkk}\muo^+<0$. With the fact that $\mut^-\geq(\muo^-)^2$, we substitute $\mut^-$ with $(\muo^-)^2$ in the RHS of (\ref{eq:NDtemp}) and get
\begin{align}\nonumber
D_k+N_k&\leq{\pi^+_{\tkk}}{\pi^-_{\tkk}}\alpha_{\tkk}(\muo^+)^2\big\{{\pi^+_{\tkk}}[\alpha_{\tkk}(\muo^-)^2-\muo^-]-{\pi_\tkk(1,1)}\alpha_{\tkk}(\muo^-)^2\big\}\nonumber\\
    &\quad+{\pi^+_{\tkk}}{\pi^-_{\tkk}}\alpha_{\tkk}(\muo^-)^2\big\{{\pi^-_{\tkk}}[\alpha_{\tkk}(\muo^+)^2-\muo^+]-{\pi_\tkk(1,1)}\alpha_{\tkk}(\muo^+)^2\big\}\nonumber\\
    &\quad +\big[{\pi_\tkk(1,1)}\alpha_{\tkk}\muo^+\muo^-\big]^2-{\pi^+_{\tkk}}{\pi^-_{\tkk}}[\alpha_{\tkk}(\muo^+)^2-\muo^+][\alpha_{\tkk}(\muo^-)^2-\muo^-]\nonumber\\
    & = \muo^+\muo^-\big[(\pi^+_{\tkk})^2\pi^-_{\tkk}\alpha_{\tkk} \muo^+(\alpha_{\tkk} \muo^--1)-2\pi^+_{\tkk}\pi^-_{\tkk}\pi_\tkk(1,1)\alpha^2_{\tkk} \muo^+\muo^-\nonumber\\
    &\qquad\qquad+\pi^+_{\tkk}(\pi^-_{\tkk})^2\alpha_{\tkk} \muo^-(\alpha_{\tkk} \muo^+-1)+\pi^2_\tkk(1,1)\alpha^2_{\tkk} \muo^+\muo^-\nonumber\\
    &\qquad\qquad-\pi^+_{\tkk}\pi^-_{\tkk}(\alpha_{\tkk} \muo^+-1)(\alpha_{\tkk} \muo^--1)\big]\nonumber\\
    &\triangleq\muo^+\muo^-\ell(\muo^+,\muo^-)\label{eq:lmuo}
\end{align}
To prove $D_k+N_k<0$, we only need to show that $\ell(\muo^+,\muo^-)<0$. $\ell(\muo^+,\muo^-)$ is a linear function in $\muo^+$. The coefficient of $\muo^+$ is
\begin{align*}
\partial_{\muo^+} \ell(\muo^+,\muo^-)&=m(\pi_{t_{k+1}}(1,1))\times \alpha^2_{\tkk}\muo^-+\alpha_{\tkk}\pi^+_{\tkk}\pi^-_{\tkk}(1-\pi^+_{\tkk})
\end{align*}
where 
\[
	m(\pi_{t_{k+1}}(1,1)):=(\pi^+_{\tkk})^2\pi^-_{\tkk}-2\pi^+_{\tkk}\pi^-_{\tkk}\pi_\tkk(1,1)+\pi^+_{\tkk}(\pi^-_{\tkk})^2+\pi^2_\tkk(1,1)-\pi^+_{\tkk}\pi^-_{\tkk}.
\]
For now we assume that $m\left(\pi_\tkk(1,1)\right)\leq 0$ holds for any $\pi_\tkk(1,1)$ in Eq.~(\ref{eq:pioo}) and we will give the prove later. Since $\muo^-\geq 0$, we plug $0$ into $\muo^-$ and get $\partial_{\muo^+} \ell(\muo^+,\muo^-)\leq \alpha_{\tkk}\pi^+_{\tkk}\pi^-_{\tkk}(1-\pi^+_{\tkk})<0$. Thus $\ell(\muo^+,\muo^-)$ decrease with $\muo^+$. Since $\muo^+\geq 0$, we have that 
\begin{align*}
\ell(\muo^+,\muo^-)\leq \ell(0,\muo^-)&=-\pi^+_{\tkk}\pi^-_{\tkk}(1-\alpha_\tkk\muo^-)-\pi^+_{\tkk}(\pi^-_{\tkk})^2\alpha_\tkk\muo^-\\
&=-\pi^+_{\tkk}\pi^-_{\tkk}+\pi^+_{\tkk}\pi^-_{\tkk}\alpha_\tkk\muo^-(1-\pi^-_{\tkk})<0.
\end{align*}
Thus $\ell(\muo^+,\muo^-)<0$ for any $\muo^\pm\geq 0$. Immediately from Eq.~(\ref{eq:lmuo}) we have $D_k+N_k<0$, which implies that  $N_k/D_k>-1$.

It remains to show that $m\left(\pi_\tkk(1,1)\right)\leq 0$ holds for any $\pi_\tkk(1,1)$ in Eq.~(\ref{eq:pioo}). From Eq.~(\ref{eq:pioo}), we know that $({\pi^+_{\tkk}}+{\pi^-_{\tkk}}-1)\vee 0\leq{\pi_\tkk(1,1)}\leq{\pi^+_{\tkk}}\wedge{\pi^-_{\tkk}}$. Since $m\left(\pi_\tkk(1,1)\right)$ is a quadratic function of $\pi_\tkk(1,1)$ opening upwards, we only need to check that the values of $m\left(\pi_\tkk(1,1)\right)$ at two end points $({\pi^+_{\tkk}}+{\pi^-_{\tkk}}-1)\vee 0$ and ${\pi^+_{\tkk}}\wedge{\pi^-_{\tkk}}$ are non-positive. Without loss of generality, we assume $\pi^-_{\tkk}\leq\pi^+_{\tkk}$.
First we check that $m\left({\pi^-_{\tkk}}\right)\leq 0$:
\begin{align*}
m\left({\pi^-_{\tkk}}\right)&=(\pi^-_{\tkk})^2-2\pi^+_{\tkk}(\pi^-_{\tkk})^2+\pi^+_{\tkk}\pi^-_{\tkk}(\pi^+_{\tkk}+\pi^-_{\tkk}-1)\\
&=(\pi^+_{\tkk}-\pi^-_{\tkk})\pi^-_{\tkk}(\pi^+_{\tkk}-1)\leq0.
\end{align*}
Next we check that  $m\left(({\pi^+_{\tkk}}+{\pi^-_{\tkk}}-1)\vee 0\right)\leq 0$. If ${\pi^+_{\tkk}}+{\pi^-_{\tkk}}-1< 0$, we immediately have $m\left(0\right) =\pi^+_{\tkk}\pi^-_{\tkk}(\pi^+_{\tkk}+\pi^-_{\tkk}-1) \leq 0$. Otherwise, if ${\pi^+_{\tkk}}+{\pi^-_{\tkk}}-1\geq 0$,
\begin{align*}
m\left({\pi^+_{\tkk}}+{\pi^-_{\tkk}}-1\right)&= ({\pi^+_{\tkk}}+{\pi^-_{\tkk}}-1)^2-{\pi^+_{\tkk}}{\pi^-_{\tkk}}({\pi^+_{\tkk}}+{\pi^-_{\tkk}}-1)\\
&=(1-\pi^-_{\tkk})(\pi^+_{\tkk})^2+(1-\pi^-_{\tkk})(\pi^-_{\tkk}-2)\pi^+_{\tkk}+(\pi^-_{\tkk}-1)^2\\
&\triangleq n(\pi^+_{\tkk}).
\end{align*}
We can see $n(\pi^+_{\tkk})$ is a quadratic function of $\pi^+_{\tkk}$ opening upwards. By assumption $\pi^-_{\tkk}\leq\pi^+_{\tkk}\leq 1$ and $\pi^+_{\tkk}+\pi^-_{\tkk}-1\geq 0$, we have the range of $\pi^+_{\tkk}$ as
\[ \begin{cases} 
      1- \pi^-_{\tkk}\leq \pi^+_{\tkk}\leq 1& \text{when } 0\leq \pi^-_{\tkk}\leq 0.5, \\
      \pi^-_{\tkk}\leq \pi^+_{\tkk}\leq 1& \text{when } 0.5\leq \pi^-_{\tkk}\leq 1.
   \end{cases}
\]
We only need to check $n(\pi^+_{\tkk})$ is non-positive at the boundary:
\[ n(1)=(1-\pi^-_{\tkk})+(1-\pi^-_{\tkk})(\pi^-_{\tkk}-2)+(\pi^-_{\tkk}-1)^2 = 0.\]
When $0\leq \pi^-_{\tkk}\leq 0.5$:
\[ n(1-\pi^-_{\tkk})=(1-\pi^-_{\tkk})^3+(1-\pi^-_{\tkk})^2(\pi^-_{\tkk}-2)+(\pi^-_{\tkk}-1)^2 = 0.\]
When $0.5\leq \pi^-_{\tkk}\leq 1$:
\[ n(\pi^-_{\tkk})=(1-\pi^-_{\tkk})(\pi^-_{\tkk})^2+(1-\pi^-_{\tkk})(\pi^-_{\tkk}-2)\pi^-_{\tkk}+(\pi^-_{\tkk}-1)^2 \leq 0.\]
Therefore $m\left({\pi^+_{\tkk}}+{\pi^-_{\tkk}}-1\right)\leq 0$ when ${\pi^+_{\tkk}}+{\pi^-_{\tkk}}-1\geq 0$. This completes the prove for the claim $m\left(\pi_\tkk(1,1)\right)\leq 0$ holds for any $\pi_\tkk(1,1)$ in Eq.~(\ref{eq:pioo}).

\subsection{Proof of Theorem~\ref{VeriThrm1} (Verification Theorem)}\label{ProofVerifyH}
{Throughout, ${W}_{t_{i}},{I}_{t_{i}}$, for $i=k,\dots,N+1$, are  the cash holding and inventory processes resulting from {adopting an} admissible placement strategy ${L}_{t_{i}}^{\pm}$, $i=k,\dots,N$. In contrast, for $i = k+1, \dots, N+1$, {${W}_{t_{i}}^{*},{I}_{t_{i}}^{*}$ and $\widehat{W}_{t_{i}},\widehat{I}_{t_{i}}$} are respectively the resulting cash holding and inventory processes starting from initial states $W_\tk,{I_\tk}$,  when setting {$L_{t_{i}}^{\pm}={L}_{t_{i}}^{\pm,{*}}$} and $L_{t_{i}}^{\pm}=\widehat{L}_{t_{i}}^{\pm}$, for some arbitrary admissible placement strategy  $\widehat{L}_{t_{i}}^{\pm}$.  
First note that, for an arbitrary admissible placement strategy $L_{t_{i}}^{\pm}$, 
$\{v(t_i,S_{t_i},W_{t_i},I_{t_i})\}_{i=k,\dots,N+1}$ is a supermartingale since
\begin{align}\nonumber
    \mathbbm{E}\big[v(t_{i+1},S_{t_{i+1}},W_{t_{i+1}},I_{t_{i+1}})|\mathcal{F}_{t_{i}}\big]&\leq\sup_{\widehat{L}^\pm_{t_{i}}}\mathbbm{E}\big[v(t_{i+1},S_{t_{i+1}},\widehat{W}_{t_{i+1}},\widehat{I}_{t_{i+1}})|\mathcal{F}_{t_{i}}\big]\\
    &=v(t_{i},S_{t_{i}},W_{t_{i}},I_{t_{i}}).
\end{align}
The last equation follows from~(\ref{eq:dprogram}) {and {Proposition} \ref{prop:optimalcontrol}. That is, $\alpha_{\tk},h_{t_k},g_{t_k}$ in $v(\tk,s,{\mathsf{w}},i)= {\mathsf{w}}+\alpha_{\tk}i^{2}+s i+h_{\tk}i+g_{\tk}$ are picked in order for (\ref{eq:dprogram}) to hold true}. We then have that 
\begin{align}
\begin{split}
    v(t_k,\sk,W_\tk,I_\tk)&\geq \sup_{{(L^\pm_{t_i})_{k\leq i\leq N}}}\E[v(T,S_T,W_T,I_T)|\F]\\
    &=\sup_{{(L^\pm_{t_i})_{k\leq i\leq N}}}\E[W_T+S_T I_T-\lambda I_T^2|\F]\\
    &=V_\tk.\label{eq:vlarge}
    \end{split}
\end{align}
 The first equality in Eq.~(\ref{eq:vlarge}) holds because $v(T,S_T,W_T,I_T)=W_T+S_T I_T-\lambda I_T^2$ by the terminal conditions $\alpha_T=-\lambda,g_T=0,h_T=0$.

Next we prove that $v(t_k,\sk,W_\tk,I_\tk)\leq V_\tk$. To this end, {recall from { Proposition} \ref{prop:optimalcontrol} that we pick $\alpha_\tk, h_\tk$, and $g_\tk$ so that}
\[
	v(t_{i},S_{t_{i}},{{W}_{t_{i}}^{*}},{{I}_{t_{i}}^{*}})=\E[v(t_{i+1},S_{t_{i+1}},{{W}^{*}_{t_{i+1}},{I}^{*}_{t_{i+1}}})|\mathcal{F}_{t_{i}}],
\]
for all $i=k,\dots,{N}$. Hence, by induction,
\begin{align*}
	v(t_k,\sk,W_\tk,I_\tk)&=v(t_k,{\sk,{W}^{*}_\tk,{I}^{*}_\tk})\\
	&=\E[v(t_{N+1},S_{t_{N+1}},{{W}^{*}_{t_{N+1}},{I}^{*}_{t_{N+1}}})|\mathcal{F}_{t_{k}}]\\
	&=\E[{{W}^{*}_T+S_T {I}^{*}_T-\lambda ({I}^{*}_T)^2}|\F].
\end{align*}
It also trivially follows that
\[
	\E{[{W}^{*}_T+S_T {I}^{*}_T-\lambda ({I}^{*}_T)^2}|\F]\leq{}\sup_{{(L^\pm_{t_i})_{k\leq i\leq N}}}\E[W_T+S_T I_T-\lambda I_T^2|\F]=V_{t_{k}}.
\]
We then conclude that} $v(t_k,\sk,W_\tk,I_\tk)\leq{}V_{t_{k}}$, which combined with (\ref{eq:vlarge}) implies that \\$v(t_k,\sk,W_\tk,I_\tk)={}V_{t_{k}}$.

\subsection{Proof of {\Black Proposition \ref{remark:admb} (Conditions for {Positive} Spread)}}\label{sec:proof_adm}
We first prove the result under the martingale condition (\ref{MrtCnd0}).
{\Black By Eq.~(\ref{eq:Ltilde}),} we need to prove  that 
$$L_{\tk}^{+,*}+L_{\tk}^{-,*} = ({}^{\scaleto{(1)}{5pt}}\!A^+_{\tk}-{}^{\scaleto{(1)}{5pt}}\!A^-_{\tk})I_\tk+({}^{\scaleto{(2)}{5pt}}\!A^+_{\tk}{\Black-}{}^{\scaleto{(2)}{5pt}}\!A^-_{\tk})+({}^{\scaleto{(3)}{5pt}}\!A^+_{\tk}{\Black+}{}^{\scaleto{(3)}{5pt}}\!A^-_{\tk}) >0.$$
Under {the Conditions (\ref{Cnd1PosSpr})-(\ref{Cnd2PosSpr})} in Proposition \ref{remark:admb}}, it is easy to see that 
\begin{align*}
    \beta_\tk^+-\beta_\tk^-=
    {\pi^+_{\tkk}}{\pi^-_{\tkk}}\alpha_{\tkk}(\mu_{c}^+\mu_{c^2}^{-}-\mu_{c}^{-}\mu_{c^{2}}^{+})-{\pi_\tkk(1,1)}\alpha_\tkk\mu_{c}^{-}\mu_{c}^{+}(\pi_{t_{k+1}}^{-}\mu_{c}^{-}-
    \pi_{t_{k+1}}^{+}\mu_{c}^{+})
    =0.
\end{align*}
This directly implies that ${}^{\scaleto{(1)}{5pt}}\!A^+_{\tk}-{}^{\scaleto{(1)}{5pt}}\!A^-_{\tk}=0$ and ${}^{\scaleto{(2)}{5pt}}\!A^+_{\tk}-{}^{\scaleto{(2)}{5pt}}\!A^-_{\tk}=0$}.
We now proceed to show that ${}^{\scaleto{(3)}{5pt}}\!A^+_{\tk}-{}^{\scaleto{(3)}{5pt}}\!A^-_{\tk}>0$. To this end, first note that, as shown in Eq.~(\ref{eq:Dneg}) {($D_k=\gamma_{t_{k}}$)}, the denominator $\gamma_{t_{k}}$ of ${}^{\scaleto{(3)}{5pt}}\!A^+_{\tk}-{}^{\scaleto{(3)}{5pt}}\!A^-_{\tk}$ is negative.  So, it remains to show that the numerator of ${}^{\scaleto{(3)}{5pt}}\!A^+_{\tk}{\Black+}{}^{\scaleto{(3)}{5pt}}\!A^-_{\tk}$ is also negative. By Condition (\ref{Cnd3PosSpr}) in Proposition \ref{remark:admb} {(i.e., $\muoo^{\pm}=\muo^{\pm}\mu_p^\pm$ and $\muto^{\pm}=\mut^{\pm}\mu_p^\pm$), the numerator of ${}^{\scaleto{(3)}{5pt}}\!A^+_{\tk}+{}^{\scaleto{(3)}{5pt}}\!A^-_{\tk}$ can be written as}
\begin{align*}
    N{\Black({}^{\scaleto{(3)}{5pt}}\!A^+_{\tk}+{}^{\scaleto{(3)}{5pt}}\!A^-_{\tk})}
    &=\Big\{{\Black \pi^+_{\tkk}}{\Black \pi^-_{\tkk}}(\alpha_{\tkk}\mut^--\muo^-)(\muo^+-2\alpha_{\tkk}\mut^+)+2\big[\alpha_{\tkk}{\Black \pi_\tkk(1,1)}\muo^+\muo^-\big]^2\\
    &\qquad-{\Black \pi^+_{\tkk}}{\Black \pi_\tkk(1,1)}\alpha_{\tkk}(\muo^+)^2\muo^-\Big\}\mu_p^+\\
    &+\Big\{{\Black \pi^+_{\tkk}}{\Black \pi^-_{\tkk}}(\alpha_{\tkk}\mut^+-\muo^+)(\muo^--2\alpha_{\tkk}\mut^-)+2\big[\alpha_{\tkk}{\Black \pi_\tkk(1,1)}\muo^+\muo^-\big]^2\\
    &\qquad-{\Black \pi^-_{\tkk}}{\pi_\tkk(1,1)}\alpha_{\tkk}(\muo^-)^2\muo^+\Big\}\mu_p^-.
\end{align*}
We can then show that the coefficients of $\mu_p^+$ is negative. {To wit,} denote the coefficients of $\mu_p^+$ as $r(\mut^+,\mut^-)$, {which is} a linear function of $\mut^-$ with coefficient $\pi^+_{\tkk}\pi^-_{\tkk}\alpha_\tkk(\muo^+-2\alpha_\tkk\mut^+)<0$. Since $\mut^-\geq(\muo^-)^2$, we have that 
\begin{align*}
r(\mut^+,\mut^-)&\leq r(\mut^+,(\muo^-)^2)\\
& = {\pi^+_{\tkk}}{\pi^-_{\tkk}}[\alpha_{\tkk}(\muo^-)^2-\muo^-](\muo^+-2\alpha_{\tkk}\mut^+)+2\big[\alpha_{\tkk}{\pi_\tkk(1,1)}\muo^+\muo^-\big]^2\\
&\qquad-{\pi^+_{\tkk}}{\pi_\tkk(1,1)}\alpha_{\tkk}(\muo^+)^2\muo^-.
\end{align*}
{Similarly,} $r(\mut^+,(\muo^-)^2)$ is linear in $\mut^+$ with coefficient $-2\alpha_\tkk\pi^+_{\tkk}\pi^-_{\tkk}[\alpha_\tkk(\muo^-)^2-\muo^-]<0$. {Therefore,}
\begin{align*}
 r(\mut^+,(\muo^-)^2)&\leq r((\muo^+)^2,(\muo^-)^2)\\
& = {\pi^+_{\tkk}}{\pi^-_{\tkk}}[\alpha_{\tkk}(\muo^-)^2-\muo^-][\muo^+-2\alpha_{\tkk}(\muo^+)^2]+2\big[\alpha_{\tkk}{\pi_\tkk(1,1)}\muo^+\muo^-\big]^2\\
&\qquad-{\pi^+_{\tkk}}{\pi_\tkk(1,1)}\alpha_{\tkk}(\muo^+)^2\muo^-\\
&=\muo^+\muo^-\{[2\alpha^2_\tkk\pi^2_\tkk(1,1)-2\alpha^2_\tkk\pi^+_\tkk\pi^-_\tkk]\muo^+\muo^-+2\alpha_\tkk\pi^+_\tkk\pi^-_\tkk\muo^+\\
&\qquad\qquad\quad+\pi^+_\tkk\alpha_\tkk[\pi^-_\tkk\muo^--\pi_\tkk(1,1)\muo^+]-\pi^+_\tkk\pi^-_\tkk\}{\Blue .}
\end{align*}
By Eq.~(\ref{eq:pioo}) we have $\pi_\tkk(1,1)\leq\pi^+_\tkk\pi^-_\tkk$ and by Lemma~\ref{lemma:alpha} we have $\alpha_\tkk<0$, thus the summation of {the} first two terms in the {brackets} above (i.e., $[2\alpha^2_\tkk\pi^2_\tkk(1,1)-2\alpha^2_\tkk\pi^+_\tkk\pi^-_\tkk]\muo^+\muo^-+2\alpha_\tkk\pi^+_\tkk\pi^-_\tkk\muo^+$) is negative. Under the Condition (\ref{Cnd1PosSpr}), the third term in the {brackets} (i.e., $\pi^+_\tkk\alpha_\tkk[\pi^-_\tkk\muo^--\pi_\tkk(1,1)\muo^+]$) can be written as $\pi^+_\tkk\alpha_\tkk(\pi^-_\tkk-\pi_\tkk(1,1))\muo<0$. Thus the coefficients of $\mu_p^+$ in $N{({}^{\scaleto{(3)}{5pt}}\!A^+_{\tk}+{}^{\scaleto{(3)}{5pt}}\!A^-_{\tk})}$ (i.e., $r(\mut^+,\mut^-)$) is negative. Similarly the coefficients of $\mu_p^-$ is also negative. Therefore $N{({}^{\scaleto{(3)}{5pt}}\!A^+_{\tk}+{}^{\scaleto{(3)}{5pt}}\!A^-_{\tk})}<0$ and 
 {${}^{\scaleto{(3)}{5pt}}\!A^+_{\tk}+{}^{\scaleto{(3)}{5pt}}\!A^-_{\tk}>0$ .}

{The proof for the general nonmartingale case of Subsection \ref{sec:DriftAnaly} is direct since, as mentioned above, Conditions (\ref{Cnd1PosSpr})-(\ref{Cnd2PosSpr}) imply { $\beta_\tk^+-\beta_\tk^-=0$} and, by (\ref{eq:LtildeNMGbb}), we have:}
\[
	{\widetilde{L}_{\tk}^{+,*}+\widetilde{L}_{\tk}^{-,*}={L}_{\tk}^{+,*}+{L}_{\tk}^{-,*}.}
\]

 \renewcommand{\theequation}{B-\arabic{equation}}
\setcounter{equation}{0}  

\section{Proofs of Section \ref{PrpOptPlc2b}}

\subsection{Proof of Corollary~\ref{cor:spread_t}}\label{pf:spread_t}
Under the Conditions (\ref{Cnd1PosSpr})-(\ref{Cnd2PosSpr}), it is easy to check that $\beta_\tk^+=\beta_\tk^-$. From (\ref{eq:LtildeNMGbb}), we can then easily see that the optimal spreads, denoted hereafter $Sprd_\tk$, are the same under the martingale and non-martingale midprice cases. Furthermore, 
\begin{align}
\begin{split}
Sprd_{\tk}= L_{\tk}^{+,*}+L_{\tk}^{-,*} = \widetilde{L}_{\tk}^{+,*}+\widetilde{L}_{\tk}^{-,*}
= {}^{\scaleto{(3)}{5pt}}\!A^+_{\tk}{\Black+}{}^{\scaleto{(3)}{5pt}}\!A^-_{\tk},\\
\end{split}
\end{align}
which proves that the optimal spreads are independent of the inventory level and the local drifts $\{\Delta_{t_k}\}_{k=0,\dots,N}$.
If we further assume Condition (\ref{Cnd3PosSpr}) and Condition (\ref{Cnd1Pioo}), the optimal spread can be written as
\begin{align}
\begin{split}
Sprd_{\tk}
& = \dfrac{\left[\pi(\muo-2\alpha_\tkk\mut)+2\alpha_\tkk\pi(1,1)\muo^2\right](\mu_p^++\mu_p^-)}{2\left[\pi(1,1)\alpha_\tkk\muo^2-\pi(\alpha_\tkk\mut-\muo)\right]},
\end{split}
\end{align}
%
%
{where $\pi=\pi^\pm$.} We show that $Sprd_\tk$ is non-decreasing with time by checking that the difference between $Sprd_{\tk}$ and $Sprd_{t_{k-1}}$ is non-negative:
\begin{align}
\begin{split}
&Sprd_{\tk}-Sprd_{t_{k-1}}
=\dfrac{\mu_p^++\mu_p^-}{2}\cdot\dfrac{(\alpha_\tkk-\alpha_\tk)\pi\muo\left(\pi(1,1)\muo^2-\pi\mut\right)}
{\prod_{\ell=k,k+1}\left[\pi(1,1)\alpha_{t_\ell}\muo^2-\pi(\alpha_{t_\ell}\mut-\muo)\right]}.
\end{split}
\end{align}
First, we show that the denominator {is positive}. Since $\alpha_\tk$ is negative and by definition  $0\leq\pi(1,1)\leq\pi$ and  $0<\muo^2\leq\mut$, we have $\pi(1,1)\alpha_{t_{\ell}}\muo^2-\pi(\alpha_{t_{\ell}}\mut-\muo)\geq \pi\alpha_{t_{\ell}}\mut-\pi(\alpha_{t_{\ell}}\mut-\muo)=\pi\muo>0$. This shows the denominator is positive. 
The numerator is also positive since $\alpha_\tk$ is decreasing with time and $\pi(1,1)\muo^2\leq\pi\mut$. Thus, $Sprd_{\tk}-Sprd_{t_{k-1}}\geq 0$. Particularly, if $\pi(1,1)\muo^2=\pi\mut$, $Sprd_{\tk}-Sprd_{t_{k-1}}=0$. 

To show that, at a fixed time point, the optimal spreads decrease with $\pi(1,1)$, note that:
\begin{align}
\partial_{\pi(1,1)} Sprd_{\tk}&=\dfrac{\alpha_\tkk\mu_c^2(\mu_p^++\mu_p^-)\left[\pi(1,1)\alpha_\tkk\muo^2-\pi(\alpha_\tkk\mut-\muo)\right]}{\left[\pi(1,1)\alpha_\tkk\muo^2-\pi(\alpha_\tkk\mut-\muo)\right]^2}\nonumber\\
&\qquad\quad-\dfrac{\alpha_\tkk\mu_c^2\left[\pi(\muo-2\alpha_\tkk\mut)+2\alpha_\tkk\pi(1,1)\muo^2\right](\mu_p^++\mu_p^-)}{2\left[\pi(1,1)\alpha_\tkk\muo^2-\pi(\alpha_\tkk\mut-\muo)\right]^2}\nonumber\\
& = \dfrac{\alpha_\tkk\mu_c^3(\mu_p^++\mu_p^-)}{2\left[\pi(1,1)\alpha_\tkk\muo^2-\pi(\alpha_\tkk\mut-\muo)\right]^2}<0.\nonumber
\end{align}
This completes the proof of Corollary~\ref{cor:spread_t}.

\subsection{Proof of Corollary~\ref{cor:ABI}}\label{pf:ABI}
Recall 
\begin{align*}
   \widetilde{a}_\tk^* &= \sk+{}^{\scaleto{(1)}{5pt}}\!A^+_{\tk}I_\tk+{}^{\scaleto{(2)}{5pt}}\!\widetilde{A}^+_{\tk}+{}^{\scaleto{(3)}{5pt}}\!\widetilde{A}^+_{\tk},\\ 
   \widetilde{b}_\tk^* &= \sk+{}^{\scaleto{(1)}{5pt}}\!A^-_{\tk}I_\tk+{}^{\scaleto{(2)}{5pt}}\!\widetilde{A}^-_{\tk}-{}^{\scaleto{(3)}{5pt}}\!\widetilde{A}^-_{\tk}.
\end{align*}
To show that $\widetilde{a}_\tk^*$ and $\widetilde{b}_\tk^*$ are strictly decreasing with $I_\tk$, we only need to show that ${}^{\scaleto{(1)}{5pt}}\!A^\pm_{\tk}<0$. By 
{Proposition}~\ref{prop:optimalcontrol}, we have that 
${}^{\scaleto{(1)}{5pt}}\!A^\pm_{\tk}=\frac{\beta^{\pm}_{t_{k}}\alpha_{t_{k+1}}}{\gamma_{t_{k}}}$ and
\begin{align*}
	\gamma_{t_{k}}&:=\big[\pi_\tkk(1,1)\alpha_{\tkk}\muo^+\muo^-\big]^2-\pi^+_{\tkk}\pi^-_{\tkk}(\alpha_{\tkk}\mut^+-\muo^+)(\alpha_{\tkk}\mut^--\muo^-),\\
	\beta_{t_{k}}^{\pm}&:=\pi^+_{\tkk} \pi^-_{\tkk}\muo^\pm(\alpha_{\tkk}\mut^\mp-\muo^\mp)- \pi^\mp_\tkk \pi_\tkk(1,1)\alpha_\tkk\muo^\pm(\muo^\mp)^2.
\end{align*} 
Since $\pi_\tkk(1,1)\leq\pi^+_{\tkk}\wedge\pi^-_{\tkk}$ and $(\muo^\pm)^2\leq\mut^\pm$, we have
\begin{align*}
	\gamma_{t_{k}}&\leq \pi^+_{\tkk}\pi^-_{\tkk}\left\{\alpha_{\tkk}^2\left[(\muo^+)^2(\muo^-)^2-\mut^+\mut^-\right]+\alpha_\tkk(\mut^+\muo^-+\muo^+\mut^-)-\muo^+\muo^-\right\}\\
	&\leq \pi^+_{\tkk}\pi^-_{\tkk}\big[\alpha_{\tkk}(\mut^+\muo^-+\muo^+\mut^-)-\muo^+\muo^-\big]<0,\\
	\beta_{t_{k}}^{\pm}&\leq \pi^+_{\tkk}\pi^-_{\tkk}\left[\alpha_{\tkk}\mut^\mp\muo^\pm-\muo^\pm\muo^\mp-\alpha_{\tkk}\muo^\pm(\muo^\mp)^2\right]\\
	&=\pi^+_{\tkk}\pi^-_{\tkk}\left\{\alpha_{\tkk}\muo^\pm\left[\mut^\mp-(\muo^\mp)^2\right]-\muo^\pm\muo^\mp\right\}< 0,
\end{align*} 
because, by lemma~\ref{lemma:alpha}, $\alpha_\tkk$ is negative. 
Thus, ${}^{\scaleto{(1)}{5pt}}\!A^\pm_{\tk}< 0$ for any $\tk$.

\subsection{Proof of Corollary~\ref{cor:InvThreshold}}\label{pf:InvThreshold}

Under the assumptions in Corollary~\ref{cor:InvThreshold}, the optimal strategies can be written as
\begin{align}
   a_\tk^* & = \sk+\overbrace{\dfrac{\alpha_{\tkk}\muo}{\muo-\alpha_{\tkk}\mut}I_\tk+\dfrac{{\muo}-2\alpha_{\tkk}{\mut}}{2[\muo-\alpha_{\tkk}\mut]}{\mu_p} +\dfrac{ h_{\tkk}\muo}{2[\muo-\alpha_{\tkk}\mut]}}^{L_{\tk}^{+,*}}\\
       b_\tk^* & = \sk+\overbrace{\dfrac{\alpha_{\tkk}\muo}{\muo-\alpha_{\tkk}\mut}I_\tk-\dfrac{{ \muo}-2\alpha_{\tkk}{\mut}}{2[\muo-\alpha_{\tkk}\mut]}{ \mu_p}+\dfrac{ h_{\tkk}\muo}{2[\muo-\alpha_{\tkk}\mut]}}^{-L_{\tk}^{-,*}}\label{eq:bestbidCor3}
\end{align}
where 
$$\alpha_{\tk}=\alpha_{\tkk}+\dfrac{2\pi_\tkk(\alpha_{\tkk}\muo)^2}{\muo-\alpha_{\tkk}\mut},\quad h_{\tk}= h_{\tkk}+\dfrac{2\pi_\tkk\alpha_{\tkk}\muo^2 h_{\tkk}}{\muo-\alpha_{\tkk}\mut},$$
and $\pi_\tkk:=\pi_\tkk^\pm$. Since $h_T = 0$, we have $h_\tk\equiv 0.$
It's easy to check that for any time $\tk$ and penalty levels which lead to different values of $\alpha_\tk$, the optimal ask {price $a^*_\tk = S_\tk +\dfrac{\mu_{p}}{2}$} when 
{ $I_\tk = \overline{I}^+ = \dfrac{\mut\mu_p}{2\muo}$}, and { optimal bid price $b^*_\tk = S_\tk -\dfrac{\mu_{p}}{2}$} when 
{ $I_\tk =\overline{I}^- = -\dfrac{\mut\mu_p}{2\muo}$}.

First we consider the scenario where the inventory level is non-negative. When $I_\tk=0$, we can see from Eq.~(\ref{eq:bestbidCor3}) that the optimal bid price equals to $\sk-\dfrac{\muo-2\alpha_{\tkk}\mut}{2[\muo-\alpha_{\tkk}\mut]}\mu_p=\sk-\dfrac{\mu_p}{2}+\dfrac{\alpha_{\tkk}\mut}{2[\muo-\alpha_{\tkk}\mut]}\mu_p<\sk-\dfrac{\mu_p}{2}$ since $\alpha_\tkk<0$. As stated in Corollary \ref{cor:ABI}, the optimal bid is strictly decreasing with inventory level. Thus when $I_\tk\geq0$, the optimal bid price is always smaller than $\sk-\dfrac{\mu_p}{2}$. The optimal ask is also strictly decreasing with inventory level as stated in Corollary \ref{cor:ABI}. From the previous discussion we have that the optimal ask $a^*_\tk = S_\tk +\dfrac{\mu_{p}}{2}$ when $I_\tk = \overline{I}^+ = \dfrac{\mut\mu_p}{2\muo}$. Thus for $I_\tk\in[0,\overline{I}^+)$, the optimal ask price is always larger than $S_\tk +\dfrac{\mu_{p}}{2}$ and for $I_\tk>\overline{I}^+$, the optimal ask price is always smaller than $S_\tk +\dfrac{\mu_{p}}{2}$.
The prove is symmetric for the scenario where the inventory level is non-positive. This completes the proof of Corollary~\ref{cor:InvThreshold}.

\bibliographystyle{jtbnew}



\end{document}